\documentclass[preprint,12pt,authoryear]{elsarticle}
\usepackage[a4paper,total={6.5in,9.5in}]{geometry}
\usepackage{amssymb}
\usepackage{amsmath}
\usepackage{xcolor}
\usepackage{lineno,hyperref}
\usepackage{subcaption}
\journal{Journal of Fluid and Structures}

\begin{document}

\begin{frontmatter}

\title{Controlling the chaotic wake of a flapping foil by tuning its chordwise flexibility} %% Article title

\author[label1]{Chhote Lal Shah}
\affiliation[label1]{organization={Department of Aerospace Engineering},
             addressline={IIT Madras},
             city={Chennai},
             postcode={600036},
             state={Tamil Nadu},
             country={India}}

\author[label2]{Dipanjan Majumdar}
\affiliation[label2]{organization={Department of Civil and Environmental Engineering},
             addressline={Imperial College London},
             city={London SW72AZ},
             country={United Kingdom}}
             
\author[label3]{Chandan Bose}
\affiliation[label3]{organization={Aerospace Engineering, School of Metallurgy and Materials},
             addressline={University of Birmingham},
             city={Birmingham, B15 2TT},
             country={United Kingdom}}
             
\author[label1,label4]{Sunetra Sarkar}
             
\affiliation[label4]{organization={Centre for Complex Systems and Dynamics},
           addressline={IIT Madras},
             city={Chennai},
             postcode={600036},
             state={Tamil Nadu},
             country={India}}

%% Abstract
\begin{abstract}
Effects of chord-wise flexibility as an instrument to control chaotic transitions in the wake of a flexible flapping foil have been studied here using an immersed boundary method-based in-house fluid-structure-interaction solver. The ability of the flapping foil at an optimum level of flexibility to inhibit chaotic transition, otherwise encountered in a similar but rigid configuration, has been highlighted. The rigid foil manifests chaotic transition through a quasi-periodic-intermittency route at high dynamic plunge velocities; whereas, increasing the level of flexibility gradually regularises the aperiodic behaviour through a variety of interesting wake patterns. If flexibility is increased beyond an optimum level, aperiodicity sets in again and robust chaos is restored at very high flexibility levels. The mechanisms of triggering the order-to-chaos transition are different between the rigid and the high flexibility cases. Along the route to order and back to chaos, the flexible foil exhibits different flow-field behaviours, including far-wake switching, primary \& secondary vortex streets, bifurcated wakes and interactive vortices between the bifurcated wakes. The underlying interaction mechanisms of the flow-field vortices responsible for the associated dynamical signatures of the wake have been closely tracked. This study further examines the optimum propulsive performance range of the flexible flapper and investigates its connection with the periodicity/regularity of the system.
\end{abstract}

%% Keywords

\end{frontmatter}

%% Use \section commands to start a section
\section{Introduction}
\label{sec:intro}
In nature, birds and insects, aquatic mammals and fishes enhance their maneuvering capabilities and augment the flapping/swimming efficiency by optimally exploiting the flexibility of their flapping appendages~\citep{fish1999performance,combes2003flexural}. Flexible wings/fins are capable of enduring deformation, actively as well as passively, generating higher aerodynamic forces~\citep{heathcote2007flexible,michelin2009resonance,kang2011effects}. Moreover, based on the ubiquitous presence of orderly wake patterns in natural bio-propulsion systems, even at high amplitude-frequency regimes, it can be conjectured that flexibility has a key role to play in retaining the periodic nature. To understand the underlying flow physics of bio-inspired propulsion, fluid-structure interaction (FSI) dynamics of flexible filaments have been studied as a potential canonical model in the existing literature. Various FSI models, involving flapping of flags/filament like structures in viscous fluids, have been used in this connection, and they are quite popular owing to their reduced level of complexity. A comprehensive review of the experimental, analytical, and numerical investigations on the coupling between bio-inspired flexible slender structures and ambient fluid is available at~\citet{wang2022fluid}. For some of the noteworthy experimental studies, readers can refer to the studies by ~\citet{zhang2000flexible,jia2007coupling,ristroph2008anomalous,shelley2005heavy,taneda1968waving,datta1975instability,huang1995flutter,yamaguchi2000flutter,watanabe2002experimental,tang2003flutter,eloy2008aeroelastic,abderrahmane2011flapping,abderrahmane2012nonlinear}. Numerical analyses, focusing on the effect of variation in relevant systems parameters on the dynamics of the system in similar flexible systems have been reported by~\citet{zhu2002simulation,alben2008optimal,connell2007flapping,michelin2008vortex,chen2014bifurcation}. In these studies, the main concern was the nonlinear dynamical stability/bifurcation behaviour and also on identifying the existence of chaos. Most of them did not undertake any systematic analysis of the flow-field or the resulting wake patterns.

The interaction mechanisms between the key vortices, underpinning the manifestation of dynamical transitions in the flow/flexible flapper  were not resolved. Recently, wake patterns around a flexible flapper were reported~\citep{kim2019wake}, where the parametric boundaries of transition from K\'arm\'an-to-reverse K\'arm\'an-to-deflected wake were compared to those from the rigid systems. The authors also observed a complex vortex wake pattern for a rigid foil and predicted it using a vortex dipole model. The difference between the effective velocities at the leading and trailing-edges and, in turn, the sum of the leading and trailing-edge circulations were reported to be the key players behind wake transitions in such systems. However, this study was limited to periodic situations only, and no periodic-to-chaotic transition was mentioned for the flexible foil. For perfectly rigid systems on the other hand, both wake transition and associated dynamics were thoroughly investigated in the literature for different kinematics, such as pure heaving~\citep{ashraf2012oscillation,badrinath2017identifying,majumdar2020capturing}, pure pitching~\citep{godoy2008transitions,schnipper2009vortex,deng2016correlation}, and simultaneous heaving and pitching~\citep{lentink2010vortex,bose2018investigating,bose2021dynamic,majumdar2022transition}. \citet{lewin2003modelling} and \citet{lentink2003influence} were among the early investigators to notice aperiodicity in the wake of rigid flapping airfoils. \citet{ashraf2012oscillation} and \citet{majumdar2020capturing} established a quasi-periodic route to chaos in the flow-field of a heaving airfoil. \citet{badrinath2017identifying} observed the presence of an intermittency route to chaos for a heaving airfoil. \citet{bose2018investigating} studied the complex vortex interactions behind a pitching-heaving foil and reported a quasi-periodic route to chaos. These studies identified the time-delay in the formations of leading-edge vortices (LEVs) in successive cycles to be the key player in triggering aperiodicity in the near-field. Chaos in the far-field was sustained by subsequent interactions of the LEVs with the trailing-edge vortices (TEVs) through fundamental vortex interaction mechanisms. Later, \citet{bose2021dynamic} found sporadic, intermittent windows of aperiodicity accompanying the quasi-periodic transition to chaos. This dynamical state of intermittency was manifested as near-wake switching of deflected vortex streets. As the dynamic plunge velocity was increased, the frequency of the jet-switching increased, eventually paving the way for robust chaos. 

Based on the above studies, in the significantly high amplitude-frequency regime, aperiodic transition (and possible chaos) seemed unavoidable. However, in the presence of wing flexibility, the LEV interaction mechanisms (formation, growth, shedding, and subsequent interaction with the TEV) are expected to be affected, possibly strongly, altering the dynamical characteristics and the bifurcation routes. Chord-wise flexibility was shown to affect some of the well-known wake patterns in flapping, like, deflected and switched wakes. However, chaotic wakes or recovery from it were neither investigated nor encountered in this context. Control of wake deflection as well as jet-switching by tuning the flexibility of the flapper was demonstrated by~\citet{marais_thiria_wesfreid_godoy-diana_2012,shinde2014flexibility,shah2020delaying,shah2022chordwise}. Symmetry-breaking (deflection) of a reverse K\'arm\'an wake was shown to be inhibited by chord-wise flexibility~\citep{marais_thiria_wesfreid_godoy-diana_2012}.  \citet{zhu2014flexibility} reported that enhanced flexibility helped decrease vorticity near the leading-edge and gave a more streamlined shape to the body deformation profile, while it increased the local vorticity at the trailing-edge. When the former effect was dominant, symmetry-breaking was inhibited, and symmetry-breaking was triggered if the latter was dominant and wake deflection took place. Another study by \citet{zhu2014numerical} investigated the role of  ﬂexibility on optimising the self-propelling performance of a fully flexible foil. Towards that effect, various mode-shapes and periodic and aperiodic bending of the foil were studied. Though the effect of flexibility in breaking the symmetry of the wake was reported, chaotic flow-field or recovery from chaos was not mentioned. The latter is the focus of the present study, where systematic role of flexibility on the detailed dynamical transitions of the flow-field while inhibiting chaos and reinstating periodicity has been investigated. Note that the self-propulsion system used in studies by~\citet{zhu2014flexibility,zhu2014numerical} is different from the tethered system used in the present study. In quiescent flow conditions, jet-switching phenomenon was suppressed by the addition of a flexible appendage at the trailing-edge~\citep{shinde2014flexibility}. Similarly, \citet{shah2022chordwise} also established the role of tail flexibility in suppressing jet-switching. It was shown that high flexibility prevented LEV-TEV interactions by trapping the LEVs near the tail, inhibiting the formation of strong symmetry-breaking couples that dictated the phenomenon of deflection and jet-switching. 

Despite the above direction of research on the role of flexibility, the potential to use it as a mechanism to control/inhibit extreme dynamical states like robust chaos or strong aperiodicity has not been highlighted in the existing literature. Though, for rigid flapping foils, jet-switching acted as the precursor to chaos~\citep{bose2021dynamic}, any investigation on the use of flexibility to control the dynamics beyond the jet-switching regime was not reported either. Only recently, adjustment of flapping kinematics through passive pitching for a spring-mounted rigid foil having a single rotational degree-of-freedom~\citep{majumdar2023passive} has been shown to be helpful in inhibiting aperiodic transition by changing the effective angle-of-attack, even in the high amplitude-frequency range. To the best of the authors' knowledge, a systematic and detailed analysis of the route to chaotic transitions and the use of flexibility to manipulate chaos-induced irregularity in the flow-field of a fully (chord-wise) flexible model is missing in the existing literature. This is addressed in the present study. 

In the chaotic regime, the flow-field and the loads lose their long-term predictability, posing significant challenges to the design and control of bio-mimetic flapping devices. In that context, chaos is not desirable. The present  work identifies the transition onsets along flexibility levels to determine an optimal operating regime. This has been undertaken for a fully flexible heave actuated system at different dynamic heave velocities. Comparison between a rigid and the flexible configurations in terms of  flow-field patterns, dynamical states and mechanisms to aperiodic triggers  have been undertaken. The primary questions that have been  addressed here are: i) what would be the overall nonlinear dynamical transition/bifurcation route in the wake pattern as a function of flexibility? ii) what flow-field mechanism would be instrumental in initiating chaos (if that exists)? iii) does an optimal flexibility level exist that can regularise the flow-field from aperiodicity (chaos-to-order)? iv) is periodicity indeed a welcome phenomenon to ensure optimal propulsion behaviour? First, the bifurcation scenario in the flow-field for a similar but rigid foil is examined as the reference case to compare with the effects of flexibility subsequently. The parameter space for chaos having been identified, flexibility is  introduced and the bending rigidity parameter is  systemically varied. The nonlinear dynamical states  at different flexibility levels is established  from the standpoint of dynamical systems theories. The flow-field vortices' interaction mechanisms during different dynamical states is also 
 discussed. 
 \textcolor{black}{It is also important to highlight the scope of present two-dimensional (2-D) study. In low aspect ratio configurations,  3-D results could become different from 2-D counterparts  regardless of the Reynolds number~\citep{dong2006wake,arranz2022flow}. However, \citet{deng2015dynamical,sun2018three} identified the 2-D to 3-D transition boundary for rigid flapping systems using Floquet stability analysis. Note that  the kinematic parameters chosen in the present study are well inside the 2-D parametric  boundary reported by them. Also, the parameters  chosen in the present study fall into the class of high amplitude-low frequency flapping where leading-edge vortices are reported to remain stable in 3-D even at high Strouhal numbers and   show good agreement with 2-D results~\citep{ashraf2012oscillation}. Whereas, for low amplitude-high frequency kinematics, near-field wake was seen to encounter  significant span-wise perturbations due to  3-D effects~\citep{visbal2009high}.}  

The rest of the paper is arranged in the following sections: computational methodology of the structure and flow solvers, along with convergence tests and validation studies, have been discussed in section \ref{sec:method}. The parametric space is outlined in section~\ref{sec:parameters}. The dynamical transition in the flow-field of a rigid foil has been reported in section~\ref{sec:rigid_dynamical_transition}. The effect of flexibility on the dynamical transitions and the corresponding wake patterns have been presented in section~\ref{sec:flexible_dynamical_transition}. 
% The structural envelopes of the flexible foil concerning the dynamical transitions in the flow-field behaviour are discussed in section~\ref{sec:flapping_mode_shape}.
Subsequently, section~\ref{sec:propulsive_analysis} highlights the role of flexibility on the overall propulsive characteristics. Finally, the salient outcomes of the present study and the conclusions have been given in section~\ref{sec:conclusions}.

\section{Numerical methodology} \label{sec:method}
A chord-wise flexible foil of chord length $L$ has been considered to model the two-dimensional cross-section of a flexible wing, as shown in Fig.~\ref{structure_heaving}(a). The thickness to chord length ratio ($t_h/L$) of the foil was chosen to be  $0.02$, and the chord-based Reynolds number was $\displaystyle Re = \frac{U_{\infty}L}{\nu} = 300$ ($U_{\infty}$ and $\nu$ denote the free-stream velocity and kinematic viscosity, respectively). The choice of a $Re=300$ stemmed from the fact that most natural flyers and small unmanned and micro air vehicles (MAVs), which were the primary motivations of this study, are characterized by a low $Re$ range~\citep{mueller1985low,shyy2008aerodynamics,ellington1996leading,ellington1999novel} (of approximately $O(10)$ to $O(10^3)$)~\citep{swartz2008aeromechanics}. \textcolor{black}{Such a choice also gives confidence to 2D simulations and to their consistency with 3D results, especially at the high amplitude-low frequency flapping regime as mentioned in the Introduction.} The flow-field around the flapping foil is governed by the incompressible Navier-Stokes (N-S) equations. To that end, a discrete forcing IBM-based in-house N-S solver~\citep{majumdar2020capturing} has been used to simulate the flow-field. Numerical details of the structural and flow solvers and the fluid-structure coupling have been presented in the following.

\subsection{Structural governing equations}
\begin{figure}
		\begin{subfigure}{.4\textwidth}
			\centering
			\includegraphics[scale=0.22]{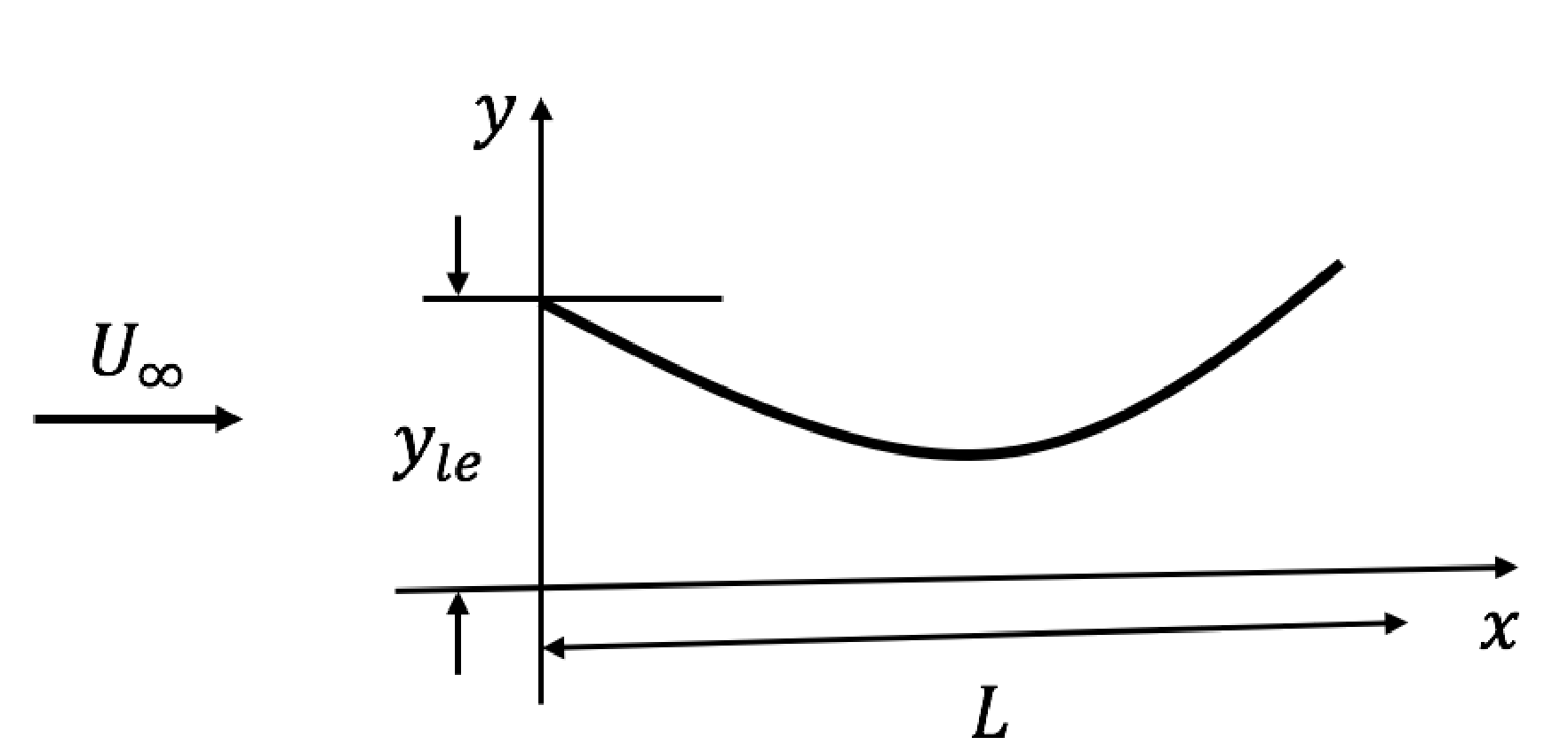}
			\caption{}
			\label{structure}
		\end{subfigure}%
		\begin{subfigure}{.6\textwidth}
			\centering
			\includegraphics[scale=0.22]{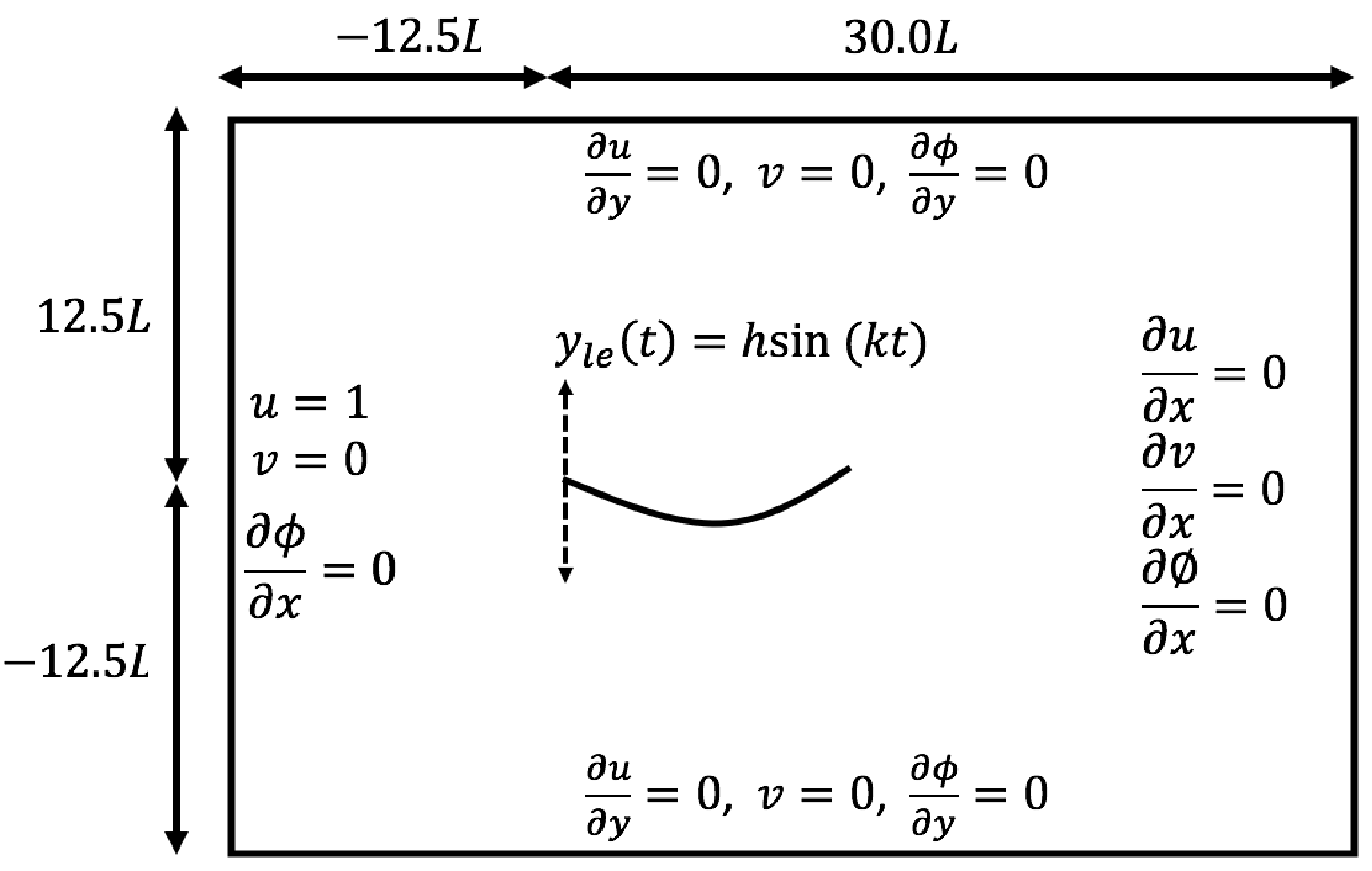}
			\caption{}
			\label{Boundary_final}
		\end{subfigure}
		\caption{(a) A schematic view of the flexible foil structure (b) the computational domain and the boundary conditions (not to scale).}
		\label{structure_heaving}
	\end{figure}

A sinusoidal heaving motion (Eq.~\ref{Eq:heave_motion}) has been imposed at the leading-edge of the foil (Fig.~\ref{structure_heaving}), and the rest of the body was allowed to oscillate passively;
	\begin{equation}
	    y_{le}(t)=h\sin(\kappa t).
	\label{Eq:heave_motion}
	\end{equation}
Here, $y_{le}(t)$ is the instantaneous position of the leading-edge of a foil in the non-dimensional form, $t$ being the time non-dimensionalised by $L/U_{\infty}$; $h=A_0/L$ and $\kappa=2\pi f_s L/U_{\infty}$ denote the non-dimensional heaving amplitude and reduced frequency, respectively, where $f_s$ is the heave frequency in $Hz$. Throughout this study, all the quantities have been presented in their non-dimensionalised forms, considering $L$ and $U_{\infty}$ to be the reference length and velocity scales, respectively.
The flexible foil was modelled as an inextensible filament, and the structural governing equation in non-dimensional form is given by
	\begin{equation}
		\beta\frac{\partial^2\mathbf{X}}{\partial t^2}=\frac{\partial}{\partial s}\left(T_s\frac{\partial\mathbf{X}}{\partial s}\right)-\frac{\partial^2}{\partial s^2}\left(\gamma \frac{\partial^2\mathbf{X}}{\partial s^2}\right)+\mathbf{F},
		\label{nonfilament}
	\end{equation}
where $s$ denotes the arc length, $\mathbf{X}=\left(X\left(s,t\right),Y\left(s,t\right)\right)$ is the position of the filament, $T_s$ is the tension coefficient along the filament, $\gamma$ is the bending rigidity, and $\mathbf{F}$ indicates the fluid force acting on the filament. These non-dimensional quantities are obtained by using characteristic scales as $\rho_f U_{\infty}^2$ for the fluid force, $\rho_f U_{\infty}^2 L$ for the tension coefficient, and $\rho_f U_{\infty}^2 L^3$ for the bending rigidity. The mass ratio is defined as $\displaystyle \beta=\frac{\rho_s t_h}{\rho_f L}$, $\rho_s$ and $\rho_f$ are the solid and fluid density, respectively. In order to maintain a constant length of the foil, the inextensibility condition has been implemented as
	\begin{equation}
		\frac{\partial \mathbf{X}}{\partial s}\cdot \frac{\partial \mathbf{X}}{\partial s} = 1.
		\label{inext}
	\end{equation}
In Eq.~\ref{nonfilament}, $\gamma$ is assumed to be constant, and $T_s$ is a function of both $s$ and $t$. Hence, $T_s$ is determined by substituting the inextensibility constraint (Eq.~\ref{inext}), in Eq.~\ref{nonfilament}. This results in a Poisson equation for $T_s$ as given below:
	\begin{equation}
		\frac{\partial \mathbf{X}}{\partial s}\cdot \frac{\partial^2}{\partial s^2}\left(T_s\frac{\partial \mathbf{X}}{\partial s}\right) = \frac{\beta}{2}\frac{\partial^2}{\partial t^2}\left(\frac{\partial \mathbf{X}}{\partial s}\cdot \frac{\partial \mathbf{X}}{\partial s}\right)-\beta \frac{\partial^2\mathbf{X}}{\partial t\partial s}\cdot \frac{\partial^2\mathbf{X}}{\partial t\partial s}-\frac{\partial \mathbf{X}}{\partial s}\cdot \frac{\partial}{\partial s}\left(\mathbf{F}_b+\mathbf{F}\right),
		\label{poissonT}
	\end{equation}
where $\displaystyle \mathbf{F}_b=-\frac{\partial^2}{\partial s^2}\left(\gamma \frac{\partial^2\mathbf{X}}{\partial s^2}\right)$ denotes the bending force. For more details of the formulation of the above equations, one can refer to the work of~\citet{huang2007simulation}. Note that the non-dimensional reference scales used in this study are different from those used by~\citet{huang2007simulation}.

At $t = 0$, the leading-edge of the foil lies at the origin, and the rest of the body aligns in the horizontal direction without any velocity. The boundary conditions for the foil are as follows; at free end $(s=L)$,
	\begin{equation}
		T_s=0, \hspace{0.1in} \frac{\partial^2 \mathbf{X}}{\partial s^2}=\left(0,0\right), \hspace{0.1in} \frac{\partial^3 \mathbf{X}}{\partial s^3}=\left(0,0\right),
		\label{BC1}
	\end{equation}
	and at fixed end $(s=0)$,
	\begin{equation}
		\mathbf{X}=\mathbf{X}_0, \hspace{0.1in}  \frac{\partial \mathbf{X}}{\partial s}=\left(1,0\right),
		\label{BC2}
	\end{equation}
where $\mathbf{X_0}=(0,y_{le}(t))$ is the position of the leading-edge. Equations~\ref{nonfilament} and \ref{poissonT} were solved using a finite difference technique~\citep{huang2007simulation,shah2022chordwise}. 
	
\subsection{Flow solver}
A momentum forcing ($\mathbf{f}$) has been applied throughout the solid domain to enforce the no slip-no penetration boundary condition exactly on the solid boundary in the present IBM framework. A source/sink term ($q$) has been added to the continuity equation (Eq.~\ref{ref:eq4}) in order to ensure rigorous mass conservation across the immersed boundary. Therefore, the flow governing equations take the form as
	\begin{equation} \label{ref:eq3}
		\frac{\partial \textbf{u}}{\partial t}+\nabla \cdot \left(\textbf{uu}\right)=-\nabla p+\frac{1}{Re}\nabla^2\textbf{u}+\textbf{f},
	\end{equation} 
	\begin{equation} \label{ref:eq4}
		\nabla \cdot \textbf{u}-q=0.
	\end{equation} 
Here, $\mathbf{u} = {u, v}$ denotes the flow velocity vector non-dimensionalised by $U_\infty$, $p$ is the pressure non-dimensionalised by $\rho_f U_\infty^2$, and Reynolds number $Re = \displaystyle \frac{U_\infty L}{\nu}$. Equations~\ref{ref:eq3} and~\ref{ref:eq4} were solved on a background Eulerian grid using a finite volume-based semi-implicit fractional step method. The diffusion term was discretised using the second-order Crank-Nicolson method, and Adams-Bashforth discretisation was used for the convection term. At every time step, the velocity was corrected using a pseudo-pressure correction term ($\phi$). Note that, for a grid cell entirely in the fluid domain, which has none of the cells faces inside the solid domain, $\mathbf{f}$ and $q$ were considered to be zero. For grid locations inside the solid domain, they take non-zero values. Please refer to~\citet{kim2001immersed,lee2015discrete,majumdar2020capturing} for further details about the flow solver. The computationally intensive parts of the flow solver were run in parallel using the OpenMP and OpenAcc techniques to enhance the simulation speed~\citep{shah2019performance,sundar2024massive}.

The fluid forces acting on each solid element were computed using the following equation~\citep{lee2011sources}
	\begin{equation} \label{Force_calc}
		\mathbf{F} = \int_{\Delta V_i}  \left(\frac{\partial \mathbf{u}}{\partial t} + \nabla.\left(\mathbf{uu}\right) \right) \,dV - \int_{\Delta V_i} \mathbf{f} \,dV,
	\end{equation}
where $\Delta V_i$ denotes the control volume of each solid element. The present formulation requires the size of the Eulerian grid near the solid domain to be much smaller than a solid element ensuring a sufficient number of fluid cells within a solid element, minimising the error involved in computing the volume integrals. The overall lift and drag coefficients, $C_L$ and $C_D$, respectively, were evaluated by performing the integration (see Eq.~\ref{Force_calc}) over the entire solid domain. Further details can be found in the studies of~\citet{lee2015discrete,shah2022chordwise}.

The present FSI framework consists of a partitioned weak coupling strategy. In this approach, the flow governing equations are solved to determine the flow-field variables around the body at every time step, and the aerodynamic loads acting on the body are evaluated. These load values are supplied to the structural solver to compute the position/shape of the body for the next time step, and the flow-field is subsequently solved with the modified position/shape of the structure. At every time step, the flow and the structural solvers exchange information in a staggered manner.
	
\subsection{Convergence study}

\begin{figure}[b!]
		\begin{subfigure}{.5\textwidth}
			\centering
			\includegraphics[scale=0.25]{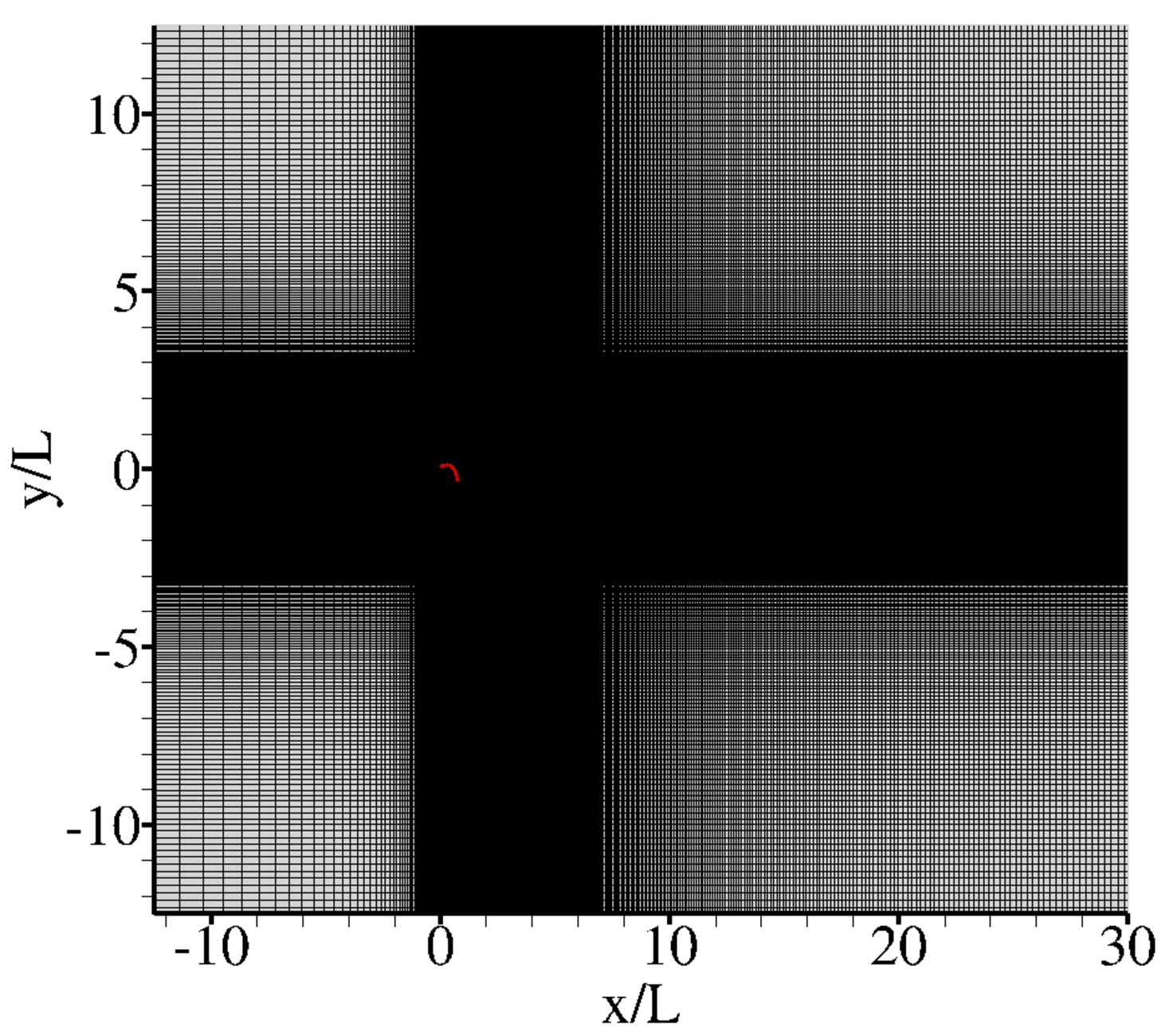}
			\caption{}
			\label{mesh}
		\end{subfigure}%
		\begin{subfigure}{.5\textwidth}
			\centering
			\includegraphics[scale=0.25]{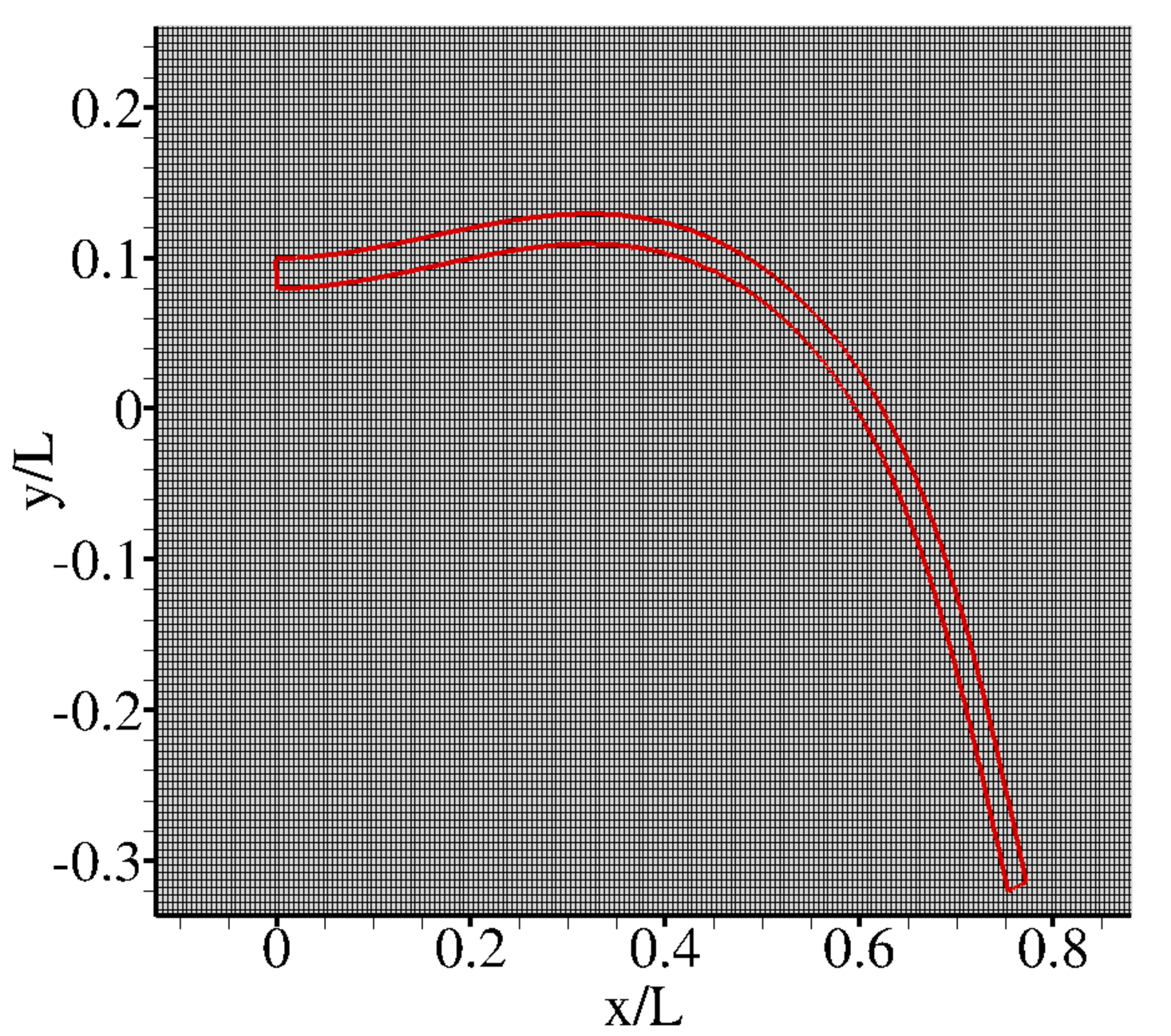}
			\caption{}
			\label{Mesh_final_zoomed}
		\end{subfigure}
		\begin{subfigure}{1.0\textwidth}
			\centering
			\includegraphics[scale=0.25]{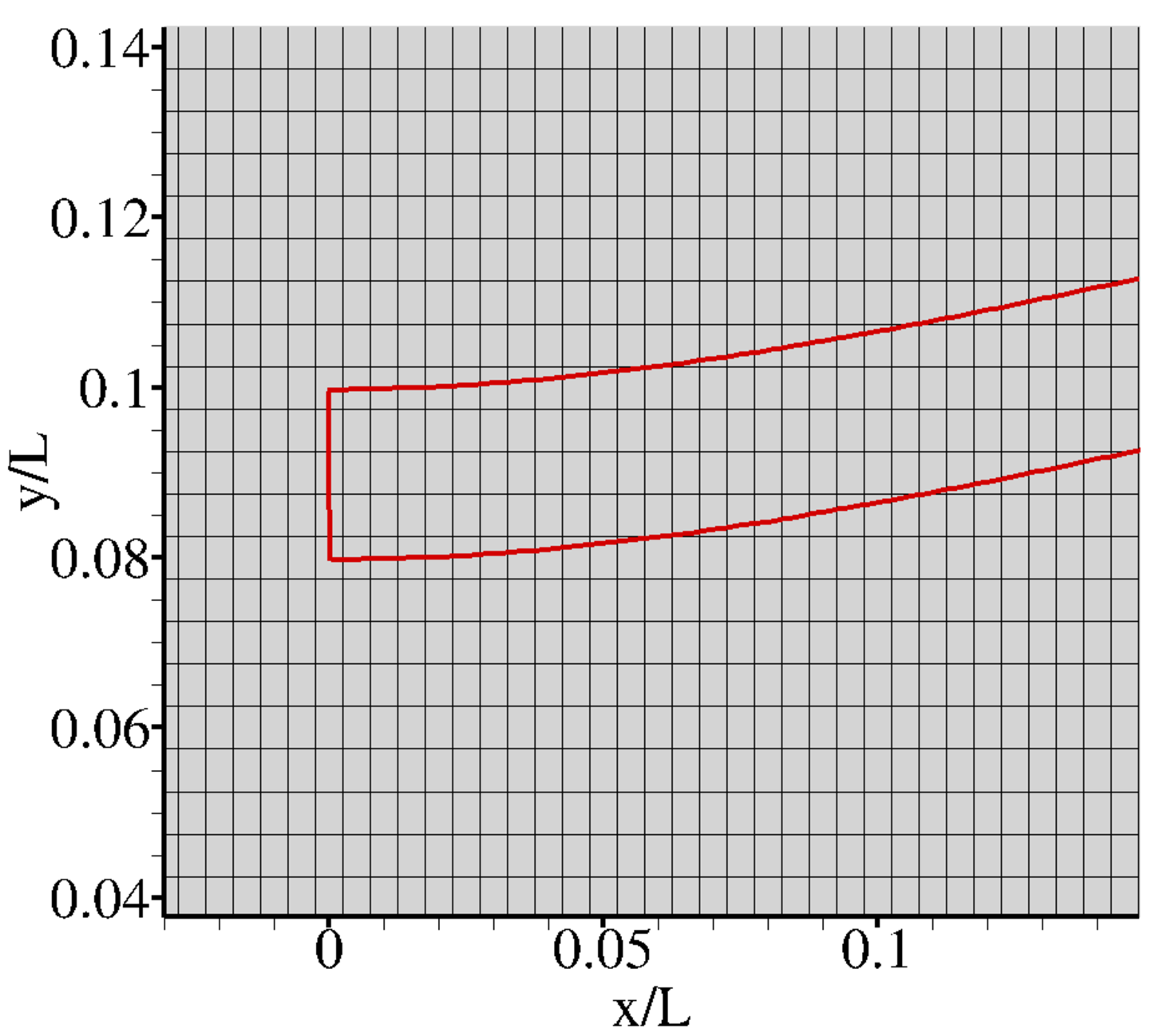}
			\caption{}
			\label{Zoomed_inset}
		\end{subfigure}
		\caption{Meshing strategy: (a) Cartesian grid, (b) zoomed section of the background grid around the flexible foil, and (c) uniform mesh grid near the body.}
		\label{mesh_zoom}
	\end{figure}
\begin{figure}
		\centering
		\includegraphics[scale=0.2]{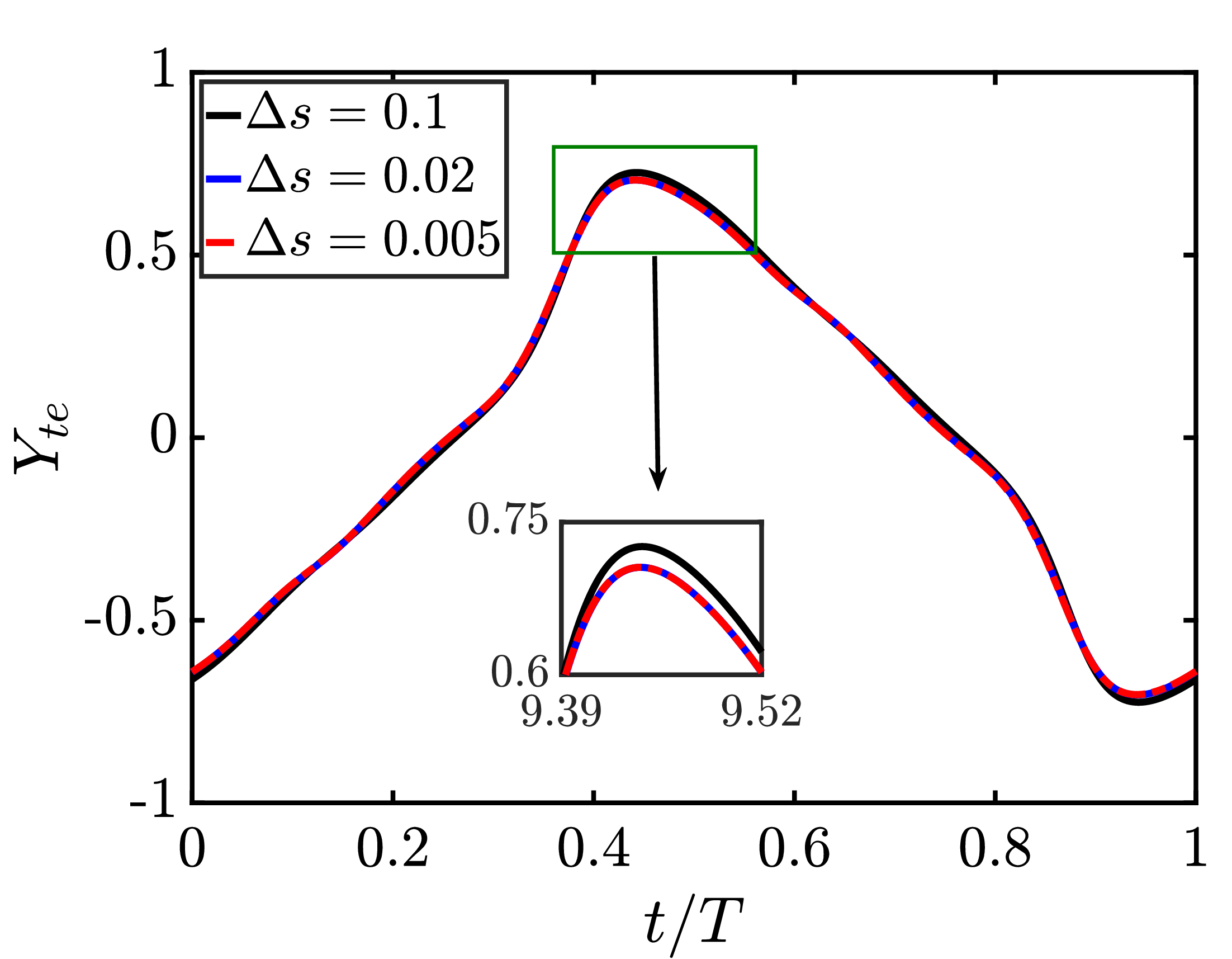}
		\vspace{6pt}
		\caption{Convergence study to select the optimum structural element length $(\Delta s)$ to discretise flexible foil.}
		\label{Nodal_covergence}
	\end{figure}

Schematic representations of the rectangular computational domain and the non-uniform structured Cartesian mesh used in this study have been shown in Figs.~\ref{structure_heaving}(b) and \ref{mesh_zoom}, respectively. The mesh size was uniform in the near-body region and then gradually stretched towards the outer boundaries. The size of the flow domain was chosen to be sufficiently large so that the boundary effects could be redundant. A Dirichlet-type boundary condition for velocity was applied at the inlet, a Neumann-type boundary condition was employed at the outlet, and a slip boundary condition was implemented at the upper and lower boundaries. A Neumann-type boundary condition was applied for the pressure-correction ($\phi$) at all the boundaries. A total of $1040$ Lagrangian markers were used to represent the solid surface. The optimum size ($\Delta s$) of the discretised element for the flexible foil structure to solve the structural equations has been selected after performing a convergence study with the following parametric combination:  $h=0.575$, $\kappa = 4.0$, $\gamma=0.5$, $\beta=1.0$, and $Re=300$. Three different sizes of the structural element, $\Delta s = 0.1, \, 0.02, \, \& \, 0.005$ have been considered. The structural element length independence test was performed with $\Delta x=\Delta y=0.005$ and $\Delta t = 0.0002$. The corresponding time histories of the free-end tip deflection of the flexible foil ($Y_{te}$) were compared in Fig.~\ref{Nodal_covergence}. The results obtained for $\Delta s = 0.02$ and $0.005$ were seen to be in very good agreement with each other. Therefore, the element size for the discretisation of the flexible foil structure was considered to be $\Delta s = 0.02$ for the rest of the simulations in this work.
	
	\begin{figure}[t!]
		\begin{subfigure}{.5\textwidth}
			\centering
			\includegraphics[scale=0.15]{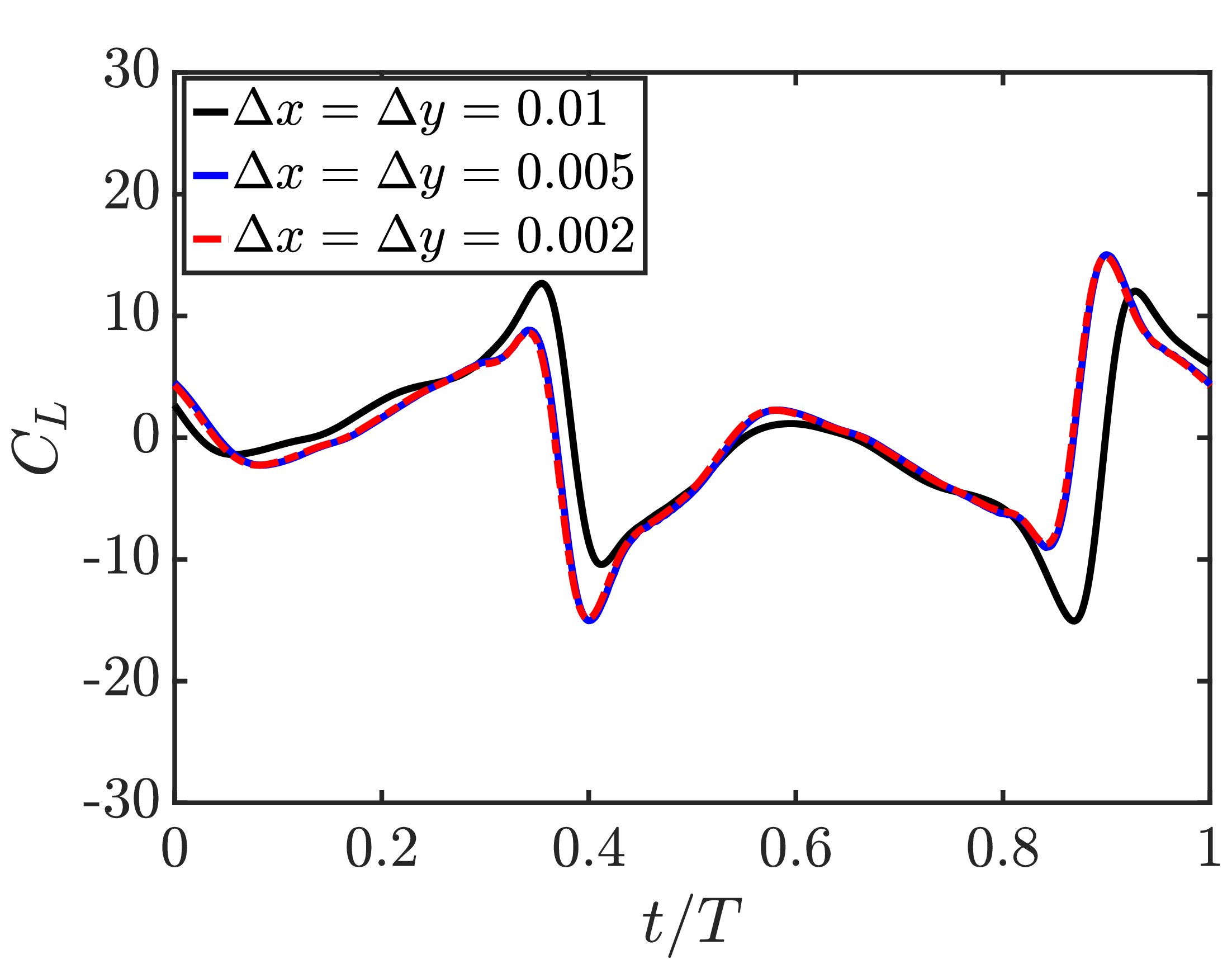}
			\caption{}
			\label{CL_grid_convergence}
		\end{subfigure}%
		\begin{subfigure}{.5\textwidth}
			\centering
			\includegraphics[scale=0.15]{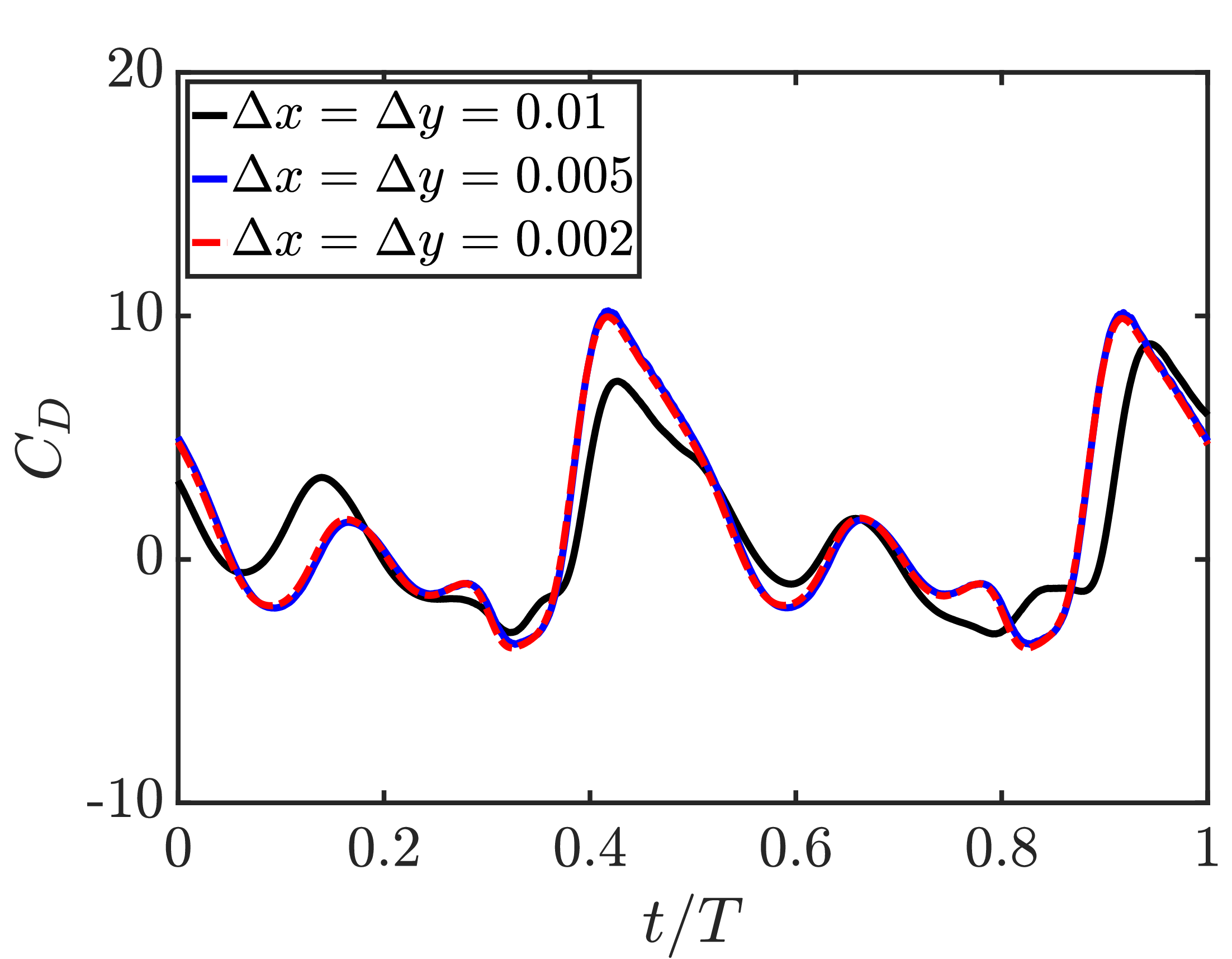}
			\caption{}
			\label{CD_grid_convergence}
		\end{subfigure}
		\caption{Grid convergence study: (a) lift coefficient, and (b) drag coefficient}.
		\label{CL_CD_grid_convergence}
	\end{figure}
	
	\begin{figure}[t!]
		\begin{subfigure}{.5\textwidth}
			\centering
			\includegraphics[scale=0.15]{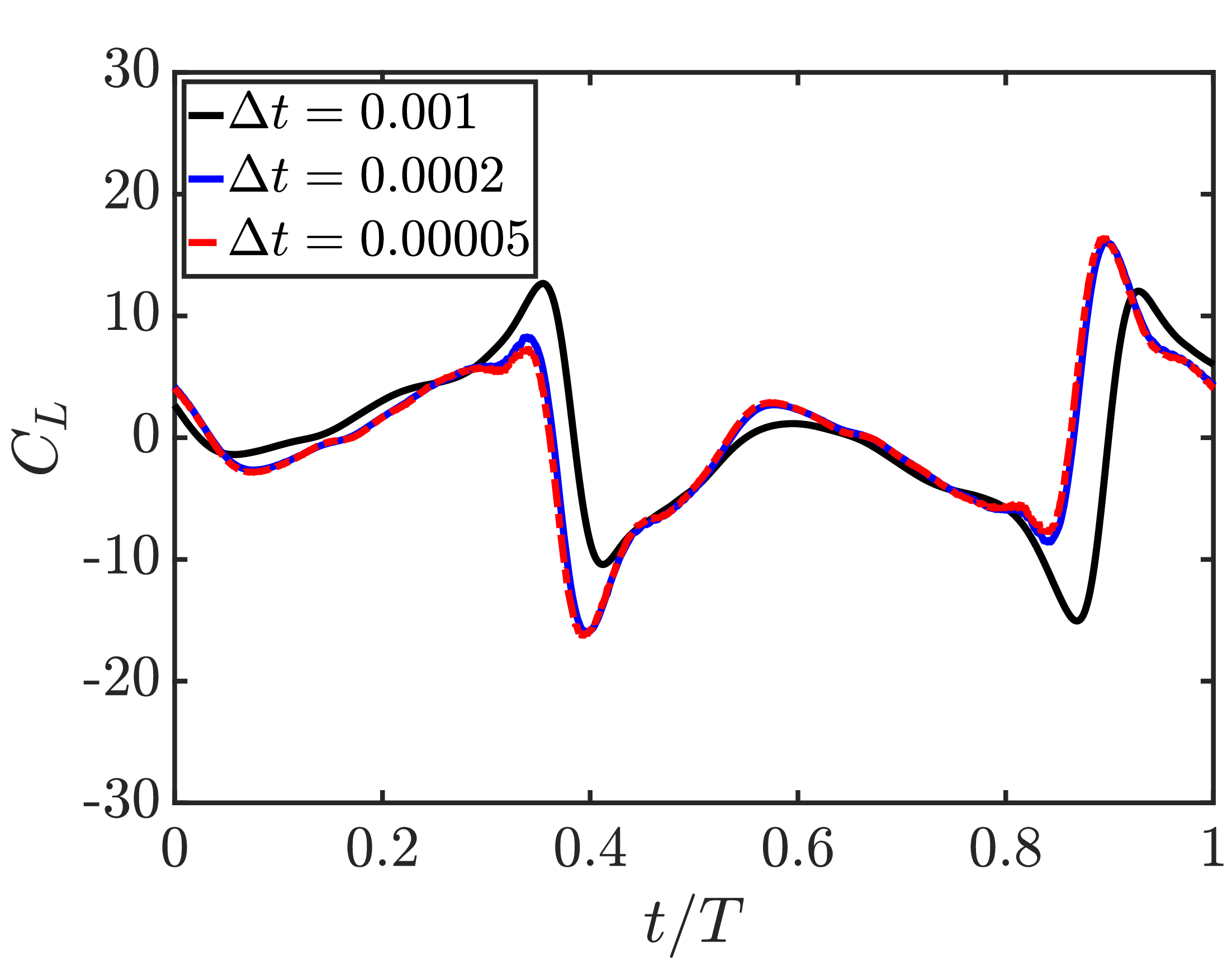}
			\caption{}
			\label{CL_time_convergence}
		\end{subfigure}%
		\begin{subfigure}{.5\textwidth}
			\centering
			\includegraphics[scale=0.15]{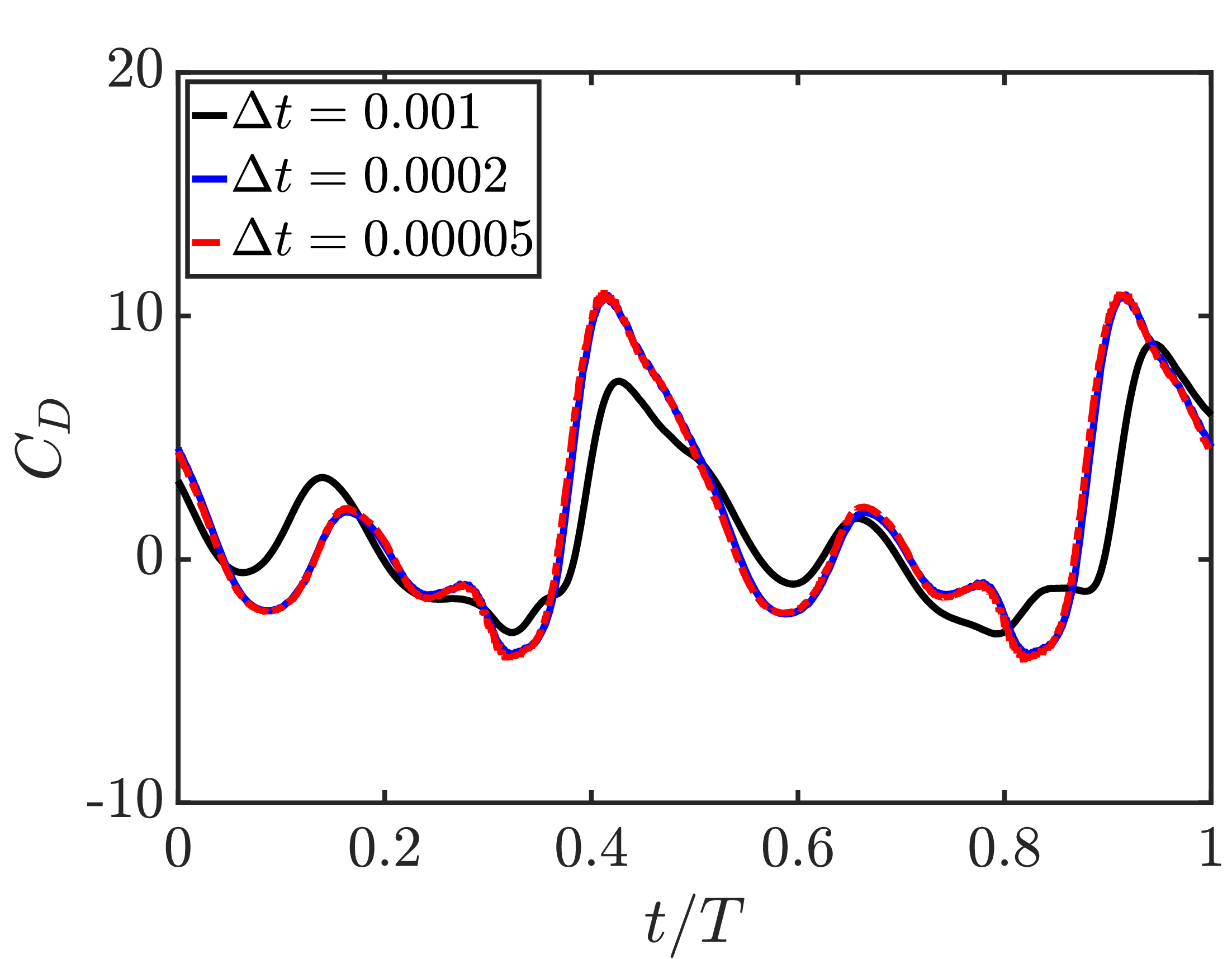}
			\caption{}
			\label{CD_time_convergence}
		\end{subfigure}
		\caption{Time step convergence study: (a) lift coefficient, and (b) drag coefficient}.
		\label{CL_CD_time_convergence}
	\end{figure}
 
The flow domain was discretised into a mesh consisting of $N_x \times N_y$ grid points, where $N_x$ and $N_y$ were the number of Cartesian grid points along the stream-wise and transverse directions, respectively. Three different meshes having minimum grid size of $\Delta x=\Delta y=0.01$, $\Delta x=\Delta y=0.005$, and $\Delta x=\Delta y=0.002$ were considered. The corresponding total number of grid points ($N_x \times N_y$) are: ($594 \times 620$), ($869 \times 988$), and ($1570 \times 1978$). The structural element size and time-step were kept at $\Delta s = 0.02$ and $\Delta t = 0.0002$, respectively. The aerodynamic load coefficients obtained from these three grids were compared in Fig.~\ref{CL_CD_grid_convergence}. Time evolution of $C_L$ and $C_D$ obtained from minimum grid size of $\Delta x=\Delta y=0.005$ matched closely with those obtained from $\Delta x=\Delta y=0.002$. Therefore, the minimum of grid size of $\Delta x=\Delta y=0.005$ was chosen for all further simulations.
	
To test the time step independence, three different time steps were considered: $\Delta t=0.001$, $0.0002$, and $0.00005$ at $\Delta s = 0.02$ and $\Delta x=\Delta y=0.005$. Comparison of corresponding $C_L$ and $C_D$ time histories in Figs.~\ref{CL_CD_time_convergence}(a) and \ref{CL_CD_time_convergence}(b), respectively, indicate very good agreements of the results obtained for $\Delta t=0.0002$ and $0.00005$. Therefore, the time-step of $\Delta t=0.0002$ was selected for the rest of the computations.
	
\subsection{Validation of the FSI solver}

\begin{figure}[b!]
		\begin{subfigure}{.5\textwidth}
			\centering
			\includegraphics[scale=0.18]{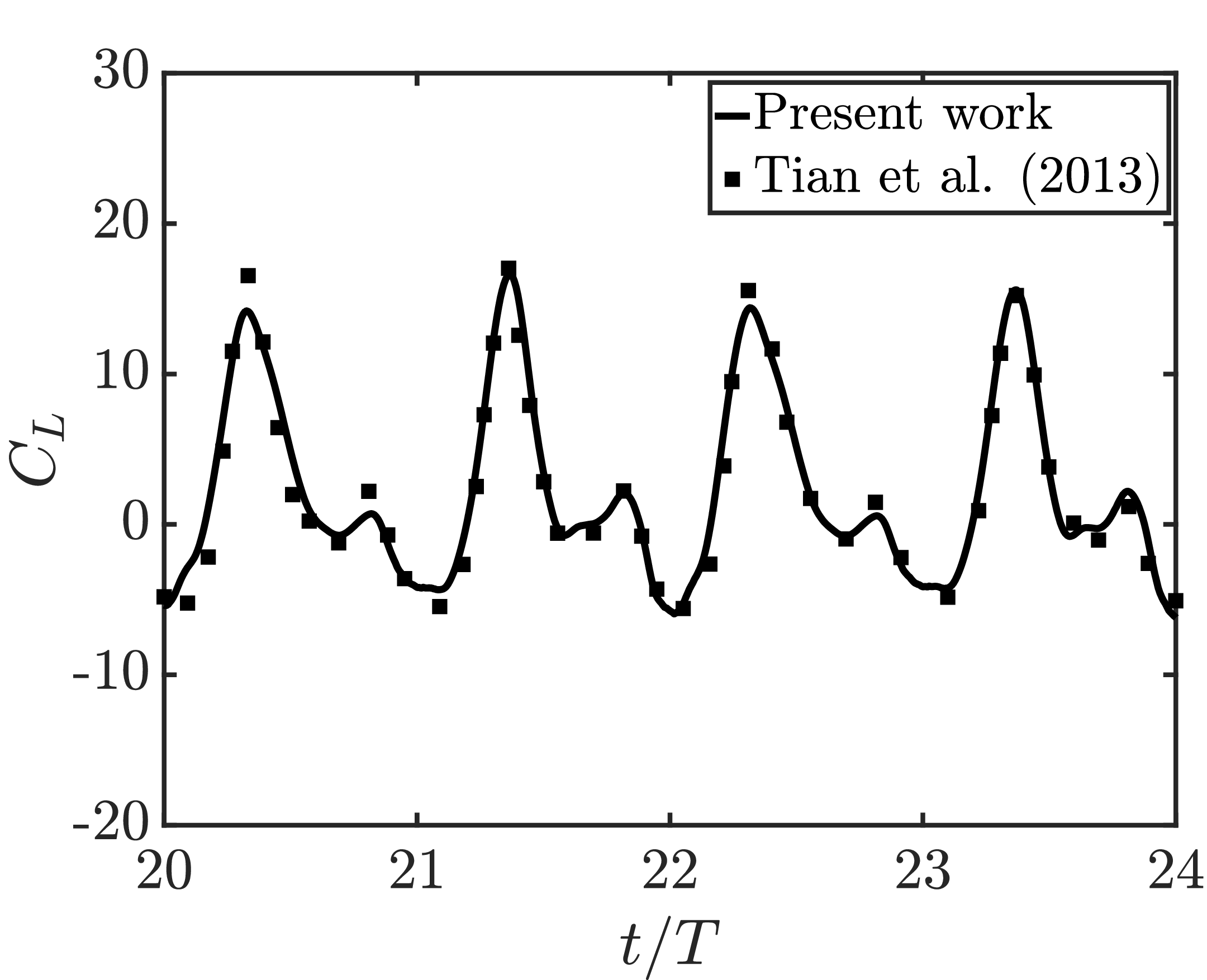}
			\caption{}
			\label{Validation_cl}
		\end{subfigure}%
		\begin{subfigure}{.5\textwidth}
			\centering
			\includegraphics[scale=0.18]{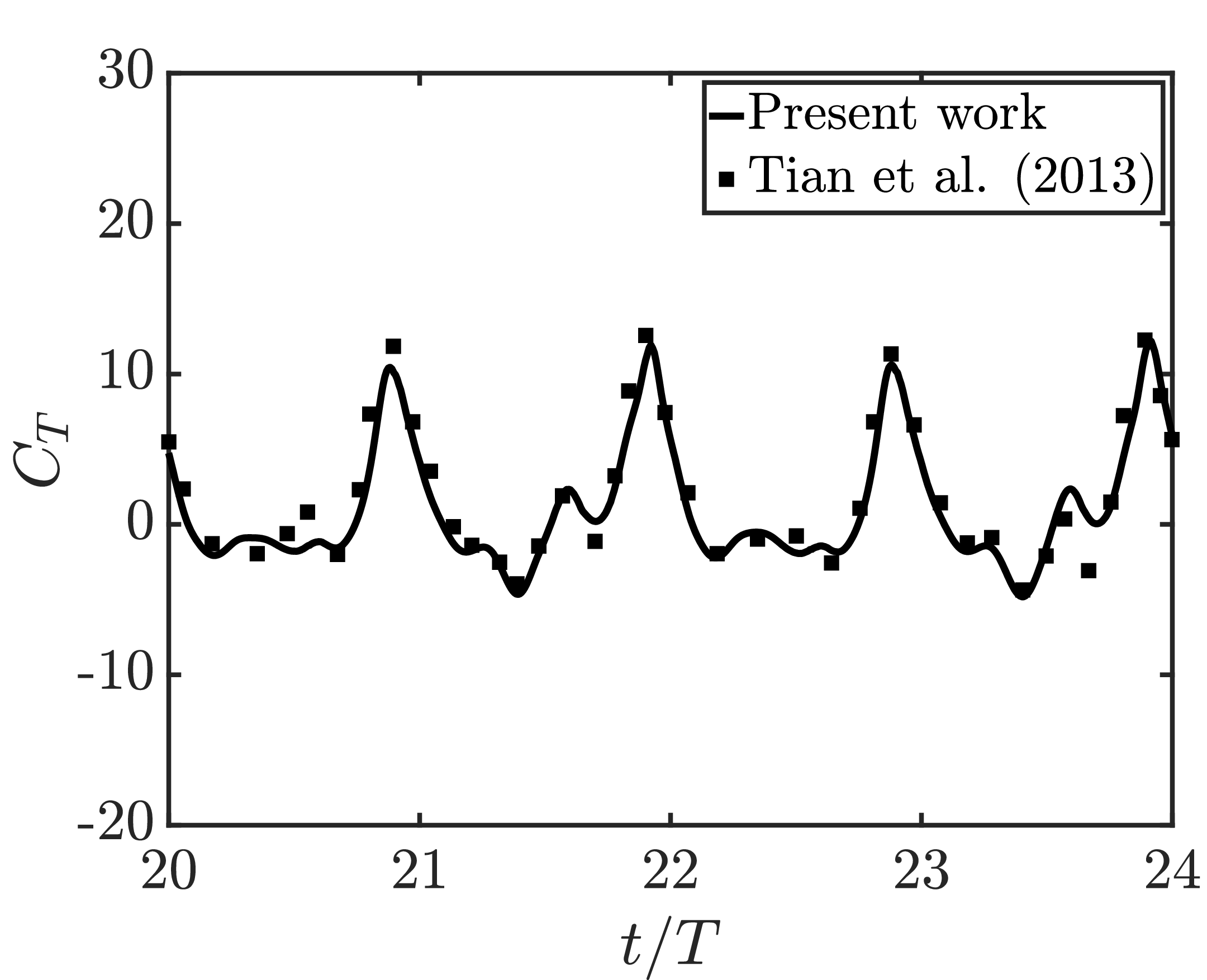}
			\caption{}
			\label{Validation_cd}
		\end{subfigure}
		\caption{Validation of the present solver set-up with \citet{tian2013force} in terms of the aerodynamic loads: (a) lift coefficient and (b) thrust coefficient time histories of a flexible flapper undergoing FSI with the surrounding free stream at $Re=100$, stroke plane angle $=45^o$, mass ratio $=1.0$, $L=1.0$, velocity ratio $U_r=0.4$ and frequency ratio $\omega^*=0.6$ (symbols follow the same definitions as given by~\citet{tian2013force}).}
		\label{Validation_cl_cd}
	\end{figure}

An extensive qualitative and quantitative validation of the flow solver has been presented by the authors in their recent study~\citep{majumdar2020capturing, shah2022chordwise,chatterjee2024energy}, thus has not been repeated here for the sake of brevity. Instead, the coupled FSI solver was validated by simulating the interaction of a flexible flapper with the surrounding free-stream at $Re=100$, whose clamped leading-edge performed a combined translational and rotational motion. The parameters considered for the validation study were: stroke plane angle $45^o$, velocity ratio $0.4$, mass ratio $1.0$, and frequency ratio $0.6$. For more details regarding the parametric definition and problem set-up, one can refer to~\citet{tian2013force}. The results obtained from the present FSI simulations have been compared with the results reported by \citet{tian2013force} in terms of the lift coefficient $(C_L)$ and thrust coefficient $(C_T)$ in Figs.~\ref{Validation_cl} and \ref{Validation_cd}, respectively. Evidently, the present results corroborated very well the reference results from the literature, thus establishing the accuracy of the current in-house FSI solver.

\section{Parameter space}\label{sec:parameters}
The unsteady flow-field around a rigid foil configuration was investigated first by varying the heave amplitude as the bifurcation parameter. The analysis was performed at a fixed reduced frequency $\kappa=4$ and a constant Reynolds number, $Re=300$. The results have been presented for $h = 0.25,\,0.375,\,0.4125,\,0.45,\,0.47$, and $0.575$, corresponding to $\kappa h = 1.0,\,1.5,\,1.65,\,1.8,\,1.88$, and $2.3$, respectively. As will be shown in the following, the flow-field around the rigid foil transitions to robust chaos at high $\kappa h$. The present study investigates the ability of flexibility in a chord-wise flexible system to inhibit chaos and bring it back to a proper periodic state. Therefore, the parametric regime in which the rigid foil showed sustained chaos has been chosen as the benchmark case to study the role of flexibility. To that end, a series of simulations were carried out, gradually decreasing the non-dimensional bending rigidity in the range of $6.0$ to $0.3$ ($\gamma = 6.0,\,3.0,\,1.0,\,0.70,\,0.50,\,0.40,\,0.39,\,0.38,\,0.37$ and $0.30$), keeping the dynamic plunge velocity fixed at $\kappa h=2.3$.

The dynamical states of the flow-fields were established using various nonlinear time series analysis measures and a detailed description of these tools can be found in our previous studies \citep{badrinath2017identifying,shah2022chordwise}. Classical topological measures from the dynamical systems theory, such as Lyapunov exponents or correlation dimensions, require significantly long-time history data to obtain converged values which may involve prohibitively large computational time and cost. Time series tools such as reconstructed phase portrait~\citep{kennel1992determining}, frequency spectra, Morlet wavelet transform~\citep{grossmann1990reading} and Recurrence plots~\citep{marwan2007recurrence} can conclusively establish the dynamics even from relatively shorter time histories. The reconstructed phase space~\citep{kennel1992determining} and Morlet wavelet transform~\citep{grossmann1990reading} results have been presented here. Recurrence-based results have been given in the Supplementary document. The flow simulations were carried out for a sufficiently long time (about 160 flapping cycles), and the first 20 cycles were discarded to ensure that the transient effects were not included in the analysis.

To derive the phase space and characterize the attractors, data time histories ($C_D$) were used for phase-space reconstruction using the delay embedding theorem~\citep{Takens1981} to compute the optimum time delay and the embedding dimension~\citep{fraser1986independent,kennel1992determining,juniper2018sensitivity}. The details of the formulation are available in our earlier studies~\citep{badrinath2017identifying,bose2021dynamic} and hence are not repeated here.  A closed-loop orbit in the reconstructed phase portrait is a signature of periodic dynamics, whereas a toroidal attractor indicates quasi-periodicity. However, reconstructed phase portraits may not be very effective in capturing the dynamical state of intermittency. For that, Morlet wavelet transforms and recurrence plots are more useful. The Morlet wavelet transform~\citep{grossmann1990reading} shows the time-frequency behaviour of a signal on a scalogram plot, in which a narrow crisp frequency band indicates periodicity; the presence of incommensurate frequency bands designates quasi-periodic dynamics, and a broad-banded frequency spectrum is signature of chaos.

\section{Rigid foil: A baseline overview of dynamical transitions}
\label{sec:rigid_dynamical_transition}

\begin{figure}[t!]
    	\centering
    		\includegraphics[scale=0.25]{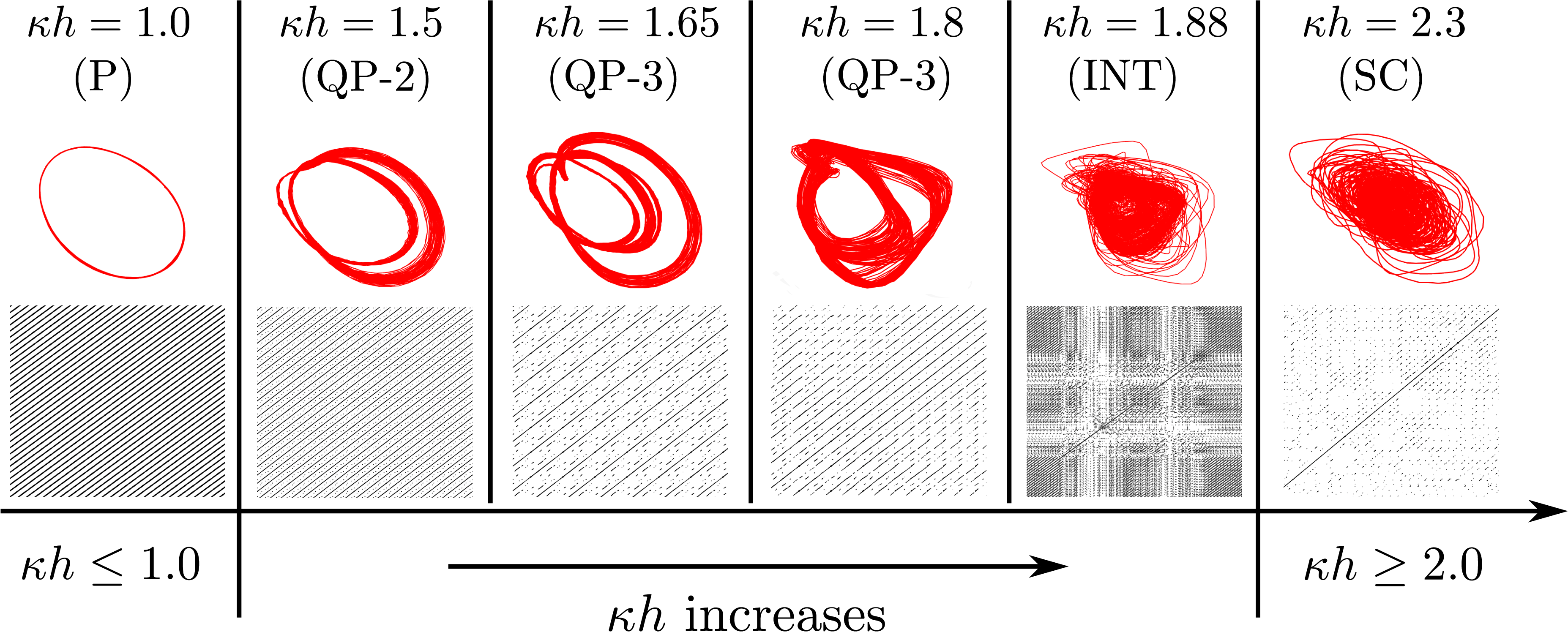}
    		\caption{Route to chaos for rigid foil: schematic representation of successive bifurcations leading to sustained chaos when $\kappa h$ is increased as the control parameter. P = periodic state; QP-2 = two-frequency quasi-periodicity; QP-3 = three-frequency quasi-periodicity; INT = intermittency; SC = sustained chaos. Reconstructed phase-portraits and recurrence plots of the $C_D$ data are given in the first and second rows, respectively. A detailed explanation of the recurrence plots is given in Section~1.1 of the supplementary document.}
    		\label{Rigid_route_choas}
    \end{figure}

The rigid system's  
%went to chaos with increasing $\kappa h$,  following a quasi-periodic and intermittency route, which is similar to our earlier study~\citep{bose2021dynamic}. Hence the description of the dynamical transition for this baseline case will be taken up briefly here. The 
transition route has been schematically shown in Fig.~\ref{Rigid_route_choas} along with reconstructed phase space and recurrence plots for the representative cases.  
    \begin{figure}[b!]
		\begin{subfigure}{.33\textwidth}
			\centering
			\includegraphics[scale=0.14]{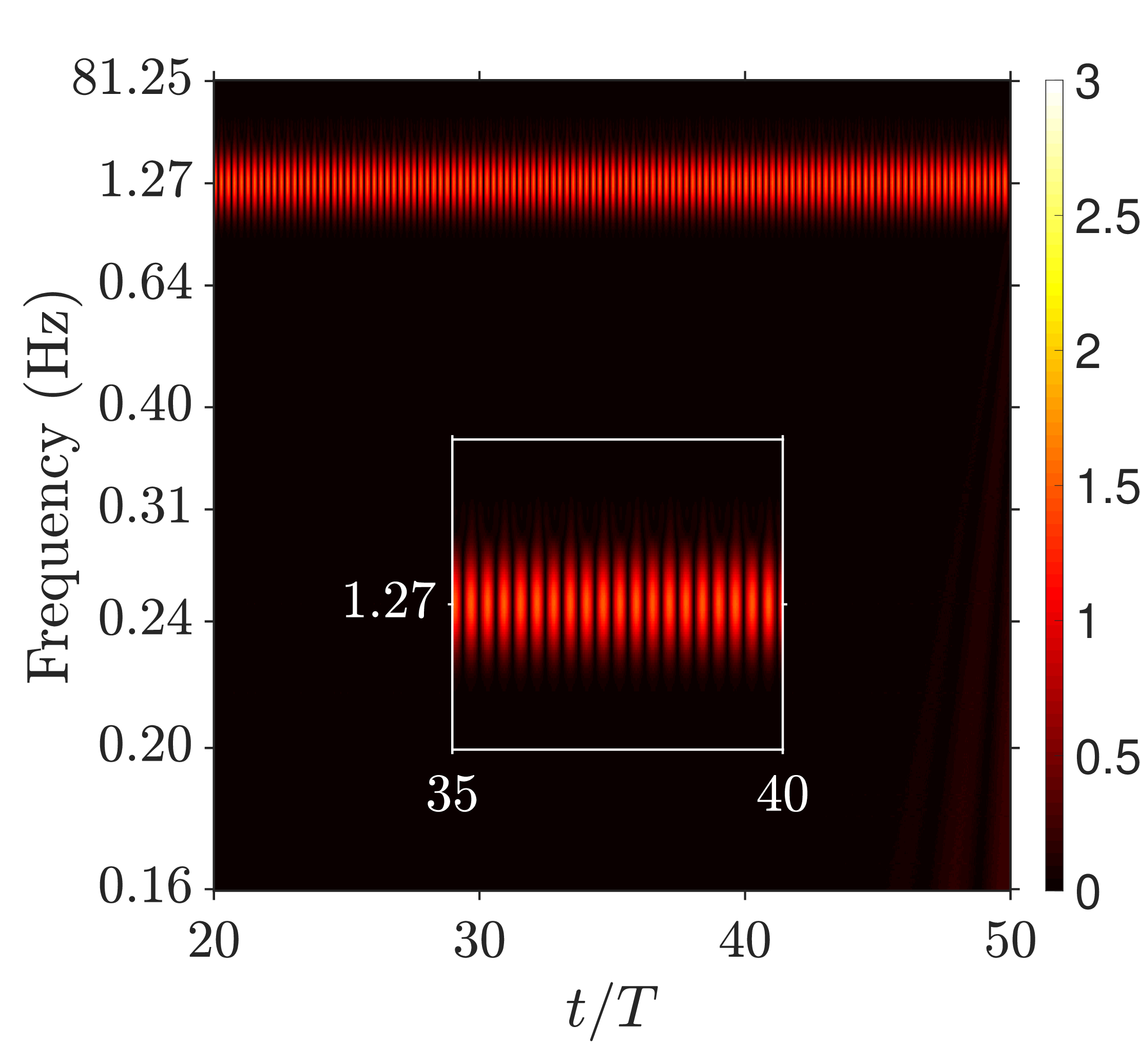}
			\caption{$\kappa h=1.0$ (P)}
			\label{wavelet_kh=1.0}
		\end{subfigure}
		%\vspace{6pt}
		\begin{subfigure}{.33\textwidth}
			\centering
			\includegraphics[scale=0.14]{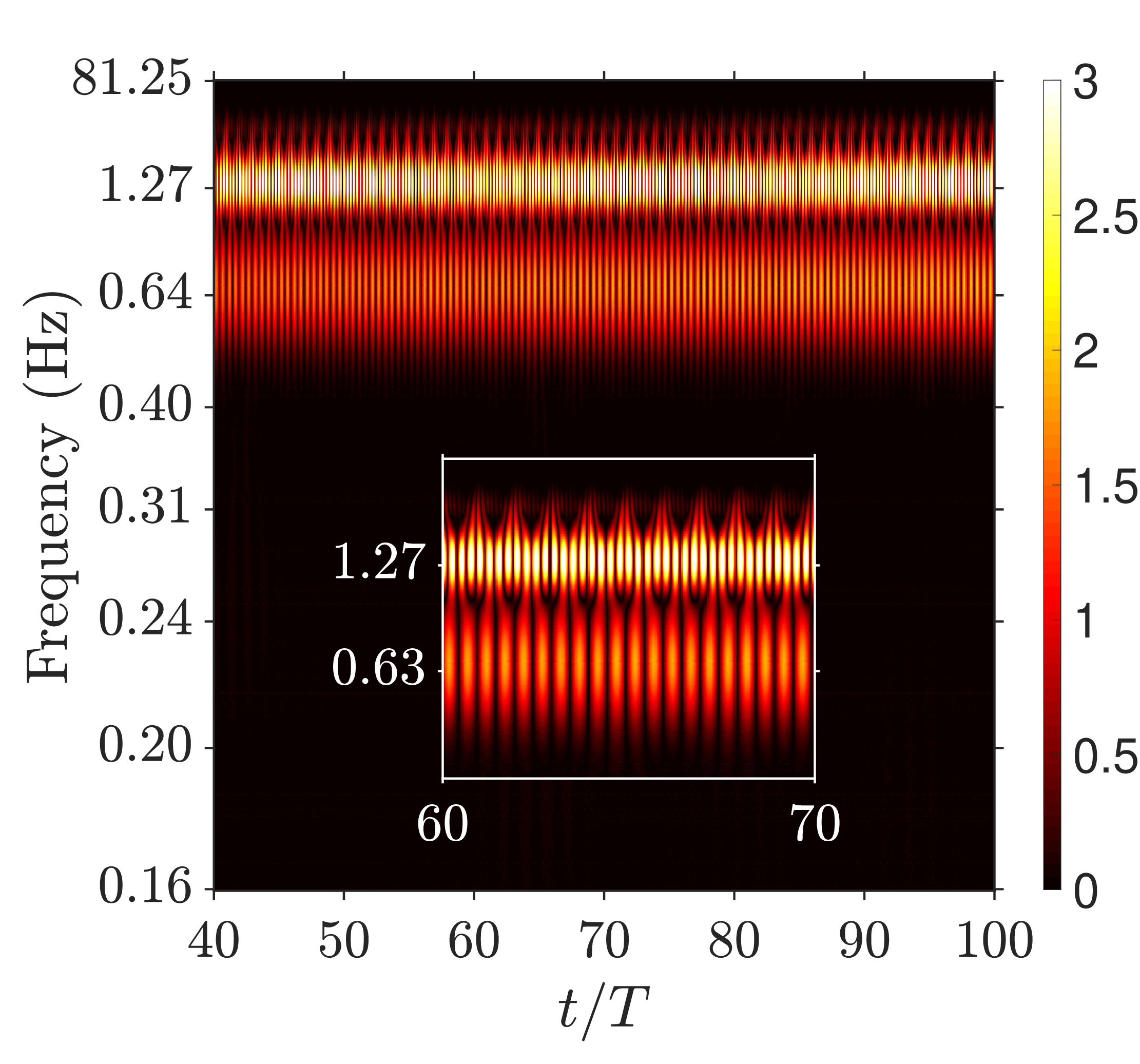}
			\caption{$\kappa h=1.5$ (QP-2)}
			\label{wavelet_kh=1.5}
		\end{subfigure}%
		%\vspace{6pt}
		\begin{subfigure}{.33\textwidth}
			\centering
			\includegraphics[scale=0.14]{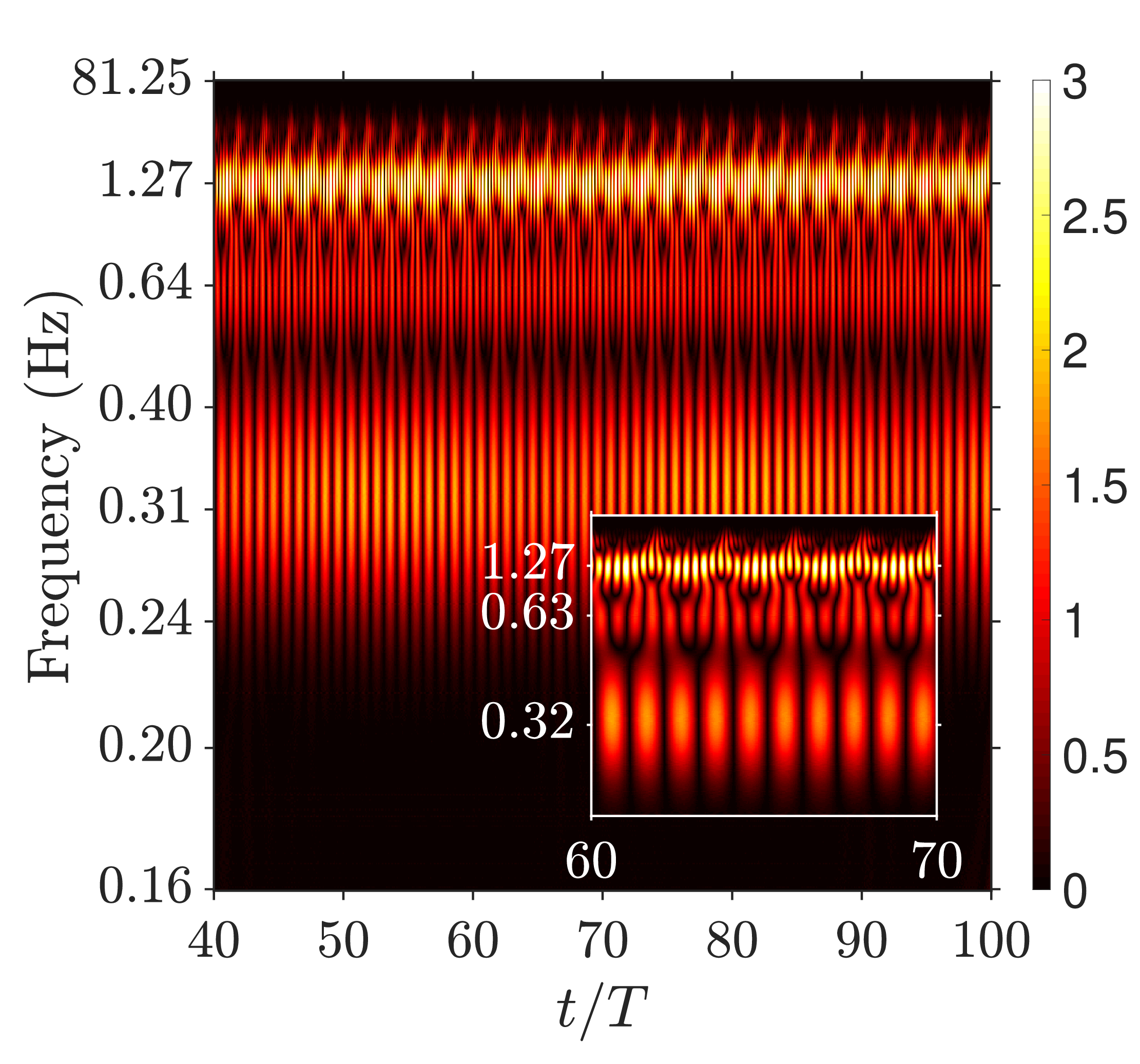}
			\caption{$\kappa h=1.65$ (QP-3)}
			\label{wavelet_kh=1.65}
		\end{subfigure}
		%\vspace{6pt}
		\begin{subfigure}{.33\textwidth}
			\centering
			\includegraphics[scale=0.14]{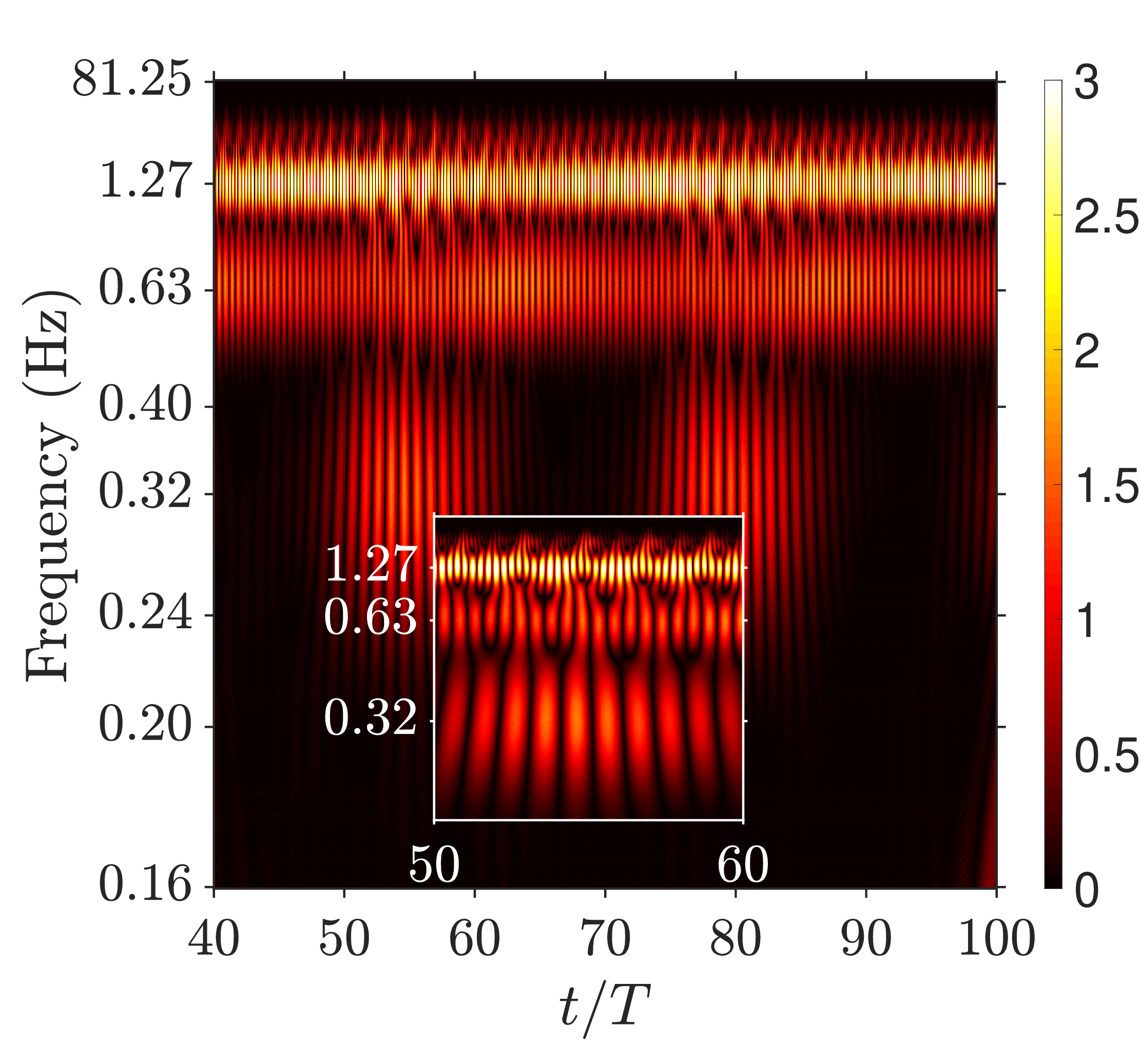}
			\caption{$\kappa h=1.8$ (QP-3)}
			\label{wavelet_kh=1.8}
		\end{subfigure}
		%\vspace{6pt}
		\begin{subfigure}{.33\textwidth}
			\centering
			\includegraphics[scale=0.14]{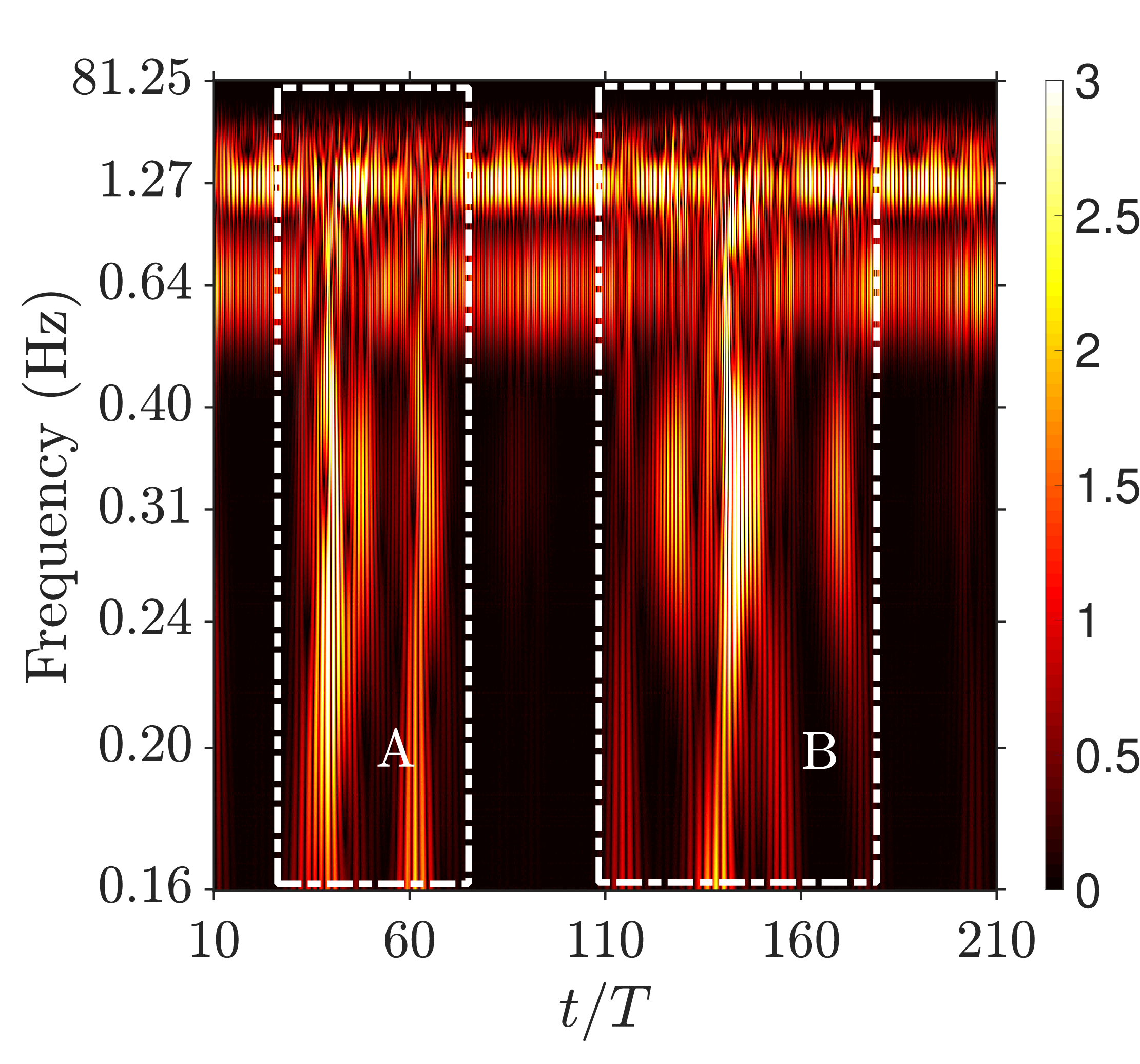}
			\caption{$\kappa h=1.88$ (INT)}
			\label{wavelet_kh=1.88}
		\end{subfigure}%
		%\vspace{6pt}
		\begin{subfigure}{.33\textwidth}
			\centering
			\includegraphics[scale=0.14]{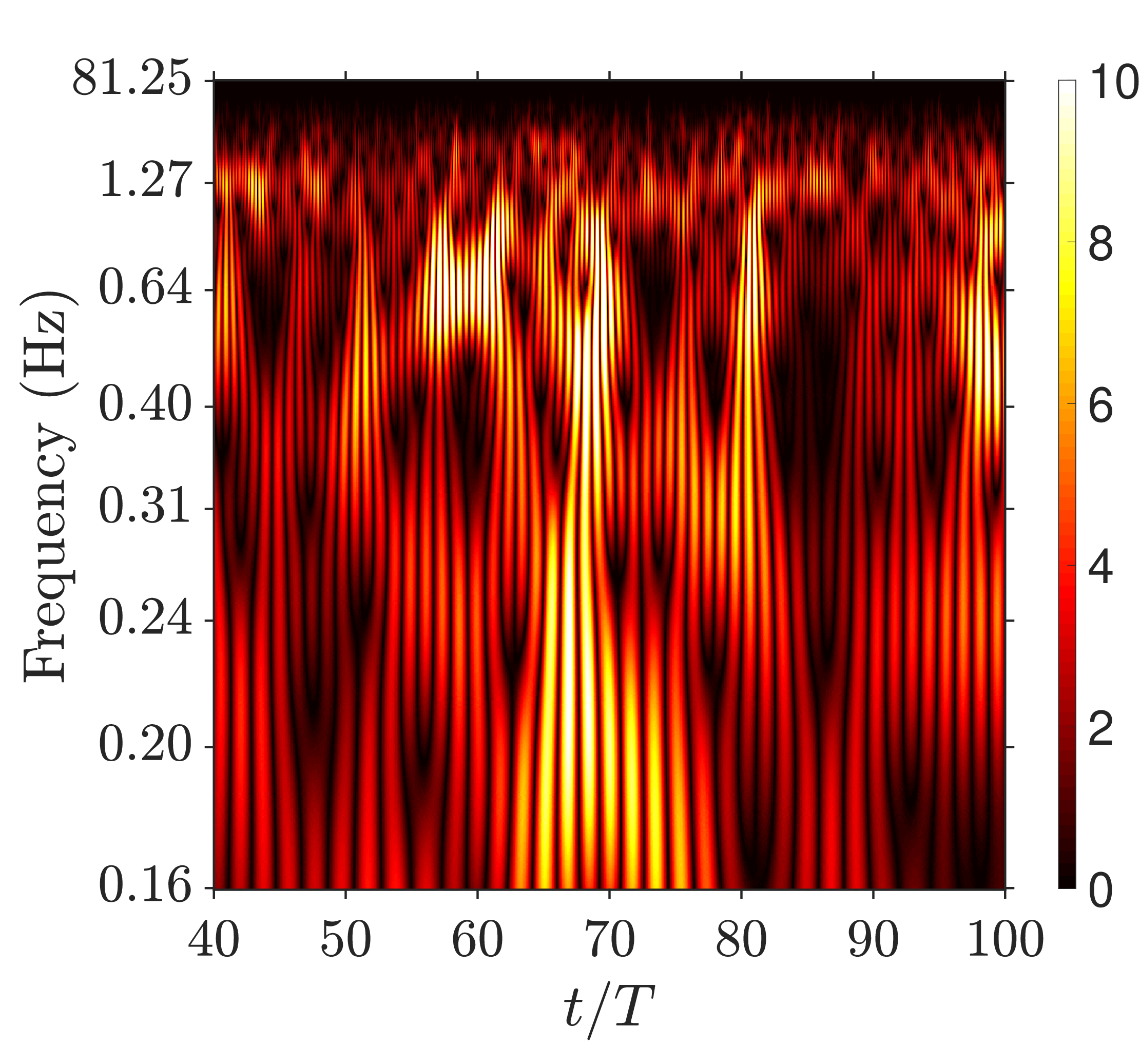}
			\caption{$\kappa h=2.3$ (SC)}
			\label{wavelet_kh=2.3}
		\end{subfigure}
		%\vspace{6pt}
		\caption{(a)-(f) Wavelet transforms of the $C_D$ data for rigid foil at different $\kappa h$ values.}
		\label{wavelet_Rigid}
	\end{figure}
 The system exhibited chaos with increasing $\kappa h$,  following a quasi-periodic and intermittency route, which was similar to our earlier observation from~\citep{bose2021dynamic}. The system showed periodic dynamics for $\kappa h\le 1.0$. A representative periodic dynamics at $\kappa h=1.0$ has been shown in Fig.~\ref{Rigid_route_choas} where the reconstructed phase space displayed a closed loop attractor, and the trajectory in each cycle repeated exactly, characterizing periodic dynamics. The corresponding wavelet spectra (Morlet wavelet transforms have been given in Fig.~\ref{wavelet_Rigid}). with a narrow frequency band of 1.27 Hz (flapping frequency) without any temporal modulation  further confirmed the periodic signature. The system exhibited quasi-periodic and intermittency dynamics when $\kappa h$ was gradually increased in the range $1.0<\kappa h<2.0$. Different qualitative changes in the dynamics such as two-frequency quasi-periodicity (QP-2), and three-frequency quasi-periodicity (QP-3) were observed.  Representative cases of QP-2 and QP-3 are shown for $\kappa h=1.5,\,1.65\,\&\,1.8$; see Fig.~\ref{Rigid_route_choas}. Since the phase space trajectory did not repeat exactly in every cycle but stayed within a close neighbourhood of the earlier position, a thick band was formed in the phase portrait. The corresponding wavelet plots also revealed the presence of two incommensurate frequency bands (1.27 Hz and 0.63 Hz) and three incommensurate frequency bands (1.27 Hz, 0.63 Hz, and 0.32 Hz) indicating QP-2 and QP-3, respectively; see Figs.~\ref{wavelet_kh=1.5},~\ref{wavelet_kh=1.65},~\ref{wavelet_kh=1.8} and their zoomed insets. The representative case for intermittency has been  presented for $\kappa h=1.88$ where the wavelet spectra show sporadic chaotic bursts (Fig.~\ref{wavelet_kh=1.88}), with broad-banded frequency spectra in between narrow frequency bands of incommensurate frequencies. The third incommensurate frequency during QP-3 gave way to the intermittent chaotic windows. Such chaotic bursts have been marked by rectangular boxes (with white dotted lines) `A' and `B' in Fig.~\ref{wavelet_kh=1.88}. The particular temporal pattern where the system alternates irregularly between the states of quasi-periodicity and chaos is known as type-II intermittency in the nonlinear dynamics literature~\citep{hilborn2000chaos}. Additional proofs for the dynamical state of intermittency have been presented in Figs.~1(i) and 1(j) of the supplementary documents. Finally, the flow-field became chaotic and  representative (for $\kappa h = 2.3$) reconstructed phase portrait (Fig.~\ref{Rigid_route_choas}) and the Morlet wavelet plot with broad-banded frequency spectra (Fig.~\ref{wavelet_kh=2.3}) confirmed it.
    \begin{figure}[t!]
		\centering
		\includegraphics[scale=0.19]{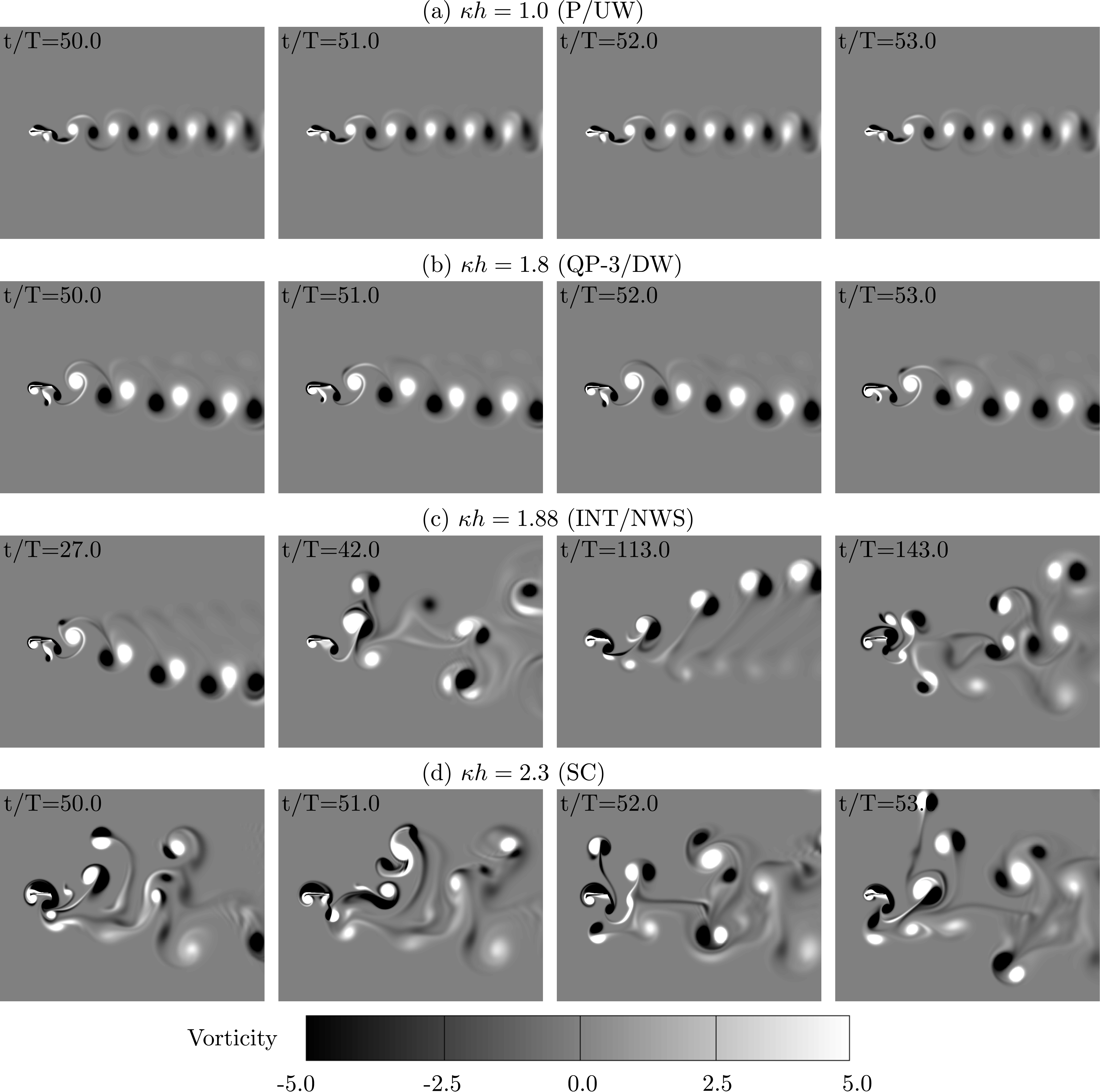}
		\caption{Rigid foil: change in wake patterns with increases in $\kappa h$; P/UW = periodic \& undeflected wake, QP-3/DW =  three-frequency quasi-periodic \& deflected wake, INT/NWS = intermittency \& near-wake switching and SC = sustained chaos.}
		\label{Rigid_wake_pattern}
	\end{figure}
 
Various wake patterns for increasing $\kappa h$ (with selected time frames) have been presented in Fig.~\ref{Rigid_wake_pattern}. The flow-field exhibited periodicity with an undeflected wake pattern (P/UW) at $\kappa h=1.0$, where the vortex structures repeated exactly in consecutive flapping cycles (Fig.~\ref{Rigid_wake_pattern}a). For $\kappa h = 1.8$, the flow-field exhibited quasi-periodic dynamics with a deflected wake pattern (QP-3/DW) (Fig.~\ref{Rigid_wake_pattern}b), in which gradual shifts in vortex core locations were seen. Note that the wake was also seen deflected for $\kappa h=1.5$ and $1.65$ where it exhibited quasi-periodic dynamics. The system exhibited intermittency \& a near-wake switching (INT/NWS) behaviour at $\kappa h=1.88$. In near-wake switching, the deflection direction of the primary vortex street switched with time (Fig.~\ref{Rigid_wake_pattern}c), and this jet-switching was seen to take place after the chaotic intervals. Interestingly, switching  did not happen after every chaotic burst. Note that jet-switching   was also observed in some of our earlier studies~\citep{bose2021dynamic,shah2022chordwise}. During the chaotic intervals of intermittency, the vortices were seen to undergo complex interactions, such as \textit{collision of vortex couples}, \textit{exchange of partners}, \textit{vortex merging}, in an irregular manner. Finally, the flow-field became completely unpredictable due to chaos where there was no correlation between the flow-field snapshots in different cycles.

In the following section, chord-wise flexibility has been introduced in a systematic manner to understand its effect on the wake patterns and also in changing the dynamics and hindering the route-to-chaos. The role of key near-field vortices in triggering aperiodicity and chaos in rigid systems was already discussed  in ~\cite{bose2018investigating,majumdar2020capturing,bose2021dynamic}. It is of interest here to investigate the possible changes to such flow-field mechanisms under flexibility and undertake a comparison of the dynamics between the rigid and the flexible configurations.

\section{Variation in flexibility and change in the dynamics}
\label{sec:flexible_dynamical_transition}

In order to investigate the modified dynamics  under flexibility, the underlying vortex mechanisms, and the feasibility of controlling aperiodicity, $\gamma$ was  varied first for  $\kappa h = 2.3$, at which the rigid foil exhibited strong sustained chaos. 
\begin{figure}[t!]
		\centering
		\includegraphics[scale=0.19]{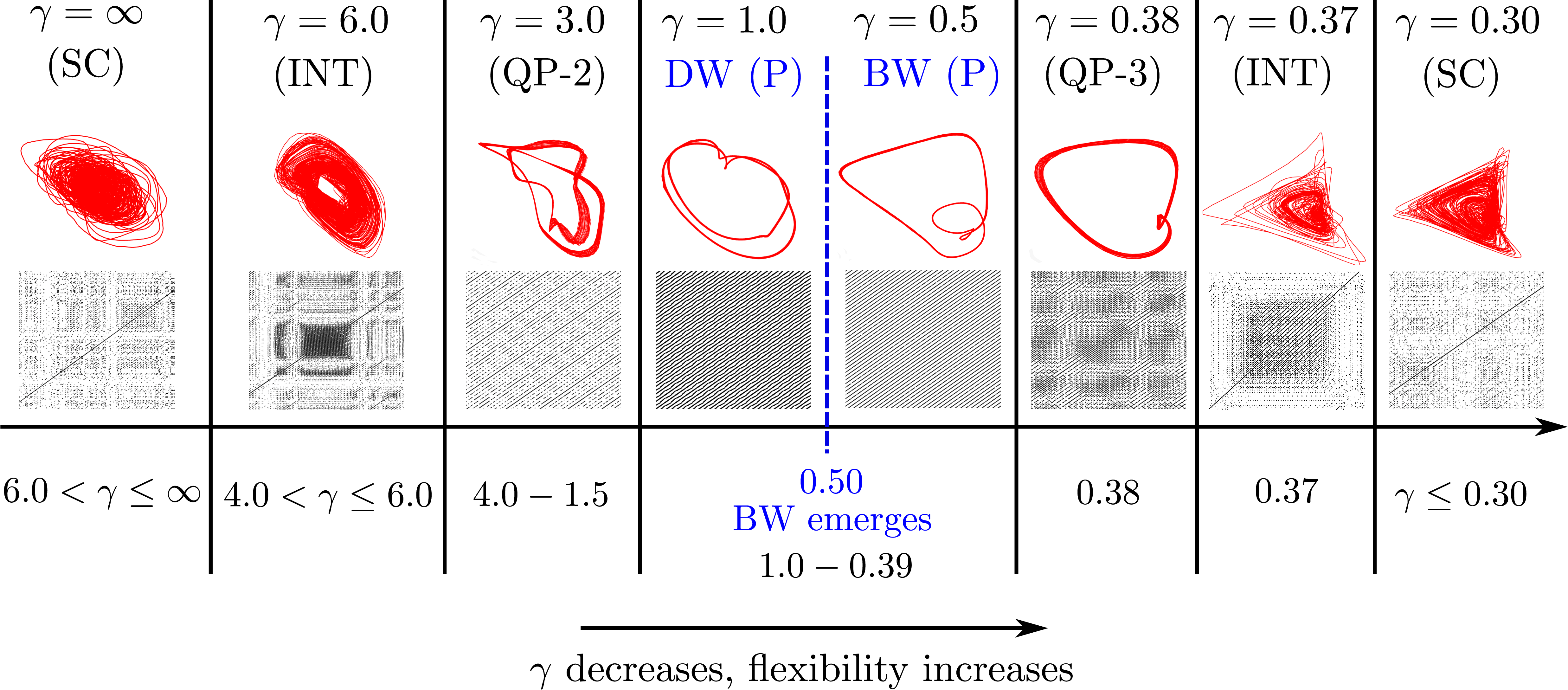}
		\caption{Dynamical transition for flexible foil at $\kappa h=2.3$: schematic representation of the bifurcation behaviour when $\gamma$ is decreased as the control parameter. SC = sustained chaos; INT = intermittency; QP-2 = two-frequency quasi-periodicity; P = periodic state; DW = deflected wake; BW = bifurcated wake; QP-3 = three-frequency quasi-periodicity. Reconstructed phase-portraits and recurrence plots of the $C_D$ data are presented in the first and second rows, respectively. A detailed explanation of the recurrence plots is given in Section~1.2 of the supplementary document.}
		\label{flexible_route_choas}
	\end{figure}
    \begin{figure}[b!]
		\begin{subfigure}{.33\textwidth}
			\centering
			\includegraphics[scale=0.14]{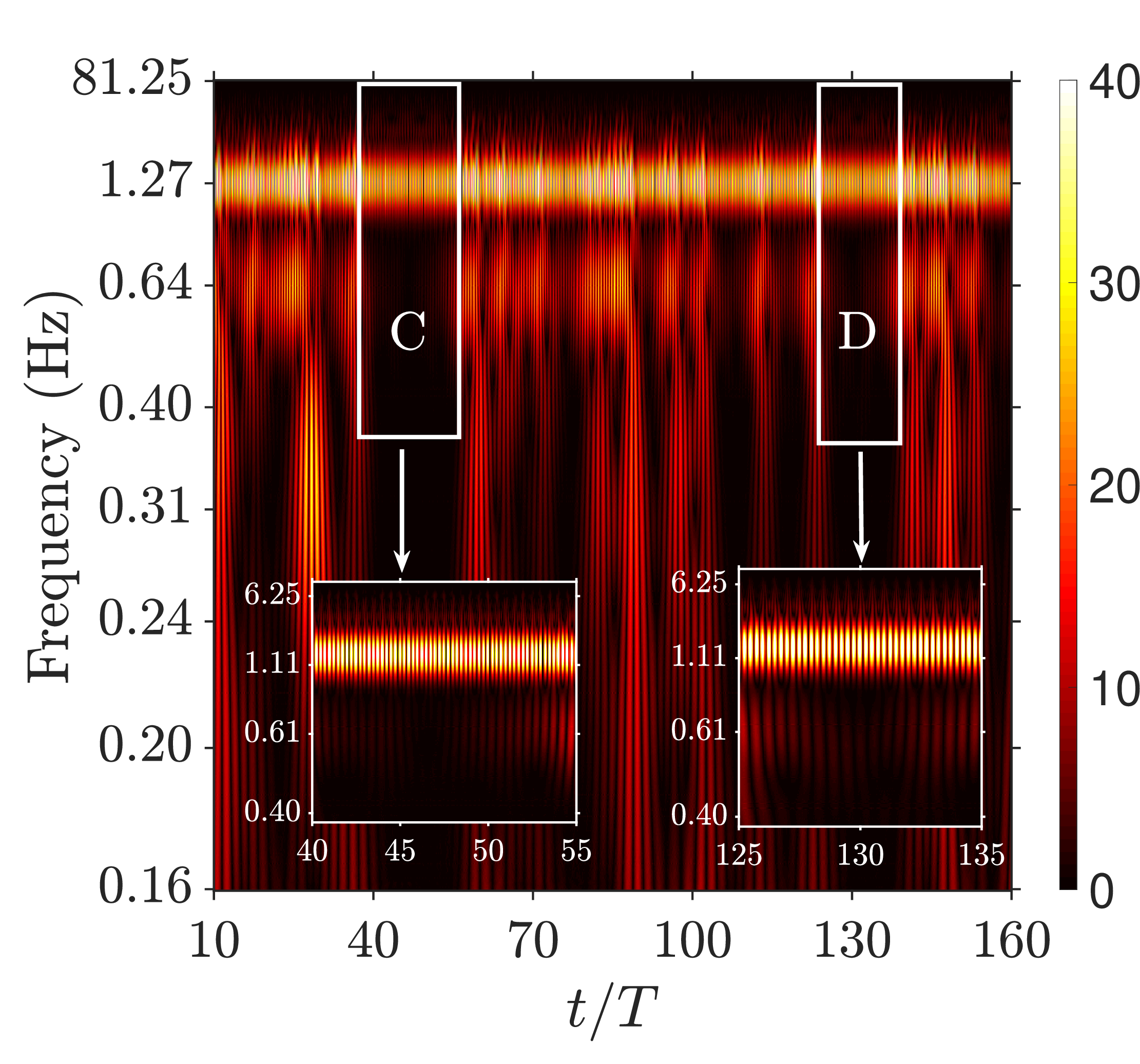}
			\caption{$\gamma=6.0$ (INT)}
			\label{wavelet_gamma=6.0}
		\end{subfigure}
		%\vspace{6pt}
		\begin{subfigure}{.33\textwidth}
			\centering
			\includegraphics[scale=0.14]{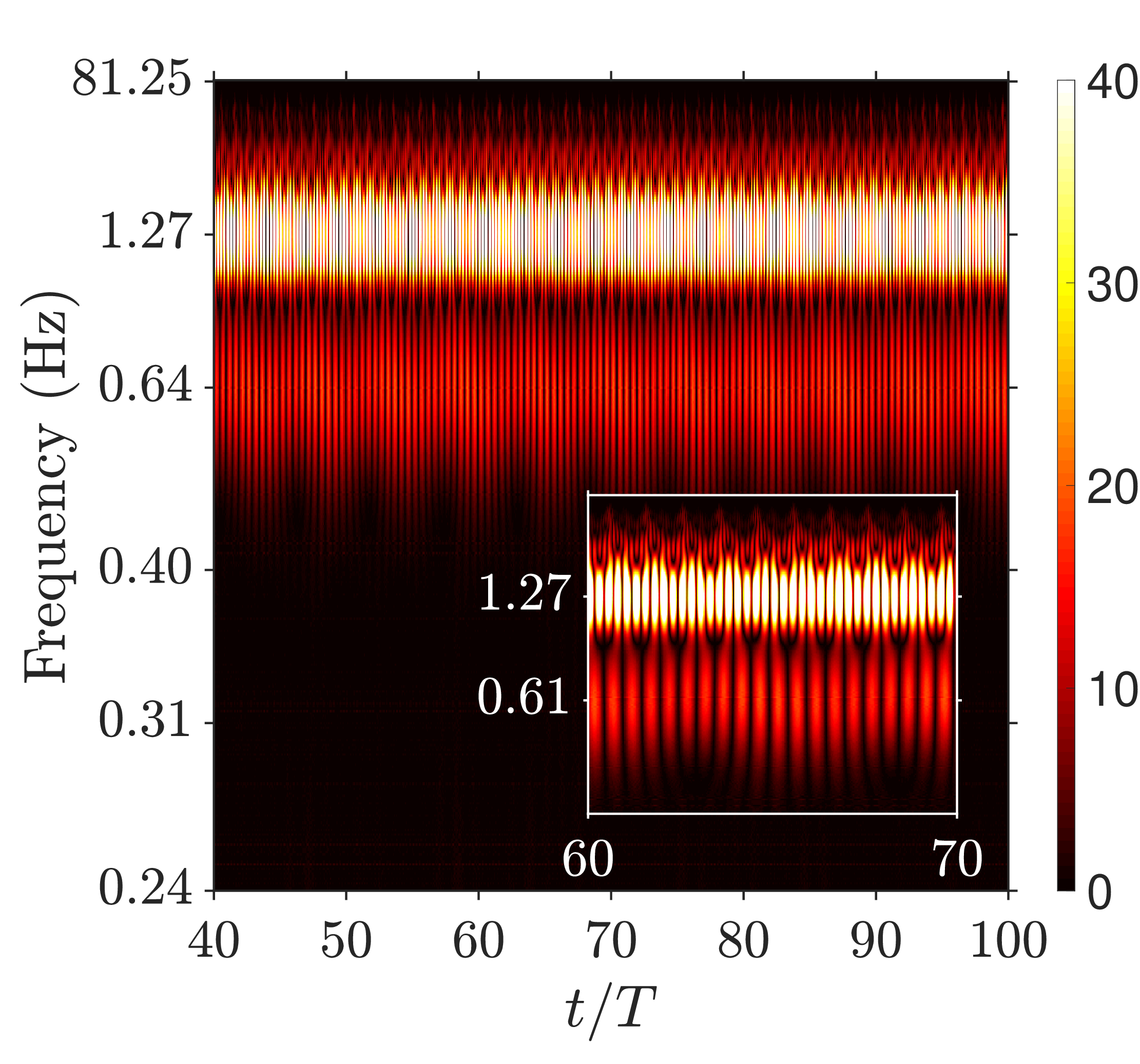}
			\caption{$\gamma=3.0$ (QP-2)}
			\label{wavelet_gamma=3.0}
		\end{subfigure}%
		%\vspace{6pt}
		\begin{subfigure}{.33\textwidth}
			\centering
			\includegraphics[scale=0.14]{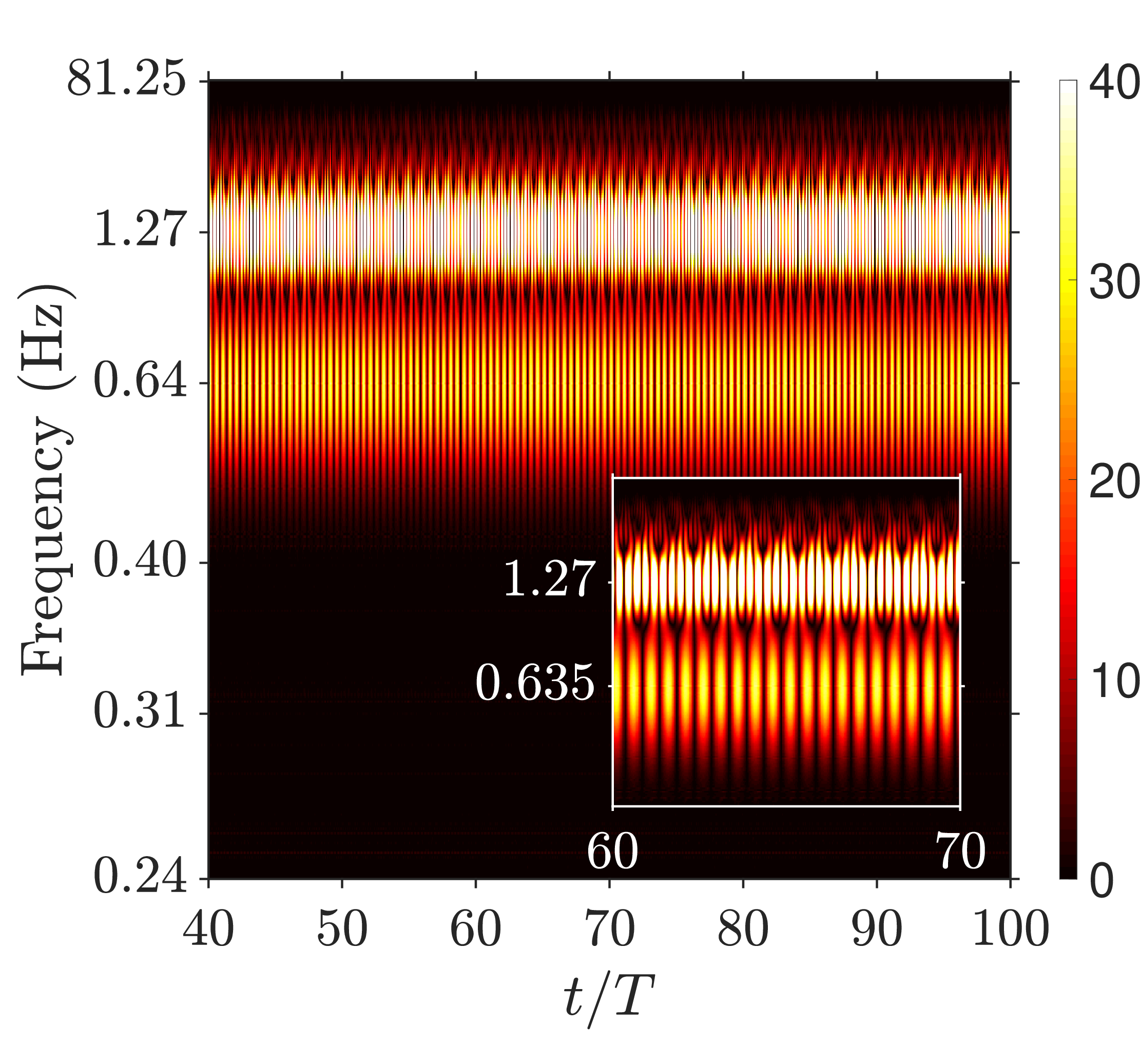}
			\caption{$\gamma=1.0$ (P)}
			\label{wavelet_gamma=1.0}
		\end{subfigure}
		%\vspace{6pt}
		\begin{subfigure}{.33\textwidth}
			\centering
			\includegraphics[scale=0.14]{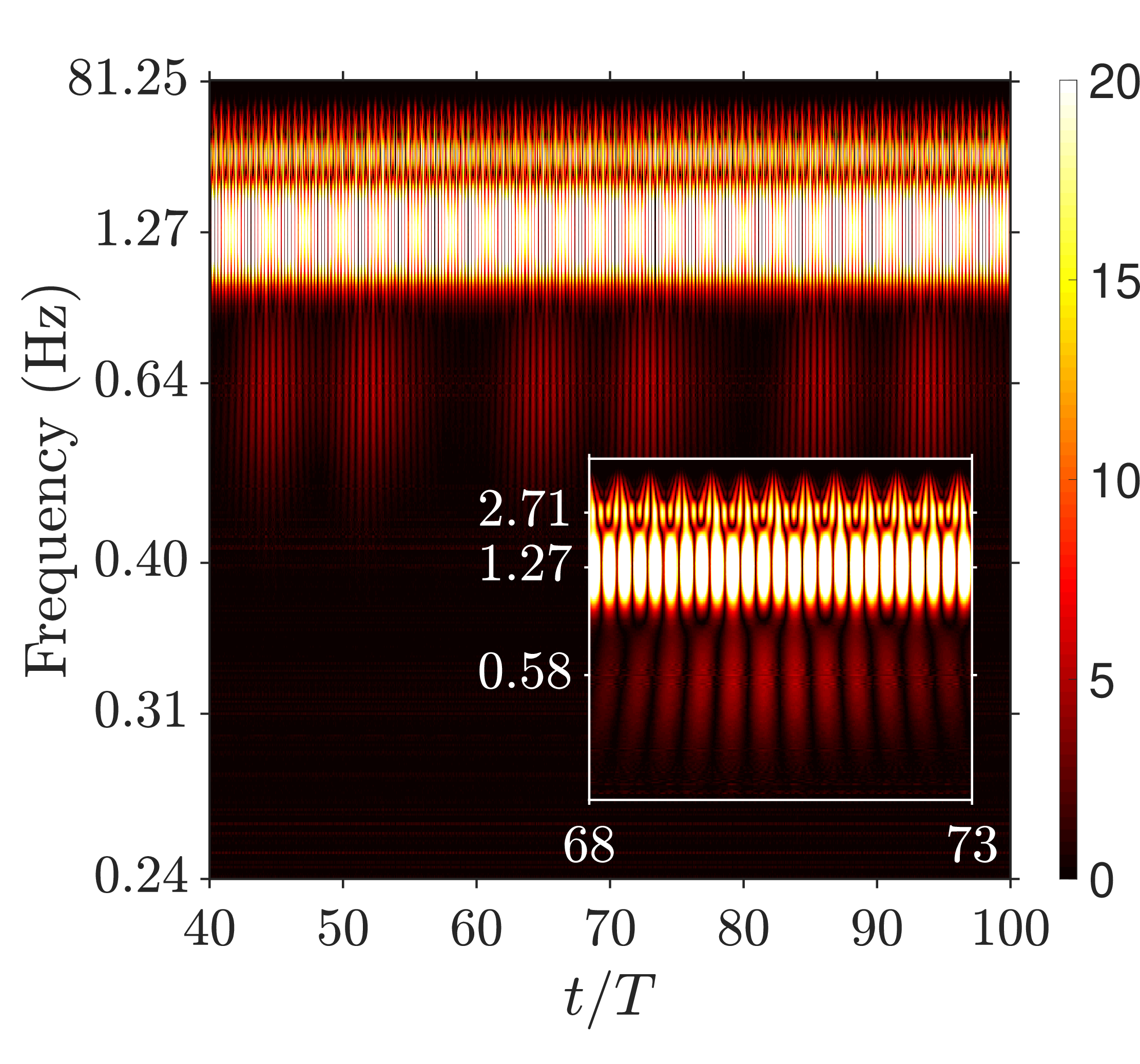}
			\caption{$\gamma=0.38$ (QP-3)}
			\label{wavelet_gamma=0.38}
		\end{subfigure}
		%\vspace{6pt}
		\begin{subfigure}{.33\textwidth}
			\centering
			\includegraphics[scale=0.14]{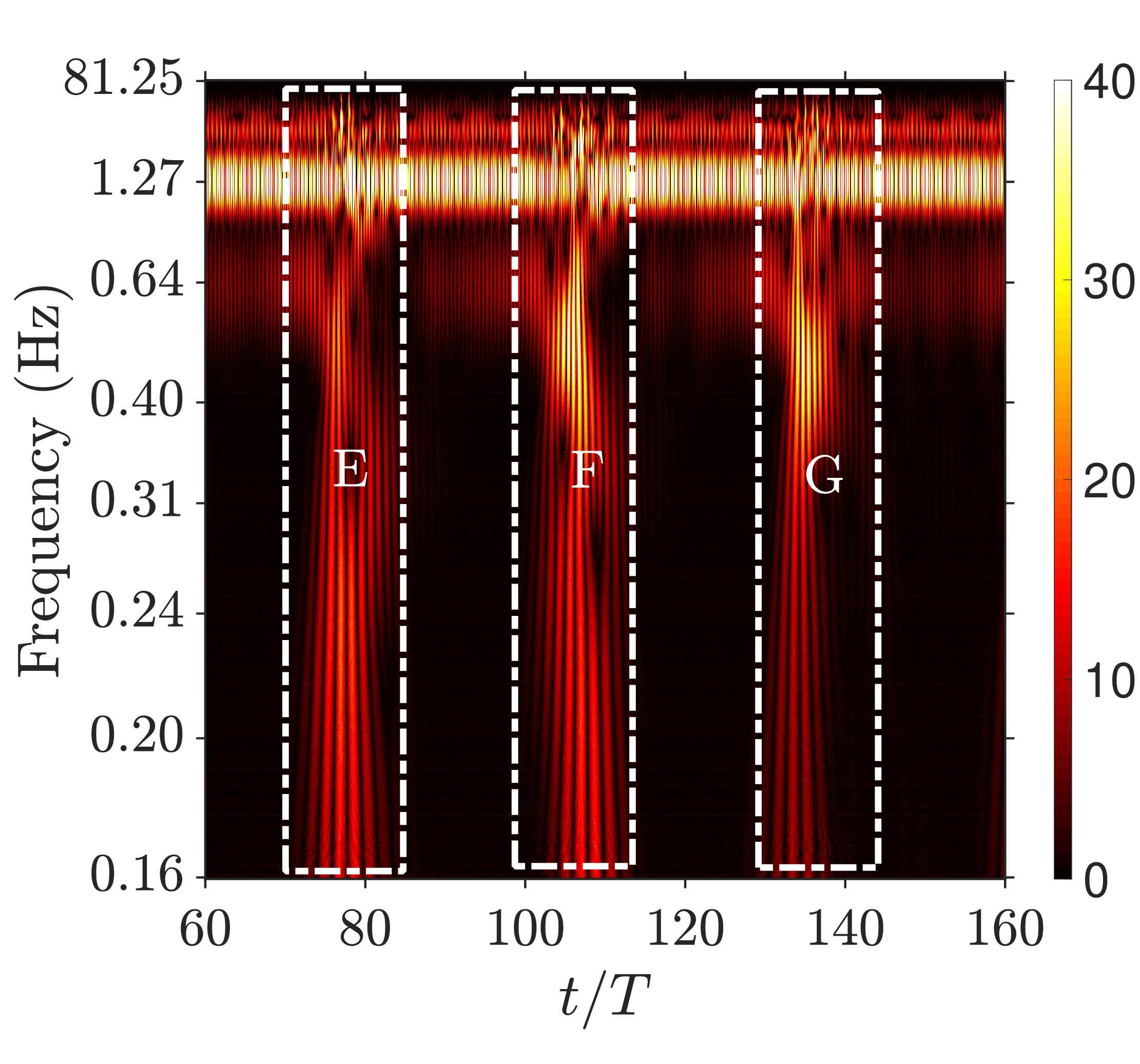}
			\caption{$\gamma=0.37$ (INT)}
			\label{wavelet_gamma=0.37}
		\end{subfigure}%
		%\vspace{6pt}
		\begin{subfigure}{.33\textwidth}
			\centering
			\includegraphics[scale=0.14]{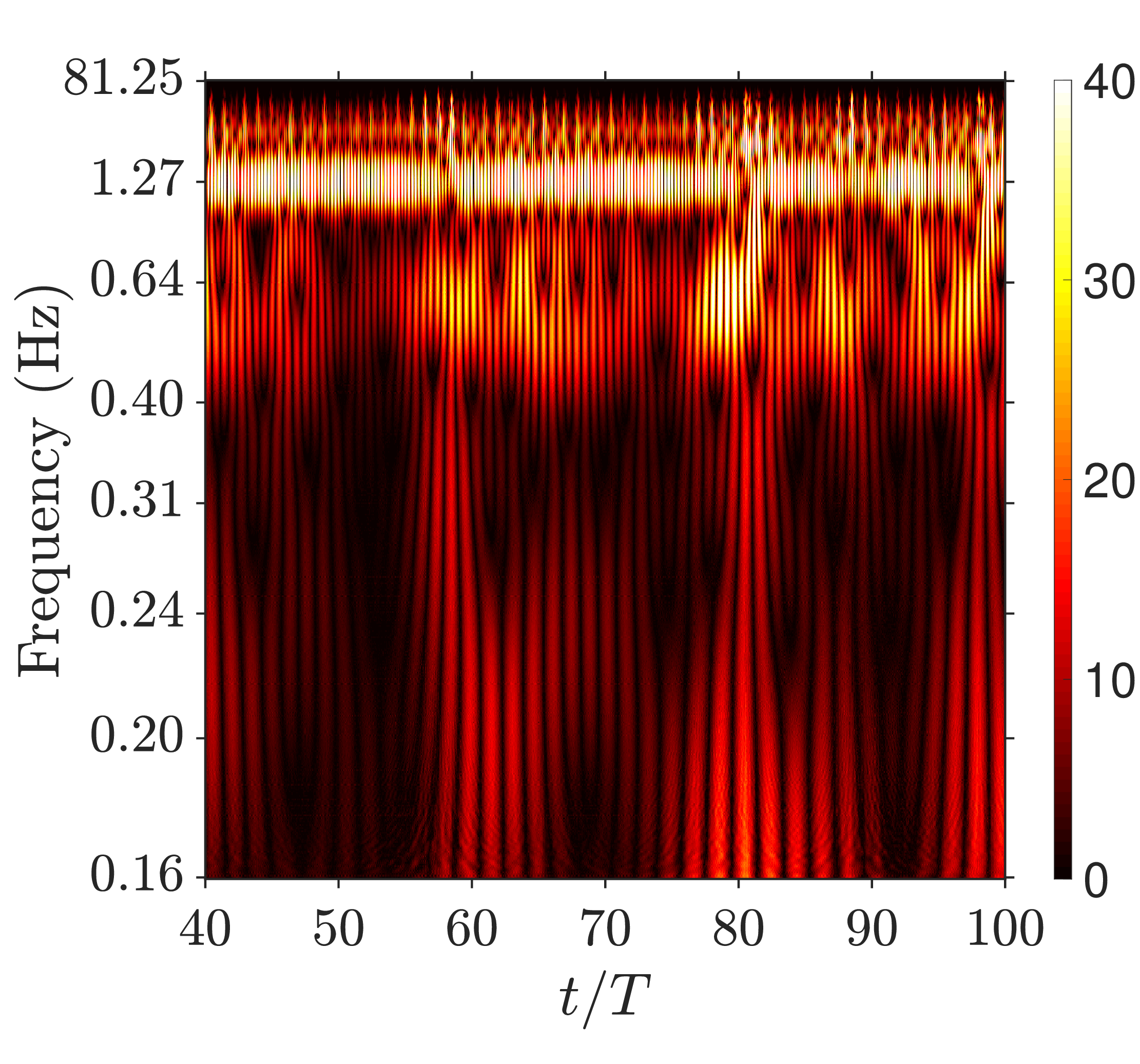}
			\caption{$\gamma=0.30$ (SC)}
			\label{wavelet_gamma=0.3}
		\end{subfigure}
		%\vspace{6pt}
		\caption{At $\kappa h=2.3:$ (a)-(f) wavelet transforms of the $C_D$ data for the flexible foil at different flexibility levels.}
		\label{wavelet_gamma}
	\end{figure}
Subsequently, the same was  done for a few typical $\kappa h$ cases  for the sake of completeness . The results for the former have been presented in the following subsections in details and the results for the other $\kappa h$ cases have been given briefly as supplementary. Recovery from chaos to periodicity was observed through a number of changes in the dynamical states, and a number of interesting wake patterns also emerged in the process. Note that these wake patterns in the flexible system during the periodic dynamics were not observed for the rigid system discussed in the previous section. The entire transition route for $kh = 2.3$ at different levels of flexibility has been summarized  in Fig.~\ref{flexible_route_choas} along with reconstructed phase portraits and recurrence plots (for few  representative cases). As the figure depicts, the respective flow-field changed from sustained chaos (for the rigid foil) to periodicity for a critical value of $\gamma$ through a series of intermediate dynamical transitions. The transitions included quasi-periodicity and intermittency. However, with further decrease in $\gamma$, the system reverted to quasi-periodicity and once again back to chaos via a dynamical state of intermittency, which is an interesting observation.

%As was done for the rigid case, the dynamical states have been established for the flexible cases using time series analysis results. 
The chaotic state seen for the perfectly rigid foil got interspersed with quasi-periodic windows for $4.0<\gamma\le6.0$. Reconstructed phase portrait (Fig.~\ref{flexible_route_choas}) and the Morlet wavelet plot (Fig.~\ref{wavelet_gamma=6.0}) have been presented for $\gamma=6.0$. In the Morlet wavelet plot in Fig.~\ref{wavelet_gamma=6.0}, intermittent quasi-periodic windows of narrow incommensurate frequency bands (marked by white rectangular boxes `C' and `D') were seen in between broad-banded frequency spectra. The first instance of quasi-periodicity appeared in the flow-field at $\gamma = 4.0$ and sustained up to $\gamma=1.50$ with further regularization in the flow-field as $\gamma$ was decreased. The results have been presented for $\gamma=3.0$ where  the reconstructed phase-space showed a thick toroidal band which is the signature of quasi-periodicity (Fig.~\ref{flexible_route_choas}). Two incommensurate frequency bands of 1.27 Hz and 0.61 Hz with temporal modulation are observed in the corresponding wavelet spectra (see Fig.~\ref{wavelet_gamma=3.0} and its zoomed inset), characterizing the QP-2 dynamics. At $\gamma=1.0$ (Fig.~\ref{flexible_route_choas}), the system returned to perfect periodicity. The corresponding wavelet spectra showed two frequency bands without any temporal modulation; one for the flapping frequency of 1.27 Hz, and the other corresponding to its exact sub-harmonic of 0.635 Hz (see Fig.~\ref{wavelet_gamma=1.0} and its zoomed inset). 
The periodic dynamics prevailed till $\gamma = 0.39$, even though the flow-field underwent substantial qualitative changes exhibiting a variety of wake patterns (at $\gamma = 1.0, \, 0.70, \, 0.5, \, 0.4 \, \& \, 0.39$). The wake patterns will be demonstrated in the later sections. 
The system went back to a  quasi-periodic state at $\gamma = 0.38$; the reconstructed phase-space and wavelet spectra confirmed this with a thick toroidal band (Fig.~\ref{flexible_route_choas}) and three incommensurate frequency bands (2.71 Hz, 1.27 Hz, and 0.58 Hz) with temporal modulations, indicating three-frequency quasi-periodicity (QP-3) (see Fig.~\ref{wavelet_gamma=0.38} and its zoomed inset), respectively. The quasi-periodic signature got disturbed by getting interspersed with irregular chaotic windows between the quasi-periodic windows at $\gamma = 0.37$. The corresponding wavelet spectra (Fig.~\ref{wavelet_gamma=0.37}) showed the insets `E', `F' and `G', the sporadic chaotic windows. The appearance of sporadic aperiodic evolution of the trajectory in the otherwise toroidal phase-space was also distinctly seen in the reconstructed phase-space (Fig.~\ref{flexible_route_choas}). More proofs regarding the dynamical state of intermittency have been presented in Figs.~2(a) \& 2(b) for $\gamma=6.0$ and Figs.~2(i) \& 2(j) for $\gamma=0.37$,in the supplementary document. With further decrease in flexibility, sustained chaos reappeared at $\gamma = 0.30$, as reflected in the evolution of phase space trajectory given in Fig.~\ref{flexible_route_choas} and the broad-banded frequency spectra given in Fig.~\ref{wavelet_gamma=0.3}. The distinct topologies of the chaotic attractors observed for $\gamma = \infty$ and $\gamma = 0.3$ are representative of how the trajectories move around the phase space, and hence are indirectly related to the degree of chaos, with the largest Lyapunov exponent~\citep{hilborn2000chaos,nayfeh2008applied} being $0.38$ and $0.3$, respectively.

These results demonstrated the large extent of changes in the dynamics due to  flexibility. The chaotic behaviour normally expected at high $\kappa h$ got  inhibited, and a perfectly periodic state was restored for a critical flexibility level. The flow patterns and vortex mechanisms responsible for this chaos-to-order recovery have been presented in the following, in comparison with the chaotic baseline case of the rigid foil at $\kappa h = 2.3$. 
%The flow-field evolution and vortex interactions were tracked at different levels of flexibility and the results have been presented for the chaotic baseline case of the rigid foil at $\kappa h = 2.3$ in the following subsections. 
For the sake of completeness, the flow-fields for different $\gamma$'s for $\kappa h=1.0,\,1.5\,\&\,1.88$ have been discussed in the supplementary document (Section~2). 
% The structural response has been discussed in Section~\ref{sec:flapping_mode_shape} to validate the flow-field dynamics coupled to it through FSI effects. 
Further, the  propulsive performance and its connection with the periodicity/regularity of the system has been examined in Section~\ref{sec:propulsive_analysis}. 

\subsection{Intermittency (Quasi-periodicity interspersed with chaotic time windows)}
\label{sec:gamma_6_flow_intermittency}

As the chord-wise flexibility was introduced in the rigid foil, regularisation started to happen, and a dynamical state of intermittency emerged from chaos. This manifested as quasi-periodic time histories interspersed with chaotic time windows appearing at irregular intervals. Recall that the rigid system exhibited jet-switching during the intermittency state %at $\kappa h=1.88$ 
as 
was presented in Fig.~\ref{Rigid_wake_pattern}(c). 
    \begin{figure}[t!]
		\centering
		\includegraphics[scale=0.19]{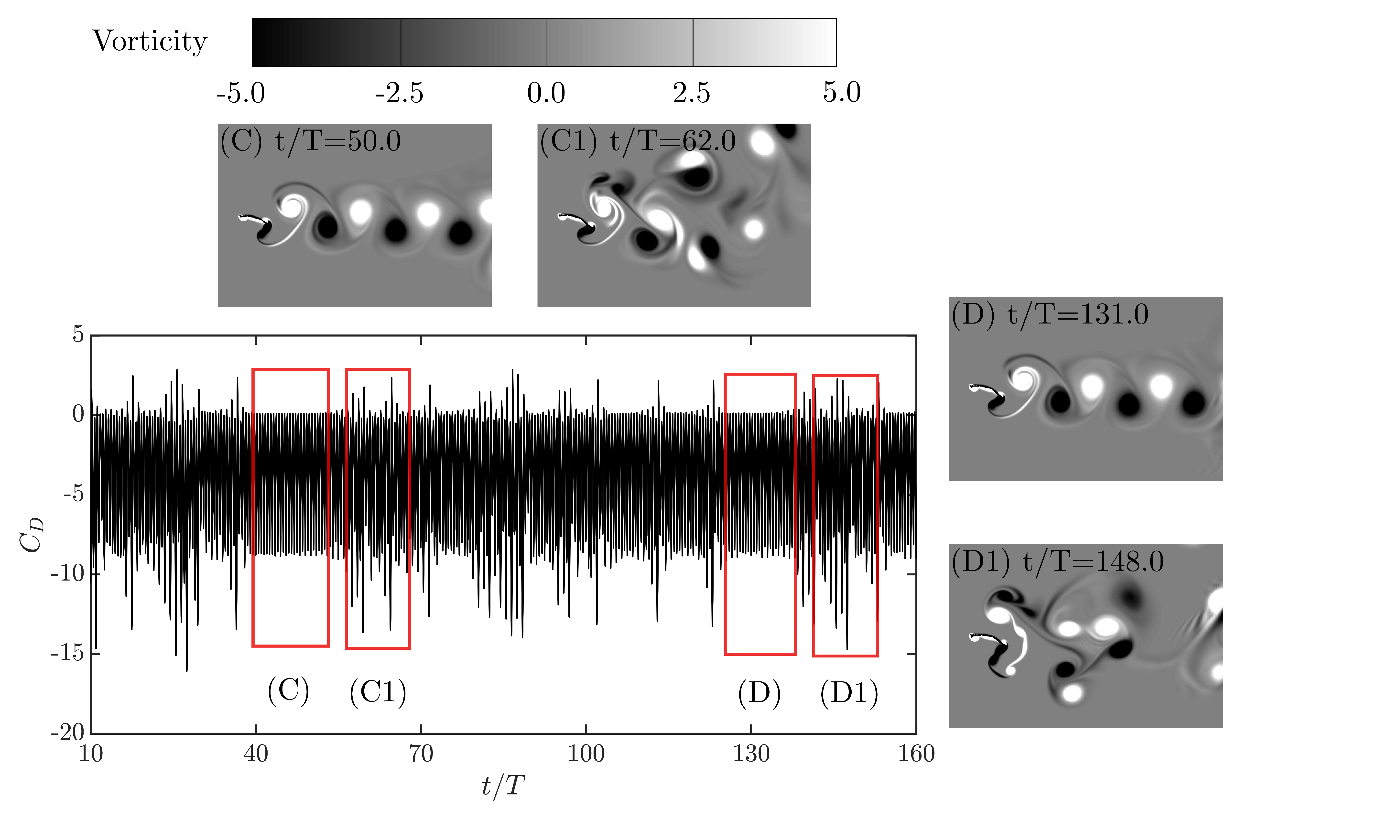}
		\vspace{6pt}
		\caption{At $\kappa h=2.3 \, \& \, \gamma=6.0\,(\rm INT):$ $C_D$ time history and instantaneous vorticity contours during quasi-period (`$\rm C$' \& `$\rm D$') and chaotic (`$\rm C1$' \& `$\rm D1$') windows. Note that the same contour level has been used throughout the paper for all the vorticity contour plots, and hence it is not repeated hereafter.}
		\label{CD_TH_gamma_6}
	\end{figure}
    \begin{figure}[t!]
		\centering
		\includegraphics[scale=0.2]{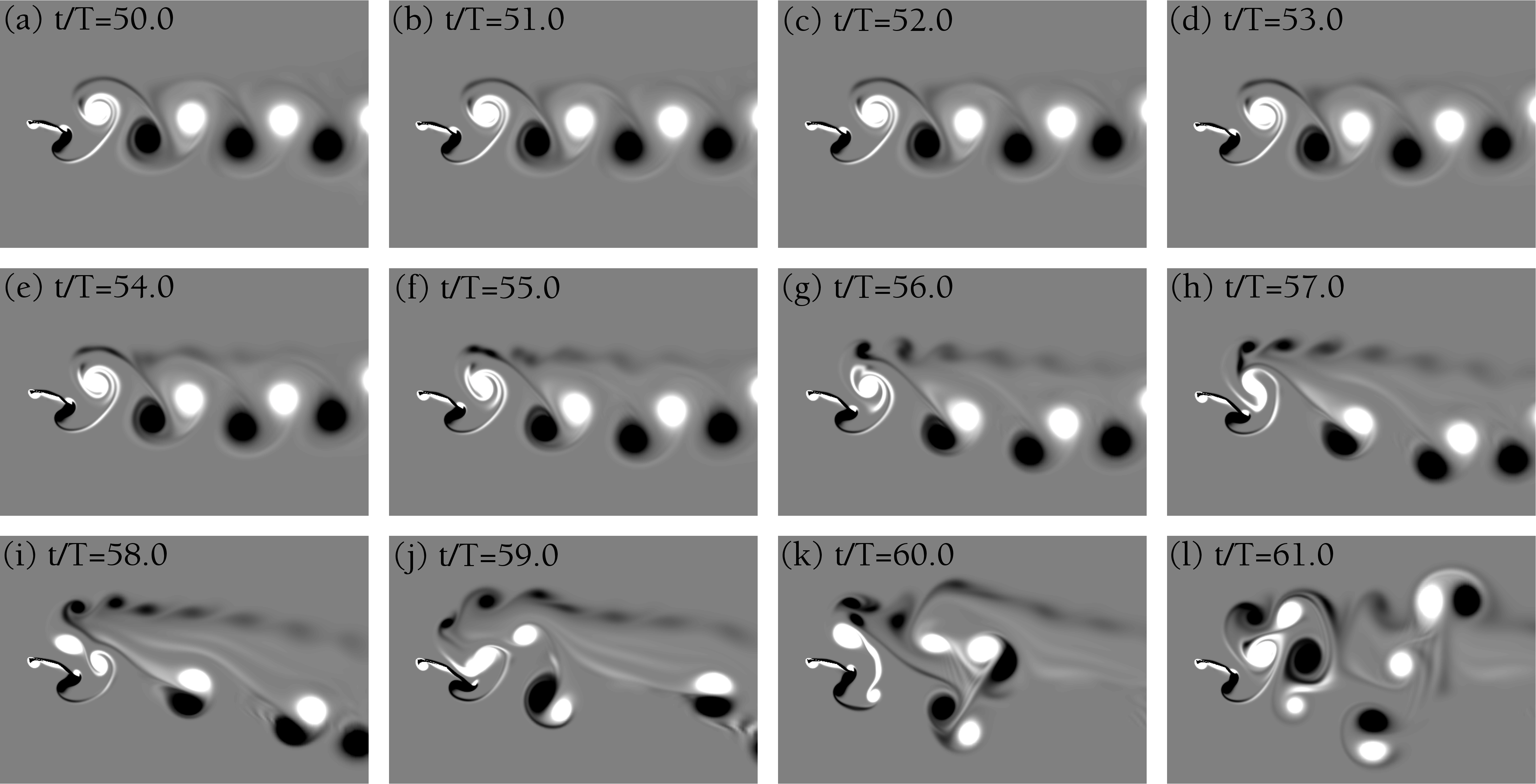}
		\vspace{6 pt}
		\caption{At $\kappa h=2.3 \, \& \, \gamma=6.0\,(\rm INT):$ instantaneous vorticity contours at different time instances depicting the growth of secondary vortices and its subsequent interactions with primary vortices.}
		\label{intermittency_flow_field_gamma_6}
	\end{figure}
However, in the  flexible case, jet switching was not observed during intermittency. 
%The nuances of the flow-field has been discussed in the following. 
A few  representative flow-field snapshots for $\gamma=6.0$ at the intermittency state have been shown in Fig.~\ref{CD_TH_gamma_6}. Typical quasi-periodic and chaotic time windows have been marked on the load time history in the figure.  During quasi-periodic windows `C' and `D',   reverse K\'arm\'an vortex streets were seen. No ordered pattern was observed during the chaotic windows `C1' and `D1'. During the chaotic intervals, the vortices were seen to undergo complex vortex interactions, similar to the rigid case. The transitions between ordered  and chaotic wakes have been presented in Fig.~\ref{intermittency_flow_field_gamma_6} and  the mechanism was observed to be similar to the rigid case. During the transition, a secondary vortex street of clockwise (CW) vortices formed, and the primary vortex street deflected downwards during $t/T = 50 - 58$, which grew stronger and influenced the primary TEVs. Interactions between the primary and the secondary streets led to spontaneous formation and destruction of multiple vortical structures, triggering  chaos (Figs.~\ref{intermittency_flow_field_gamma_6}(i) --~\ref{intermittency_flow_field_gamma_6}(l)). This sequence of interactions could be seen during every such transition over time (\textit{e.g.}, `$\rm C$' $\rightarrow$ `$\rm C1$' and `$\rm D$' $\rightarrow$ `$\rm D1$'). A video on the sequence of vortex interactions for this parametric case has been presented as supplementary video SV1. Note that at this level of flexibility, the system was seen to be periodic with undeflected wake at $\kappa h=1.0$ and quasi-periodic with far-wake switching at $\kappa h=1.5\,\&\,1.88$ (presented in the supplementary document in Figs.~3 and 5(b)).

\subsubsection{\textcolor{black}{Vortex interaction mechanisms behind the transition from chaotic to quasi-periodic intervals:}}

\begin{figure}[b!]
		\centering
		\includegraphics[scale=0.235]{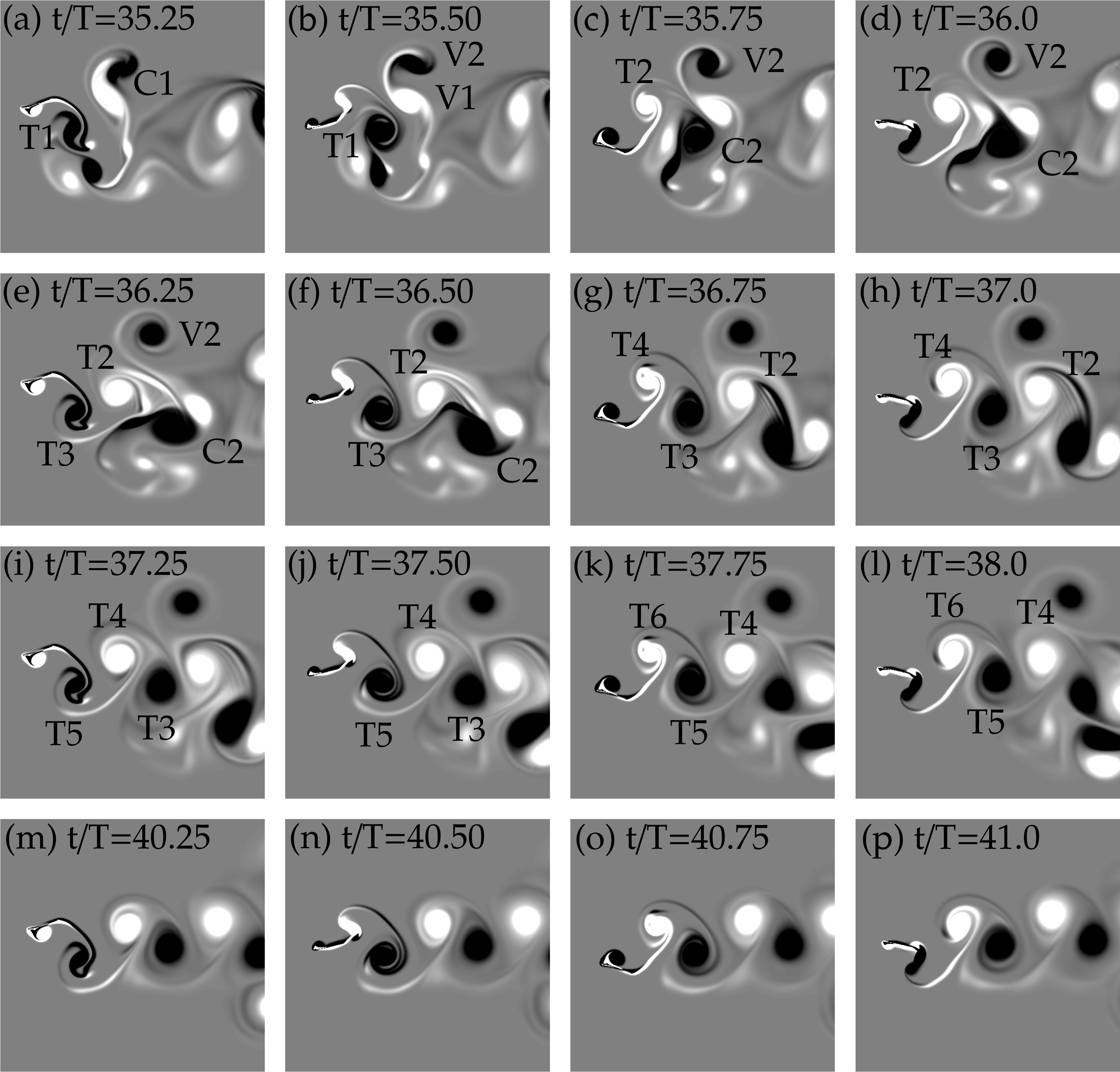}
		\vspace{6pt}
		\caption{\textcolor{black}{At $\kappa h=2.3 \, \& \, \gamma=6.0:$ instantaneous vorticity contours during $36^{\rm th}$--$38^{\rm th}$ and $41^{\rm st}$ cycles depicting the mechanisms of near field interactions during the transitions from chaos to quasi-periodic interval.}}
		\label{Mechanism_intermittency_flow_field_chaos_QP_gamma_6}
    \end{figure}

\textcolor{black}{The mechanisms behind transition from chaotic to quasi-periodic time window has been presented in Fig.~\ref{Mechanism_intermittency_flow_field_chaos_QP_gamma_6} during $36^{\rm th}$--$41^{\rm st}$ cycles. A couple $\mathrm{\mathbf{C1}}$ from the previous cycle was seen to get split into counterclockwise (CCW) $\mathrm{\mathbf{V1}}$ and clockwise (CW) $\mathrm{\mathbf{V2}}$ vortices (Fig.~\ref{Mechanism_intermittency_flow_field_chaos_QP_gamma_6}(b)). $\mathrm{\mathbf{V1}}$ formed couple $\mathrm{\mathbf{C2}}$ with CW trailing-edge vortex (TEV) $\mathrm{\mathbf{T1}}$, whereas the isolated CW $\mathrm{\mathbf{V2}}$ did not further interact with any other vortices which helped in the regularization of the wake in subsequent cycles (Figs.~\ref{Mechanism_intermittency_flow_field_chaos_QP_gamma_6}(c) and \ref{Mechanism_intermittency_flow_field_chaos_QP_gamma_6}(d)).  $\mathrm{\mathbf{C2}}$ pulled the newly formed  $\mathrm{\mathbf{T2}}$ and $\mathrm{\mathbf{T3}}$ because of its self-induced velocity and $\mathrm{\mathbf{T2}}$ and  $\mathrm{\mathbf{T3}}$ did not participate in any interaction (Figs.~\ref{Mechanism_intermittency_flow_field_chaos_QP_gamma_6}(e) -- \ref{Mechanism_intermittency_flow_field_chaos_QP_gamma_6}(h)). Therefore, the nearby vortices did not undergo any further complex interaction such as collision of vortex couples, exchange of partners and vortex merging which were mainly responsible for irregular and chaotic interactions leading to chaotic dynamics. In the subsequent cycles, TEVs $\mathrm{\mathbf{T4}}$ and  $\mathrm{\mathbf{T5}}$ were further inhibited from taking part in the near-field interactions and hence further avoiding any complex vortex interactions responsible for chaotic dynamics. Similar phenomena repeated in the subsequent cycles  as well. Gradually, the wake got organized  displaying a reverse K\'arm\'an vortex street (Figs.~\ref{Mechanism_intermittency_flow_field_chaos_QP_gamma_6}(i) -- \ref{Mechanism_intermittency_flow_field_chaos_QP_gamma_6}(p)) and remained so until a strong aperiodic trigger appeared from the LEVs. Note that a similar vortex mechanism described above took place during each chaotic to quasi-periodic transition to regularise the flow-field in the intermittancy regime. The flow-field mechanisms behind the complete recovery from chaos under flexibility and then again transitioning back to the aperiodic dynamics is discussed in later subsections.}

\subsection{Quasi-periodicity (Deflected wake with a secondary vortex street)}

\begin{figure}[h]
		\centering
		\includegraphics[scale=0.25]{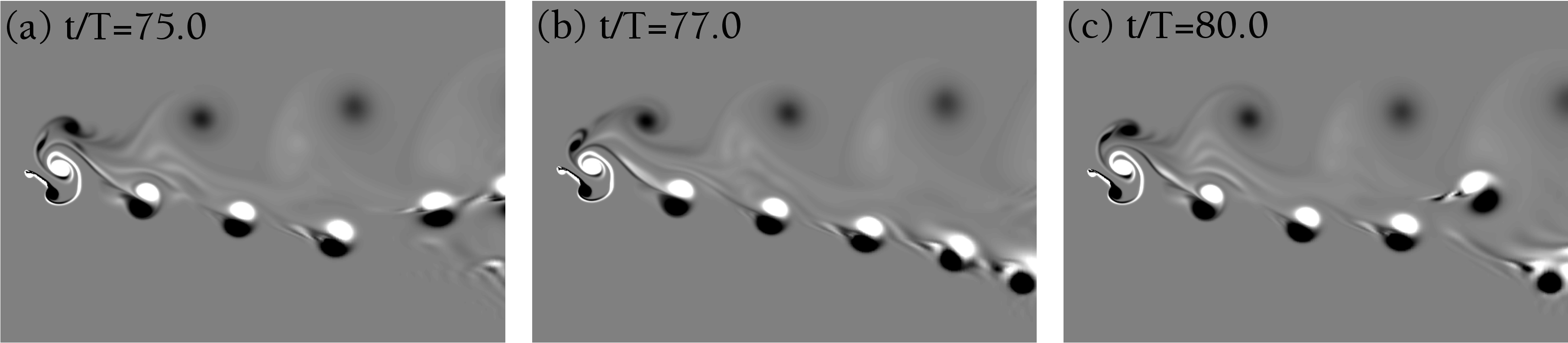}
		\vspace{6pt}
		\caption{At $\kappa h=2.3 \, \& \, \gamma=3.0\,(\rm QP-2):$ instantaneous vorticity contours at different time instances showing a deflected wake with secondary clockwise vortex street. The deflected wake shows far-wake switching tendency ((a) and (c)).}
		\label{flow_field_gamma_3}
	\end{figure}

With an increase in flexibility level, further regularization of the flow-field took place. The chaotic temporal bursts disappeared completely, and the dynamics revealed a pure quasi-periodic signature throughout. Recall that for the rigid case, quasi-periodicity was also identified (was shown in Fig.~\ref{Rigid_wake_pattern}(b)). Even though the dynamical state is the same,  some differences in the wake behaviour were observed in the flexible case. A secondary street was noticed in the flexible case which was absent in the rigid case. Also there was a tendency for wake switching which was not observed for the rigid case. Flow-field snapshots of three typical flapping cycles ($t/T = 75.0, 77.0$ and $80.0$) have been presented in Fig.~\ref{flow_field_gamma_3} for $\gamma = 3.0$ to describe the switching process in the deflection direction of the primary vortex street. A secondary street comprising CW vortices was also formed and interactions with which led to far-wake switching. The CW vortex did not get shed in every flapping cycle but merged with its counterparts from the earlier cycles. This process continued for a few cycles and only when the merged vortex reached a certain size/strength it was shed and contributed towards the formation of the secondary street. Such discrepancies in the shedding behaviour as well as the vortex locations changing marginally from one cycle to another contributed towards quasi-periodicity. The small  shift in the vortex core locations facilitated the process of formation of vortex couples in the primary street, having the resultant dipole velocity towards the upward direction in the far-wake in contrast to the downward direction in the near-wake. This gave an arc-shaped wake pattern, known as far-wake switching \citep{wei2014mechanisms,bose2021dynamic}, shown in Figs.~\ref{flow_field_gamma_3}(a) and \ref{flow_field_gamma_3}(c). Far-wake switching in a fully flexible flapping configuration has not been reported  in the literature, though it  has been known in the context of rigid systems. 
The continuous temporal evolution highlighting far-wake switching has been presented in supplementary video SV2.
At this level of flexibility the system exhibited perfect periodicity with a deflected wake and a secondary street for $\kappa h=1.0,\,h=1.5\,\&\,1.88$. The secondary vortices underwent merging at $\kappa h=1.88$, in contrast to no-merging  at $\kappa h=1.0\,\&\,1.5$ (Figs.~3, 4(b), 5(c) \& 6(b) in the supplementary document). 

\subsection{Periodic dynamics (Miscellaneous wake patterns)}
\label{periodic_flow_flexible_foil}

\begin{figure}
		\centering
		\includegraphics[scale=0.25]{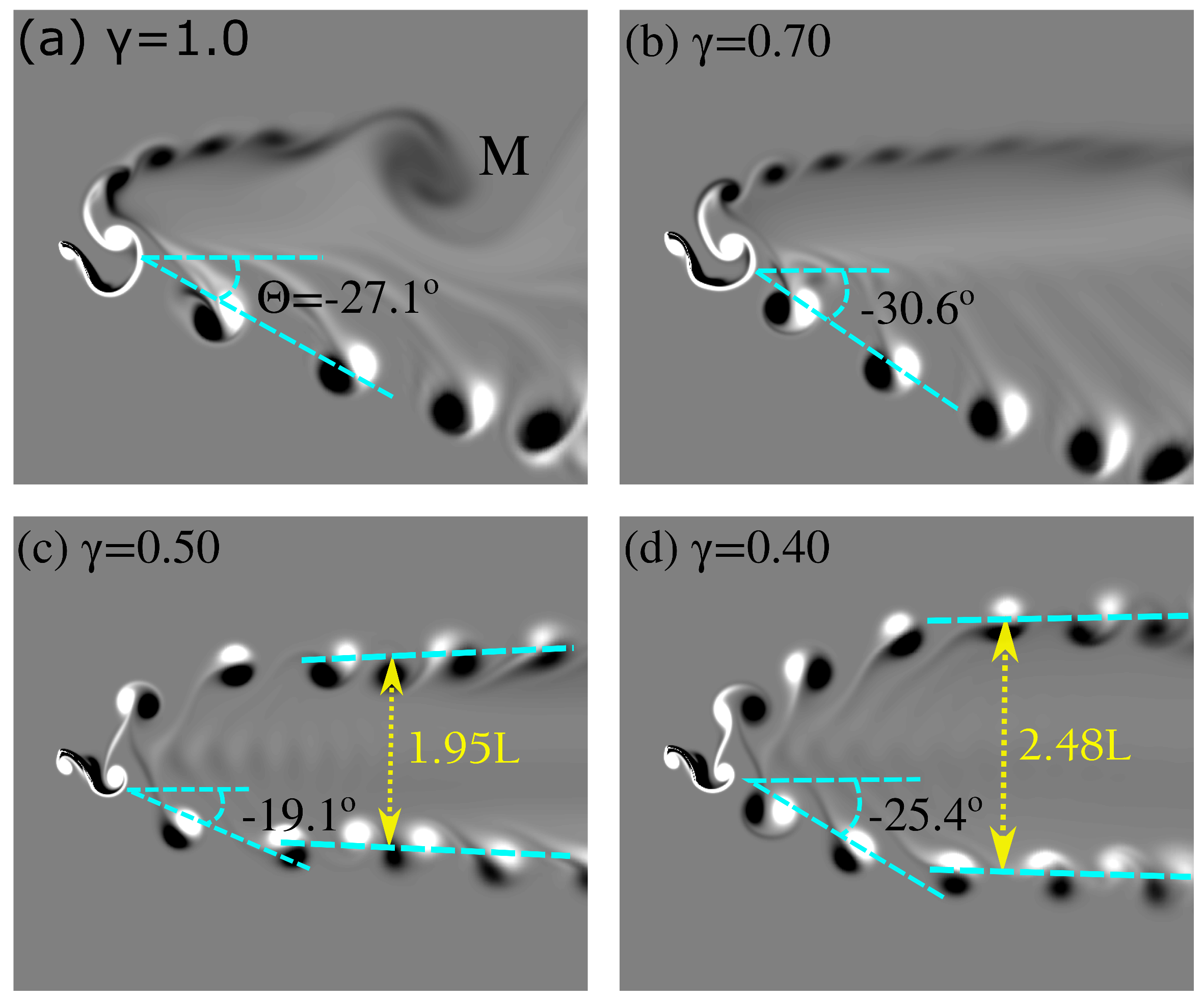}
		\vspace{6pt}
		\caption{At $\kappa h=2.3\,(\rm P):$ instantaneous vorticity contours at a typical time instant of $t/T=85.0$ showing (a) downward deflected wake with merging of secondary clockwise vortices at $\gamma=1.0$, (b) downward deflected wake without the merging of secondary vortices at $\gamma=0.7$, (c) and (d) bifurcated wake with two symmetric vortex streets at $\gamma=0.50$ and $0.40$, respectively. The inter-gap distances between the bifurcated vortex streets are marked at a stream-wise location of about $7L$.}
		\label{flow_field_gamma_1_05}
	\end{figure}

Continuing with the increase in flexibility, 
%it was possible to restore perfect periodicity in the flow even for the high $\kappa h$ value of 2.3. In
in the range of $\gamma=1.0$ -- $0.39$, periodicity was observed. Significant variation in the wake patterns was observed in this periodic range, whereas the rigid case had shown only undeflected reverse K\'arm\'an pattern (was shown in Fig.~\ref{Rigid_wake_pattern}(a)). Representative vorticity snapshots for different $\gamma$ have been given in Fig.~\ref{flow_field_gamma_1_05}. 
Deflected reverse K\'arm\'an wakes with secondary vortex streets (with merging and no-merging) and bifurcated wakes were observed as can be seen in the figure. 
%at different flexibility levels. 
%They have been discussed in more details in Subsections~\ref{sec:deflected_reverese_karman_wake} and \ref{sec:bifurcated_wake}.
% 5.3.2 and 5.3.3. 
% Within the optimal flexibility, the bending of the flexible foil facilitated a near-streamlined shape enabling the decrease in the size and strength of the LEV significantly, compared to rigid foil configuration at the same $\kappa h$. 
The flow-field mechanisms behind the recovery from chaos under flexibility as well as the other prominent wake patterns under periodic dynamics  have been discussed in the following.

\subsubsection{Flow-field mechanism leading to inhibition of chaos} \label{LEV_mechanism_periodic}

\begin{figure}
    	\centering
    		\includegraphics[scale=0.11]{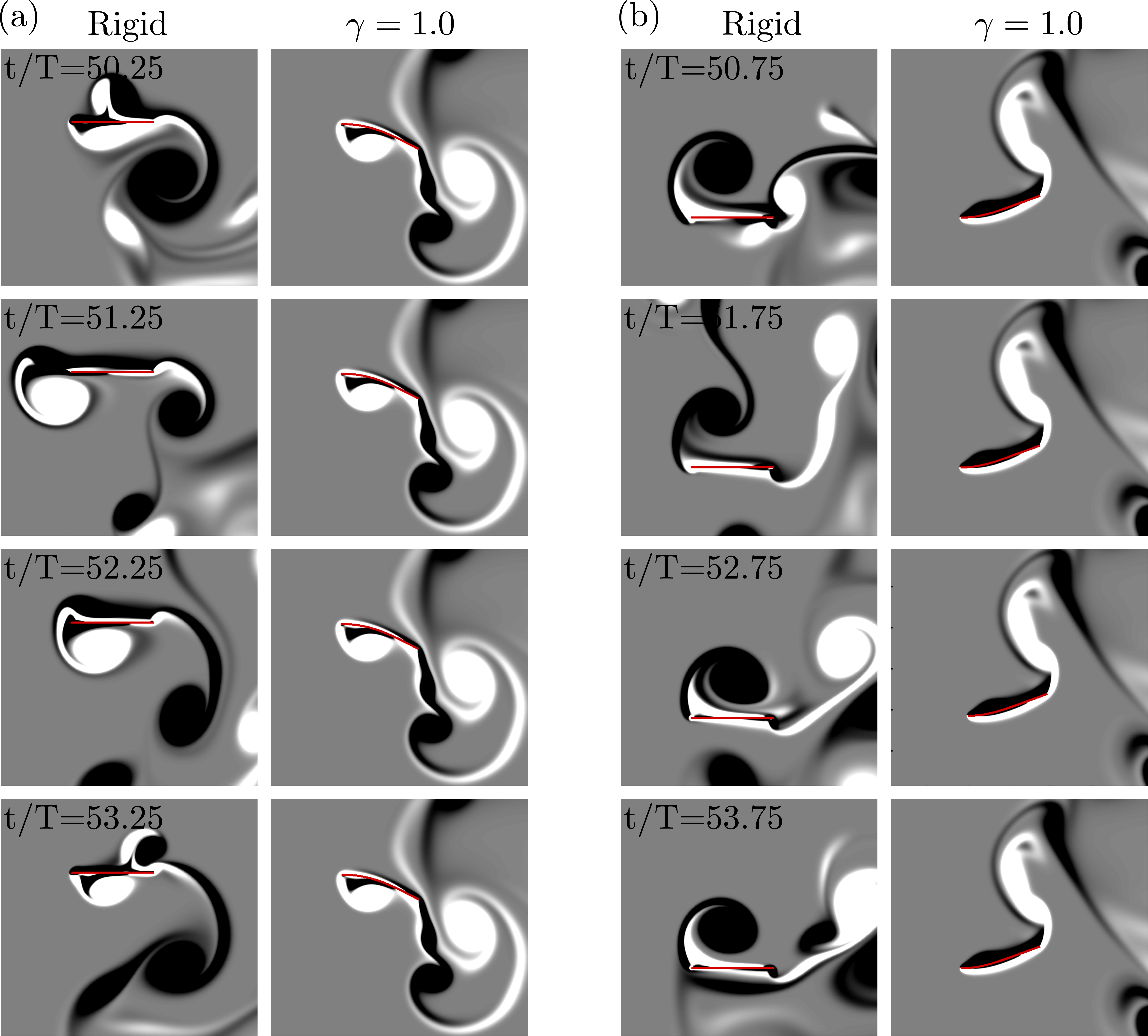}
    		\caption{Near-wake comparison between the rigid and the flexible  configuration for $\gamma=1.0$ at $\kappa h=2.3$: instantaneous vorticity fields during $51^{\rm st}$--$54^{\rm th}$ cycles (a) at the end of upstrokes and (b) at the end of downstrokes.}
    		\label{Vortex_compariosn_rigid_g_1}
    \end{figure}

LEV has been known to be the key vortex to usher aperiodicity in rigid flapping foils~\citep{bose2018investigating,bose2021dynamic,majumdar2020capturing}. Acknowledging that flexibility can influence the LEV-TEV interactions~\citep{shah2022chordwise} and modify the LEV behaviour, the  evolution of the LEVs and their interaction with TEVs in the near-field between the rigid and the flexible cases have been compared and presented for $\gamma=1.0$, during $51^{\rm st}-54^{\rm th}$ cycles at the end of  the upstrokes and downstrokes in Fig.~\ref{Vortex_compariosn_rigid_g_1}. Please note that $\gamma=1.0$ marked the onset of periodic wakes and hence has been chosen as a representative case. 
 The formation and the growth of the primary LEV took place aperiodically and the size and strength of the primary LEVs varied considerably from one cycle to another in the case of the rigid foil. The fully developed LEV eventually got separated and interacted with the TEV  in an aperiodic manner propagating aperiodicity in the near-field by forming spontaneous vortex couples. In contrast, the bending of the flexible foil facilitated a near-streamlined shape, enabling a decrease in the size and strength of the LEV significantly (the average circulation of the primary LEV decreases from 8.23 ( for rigid) to 2.92 (for $\gamma=1.0$)). This in turn, prevented its detachment and interaction with the TEV. Thus, an important agency for chaotic transition as observed in the flow-field of the rigid case, was absent. A downward vortex couple was generated at the trailing-edge periodically in every flapping cycle, leading to the eventual formation of a downward deflected wake. Moreover, during the downstroke, the LEV was completely suppressed. Therefore, the shape of the structure arising from the optimal bending and the suppression of the LEV formation were instrumental in reinstating the periodicity in the system. 

\subsubsection{Deflected reverse K\'arm\'an wake with a secondary vortex street}
\label{sec:deflected_reverese_karman_wake}

A primary wake of deflected reverse K\'arm\'an street and a secondary street with CW vortices were seen at $\gamma$ =1.0 but far-wake switch was no longer visible. Since this is a periodic regime, the CW vortices were seen to shed at every cycle (Fig.~\ref{flow_field_gamma_1_05}(a)). Multiple secondary vortices underwent complete merging in the far-wake region forming a large vortex (\textbf{M}) before convecting downstream. As the flexibility of the foil was increased, at $\gamma = 0.70$, the trailing wake still exhibited a deflected primary reverse K\'arm\'an street along with a secondary street of CW vortices; see Fig.~\ref{flow_field_gamma_1_05}(b). However,  merging of the vortices in the secondary street was not seen. It is due to the lower streamwise velocity at $\gamma=1.0$ in the region of the secondary vortex street, the secondary vortices got sufficient time to get close, interact and eventually merge together forming a stronger vortex before convecting downstream. On the contrary, the secondary vortices for $\gamma=0.7$ did not get enough time to interact as they got flushed away downstream quicker due to a higher streamwise velocity. The detailed explanation has been presented in Section 4 of the supplementary document. While characterizing the primary wake in terms of the overall wake deflection angle, $\Theta$~\citep{majumdar2020effect,shah2022chordwise}, it was observed that flexibility affected $\Theta$ significantly. To appreciate this, $\Theta$ has been plotted for $\gamma=1.0$ and $0.7$ in Fig.~\ref{flow_field_gamma_1_05}(a) and \ref{flow_field_gamma_1_05}(b), respectively, where the wake deflection increased with the increase in flexibility, from $\Theta=-27.1^o$ at $\gamma=1.0$ to $\Theta=-30.6^o$ at $\gamma=0.7$. The deflection angle of the wake depends on the dipole velocity of vortex couples that contribute to the formation of the streets. The detailed description has been presented in the supplementary document in Section 3.1. The continuous temporal evolution of the flow-field for $\gamma=1.0$ and $0.70$ have been presented in supplementary videos SV3 and SV4, respectively. At this flexibility level the system was periodic showing an undeflected wake at $\kappa h=1.0$; it showed periodicity with a deflected wake at $\kappa h=1.5\,\&\,1.88$ (Fig.~3 in the supplementary document).

\subsubsection{Bifurcated wake}
\label{sec:bifurcated_wake}

Further increase in flexibility within the periodic regime gave two symmetric vortex streets resulting in a bifurcated wake.  Such wake patterns have been reported in the literature~\citep{buchholz2006evolution,verma2022characterization}, though the systems were rigid and undergoing prescribed flapping motions. The present $\kappa h$ of $2.3$ is equivalent to a Strouhal number of $0.7$ which is close to that reported in~\citep{buchholz2006evolution}. Note that bifurcated patterns were not observed in the present study for the rigid system.  

The wakes at $\gamma=0.50,\,\&\,0.40$, presented  in Figs.~\ref{flow_field_gamma_1_05}(c) and  \ref{flow_field_gamma_1_05}(d), respectively, showed  two symmetric streets and the effect flexibility had on their deflection as well as their inter-gap distance. The deflection angle of the bifurcated streets was seen to increase with increasing flexibility, and so was the distance between them. The reasoning behind the increase in the deflection angle and the inter-gap distance between the streets can be made in terms of the self-induced velocity of the immediate/first vortex couple. For the sake of brevity, this discussion has been included in the supplementary document (Section 3.2). 
%Beyond the periodic range, a further increase in flexibility made the upper and lower vortex streets unstable, resulting in an intermingling of the bifurcated streets leading to aperiodicity once again, which will be discussed in Section~\ref{sec:quasiperiodic_intermingling_streets}.

\begin{figure}
		\centering
		\includegraphics[scale=0.19]{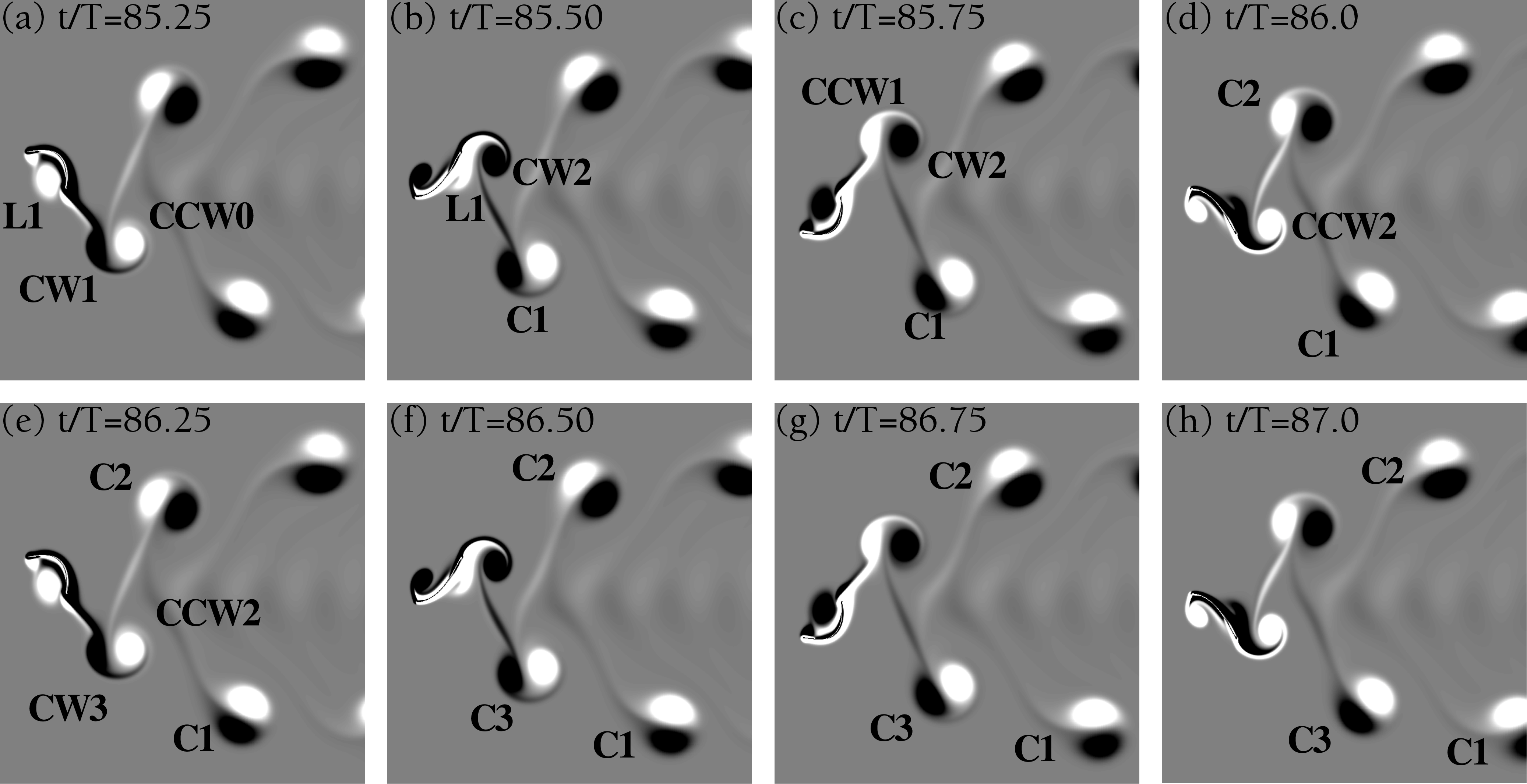}
		\vspace{6pt}
		\caption{At $\kappa h=2.3 \, \& \, \gamma=0.50 \,(\rm P):$ instantaneous vorticity contours during $86^{\rm th}$ and $87^{\rm th}$ cycles showing the mechanism of the formation of the bifurcated vortex street.}
		\label{flow_field_gamma_50}
	\end{figure}

To look into the formation of two symmetric vortex streets of the bifurcated wake, the vorticity contours of two consecutive cycles ($86^{\rm th}$--$87^{\rm th}$) at $\gamma = 0.50$ have been shown in Fig.~\ref{flow_field_gamma_50}. Unlike the cases of $\gamma = 1.0$ and $0.70$, where a couple and a single CW vortex were shed in every cycle, a pair of CW and CCW vortices were shed in each cycle for $\gamma = 0.5$ to $0.39$. Note that, in Figs.~\ref{flow_field_gamma_50}(a) -- \ref{flow_field_gamma_50}(b), the CW pair \textbf{CW1} and  \textbf{CW2} were shed during the first half of the cycle. Similarly in the second half, the CCW pair \textbf{CCW1} and \textbf{CCW2} were shed (Figs.~\ref{flow_field_gamma_50}(c) -- \ref{flow_field_gamma_50}(d)). \textbf{CW1} combined with \textbf{CCW0} (shed in the previous cycle) to form a downward deflecting couple \textbf{C1} (Figs.~\ref{flow_field_gamma_50}(a) -- \ref{flow_field_gamma_50}(b)) which eventually contributed to the formation of the downward deflected street. On the other hand, \textbf{CW2} formed the upward deflecting couple \textbf{C2} by combining with the immediate next CCW vortex \textbf{CCW1} (Fig.~\ref{flow_field_gamma_50}(c) -- \ref{flow_field_gamma_50}(d)). \textbf{C2} moved upwards as dictated by the combined effect of its self-induced velocity and the free-stream to form the upward-deflected street. The same chronology of events took place periodically in every cycle. Note that the mechanism behind the formation of bifurcated wake as reported  by~\citet{verma2022characterization} for a rigid foil was different from the present observations. In every half cycle, LEV formed a couple with the TEV. An approximate equality in their circulation strength and size were observed before they detached from the trailing-edge in every half-cycle. This unique characteristic resulted in such mutually induced velocity of the dipole-like pair, that helped in its movement from the wake center-line resulting in a bifurcated vortex street. In the present flexible case, the formation of two similar signed vortices in each half cycle was associated with two main phenomena: the trapping of the LEV due to excessive bending of the structure and the fast movement of the trailing part of the foil at this high flexibility, as explained in the following.

To continue the discussion with the help of the above presented typical cycles, a CCW LEV \textbf{L1} formed during $85^{\rm th}$ cycle was trapped under the lower surface of the foil due to its excessive bending (Fig.~\ref{flow_field_gamma_50}(a)). Delayed \textbf{L1} rolled over the foil surface to eventually get released from the trailing-edge as \textbf{CCW1} (Fig.~\ref{flow_field_gamma_50}(c)). On the other hand, fast movement of the trailing part changing its curvature resulted in the formation and shedding of \textbf{CCW2} (Fig.~\ref{flow_field_gamma_50}(c) -- \ref{flow_field_gamma_50}(d)). Note that higher flexibility/larger deformation meant a quicker movement of the trailing part of the body. Subsequently, a similar set of events occurred following the shedding of a CW LEV as well but with an opposite sense of rotation of the similar vortices. Notably, for $\gamma=1.0\,\&\,0.7$, such LEV trapping was not present as the foil did not undergo as significant bending as for $\gamma \in 0.50 - 0.39$. Moreover, the LEV formation occurred asymmetrically (only CCW LEV during the upstroke) for $\gamma \in 1.0-0.7$, leading to the formation of a single primary vortex street in the wake. On the other hand, for $\gamma \in 0.50 - 0.39$, LEVs were formed symmetrically (CCW and CW LEVs at up and down-strokes, respectively), which gave the bifurcated wake. The continuous temporal evolution of the flow-field for $\gamma = 0.50$ showing the formation of bifurcated vortex streets has been presented in supplementary video SV5. For this flexibility the system exhibited periodicity with an undeflected wake for  $\kappa h=1.0\,\&\,1.5$ and periodicity with a bifurcated wake for $\kappa h=1.88$, similar to the $\kappa h=2.3$ case  discussed above; see Figs.~3 and 6(c) of the supplementary document.

\subsection{Loss of order and return to chaos}

%To recall, a quasi-periodic intermittency route to chaos was observed in a rigid flapping system as $\kappa h$, the bifurcation parameter, was increased. With the introduction of flexibility at a high $\kappa h$ case showing chaotic behaviour, a gradual emergence of an ordered pattern was observed that brought back an organized wake. Hence f
Flexibility was found to be a useful instrument to establish order in an otherwise chaotic system as the above discussion has shown. However, with further increase in  flexibility, the system was seen to go back to aperiodic dynamics. The following subsections highlight the near-field interaction mechanisms through which the wake entered into robust chaos yet again.

\subsubsection{Quasi-periodicity again (Intermingling of vortex streets)} 
\label{sec:quasiperiodic_intermingling_streets}

\begin{figure}[t!]
		\centering
		\includegraphics[scale=0.25]{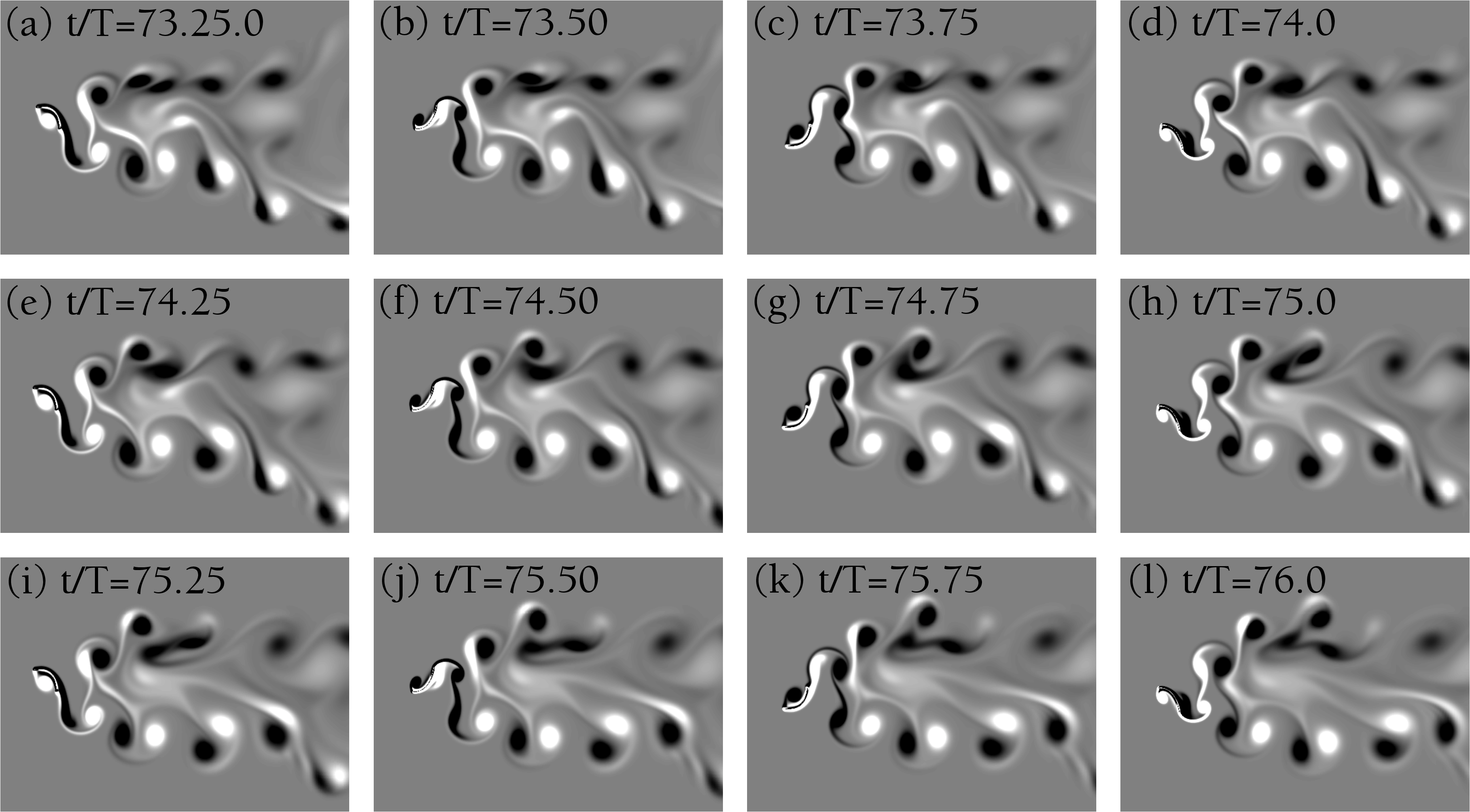}
		\vspace{6pt}
		\caption{At $\kappa h=2.3 \, \& \, \gamma=0.38\,(\rm QP-3):$ instantaneous vorticity contours during $74^{\rm th}$--$76^{\rm th}$ cycles showing a quasi-periodic flow-field.}
		\label{flow_field_gamma_38}
	\end{figure}

\begin{figure}
		\begin{subfigure}{.33\textwidth}
			\centering
			\includegraphics[scale=0.18]{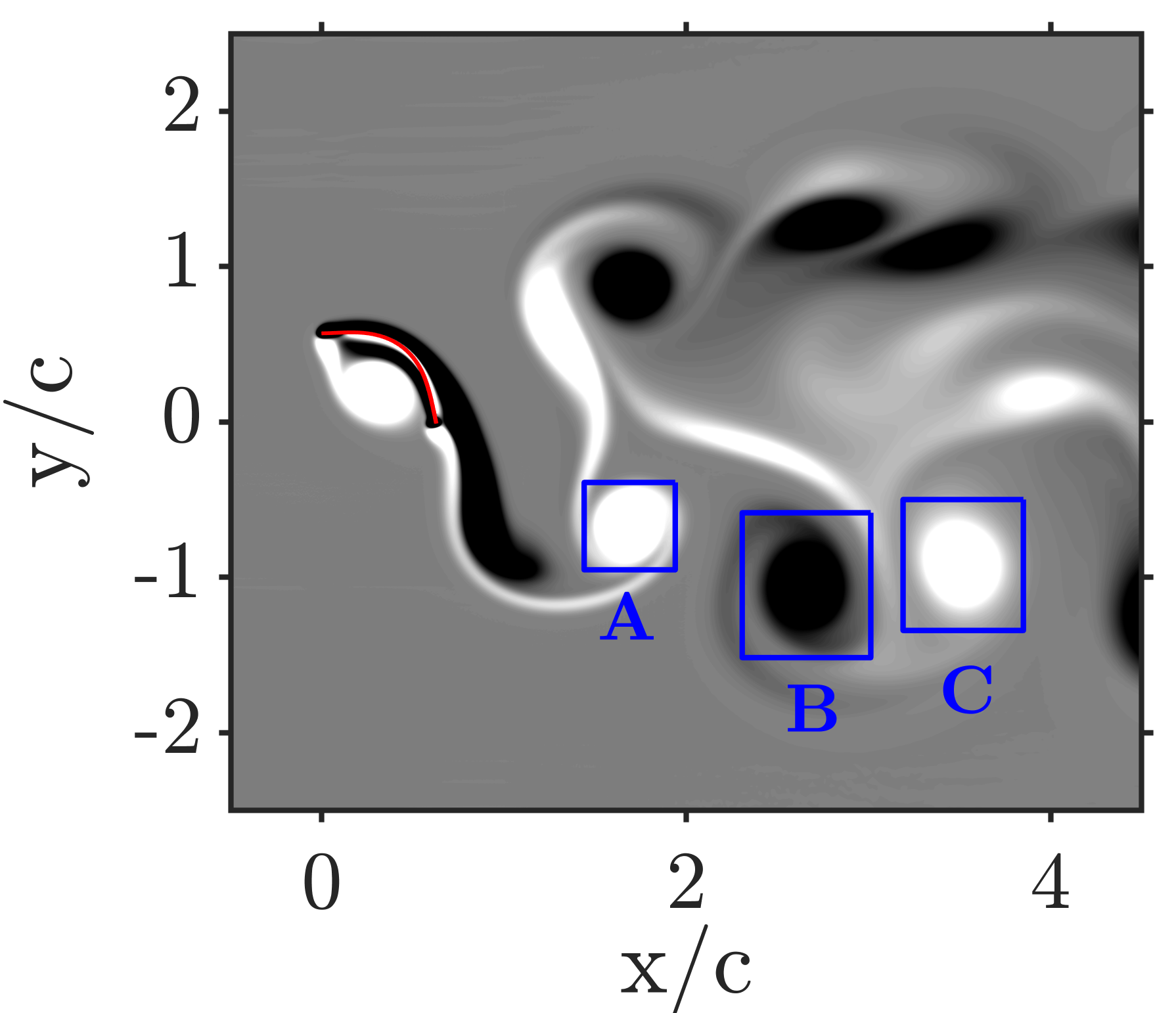}
			\caption{}
			\label{Vortex_cure_mark}
		\end{subfigure}
		\begin{subfigure}{.33\textwidth}
			\centering
			\includegraphics[scale=0.2]{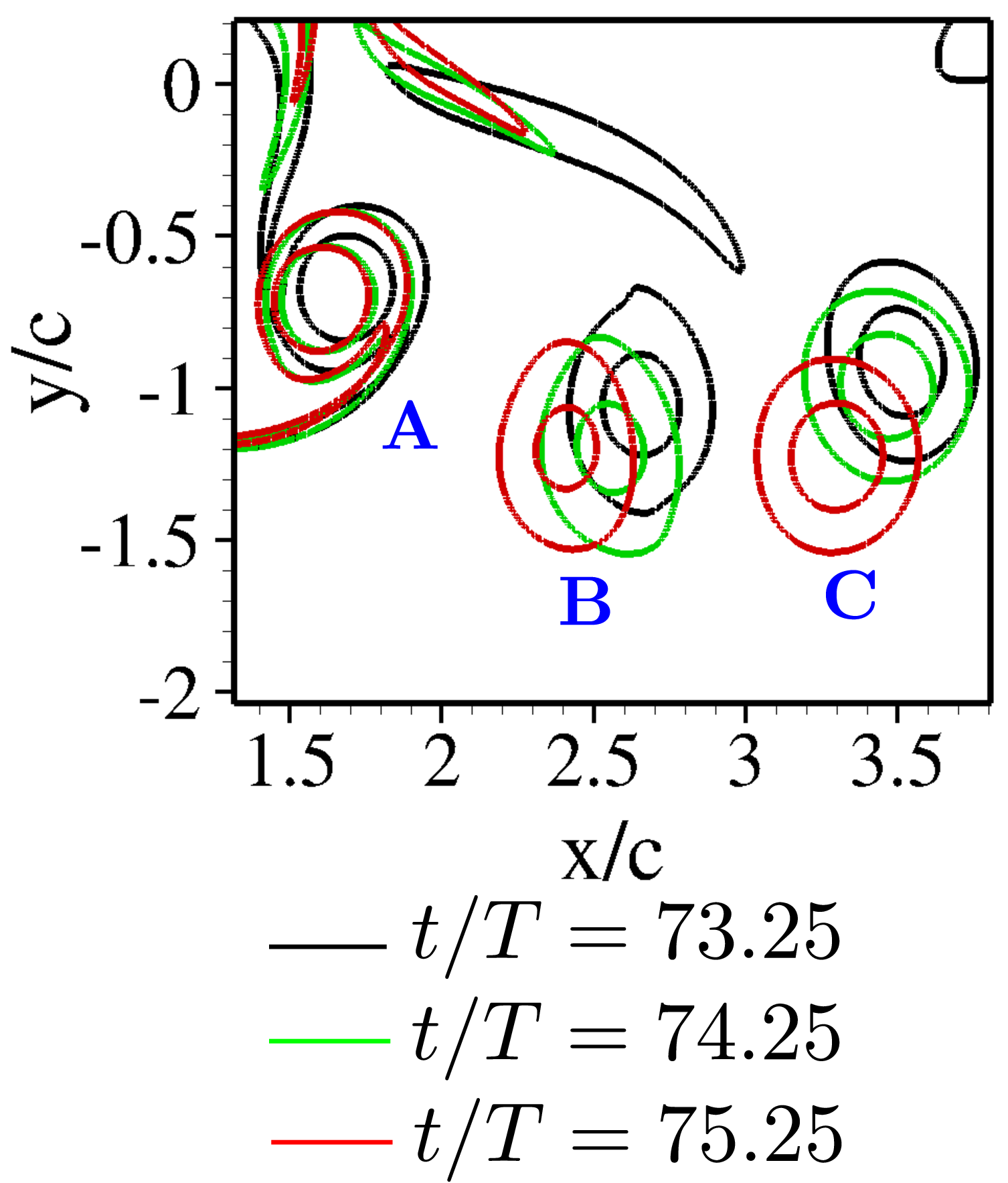}
			\caption{}
			\label{Vortex_core_line}
		\end{subfigure}%
		\begin{subfigure}{.33\textwidth}
			\centering
			\includegraphics[scale=0.18]{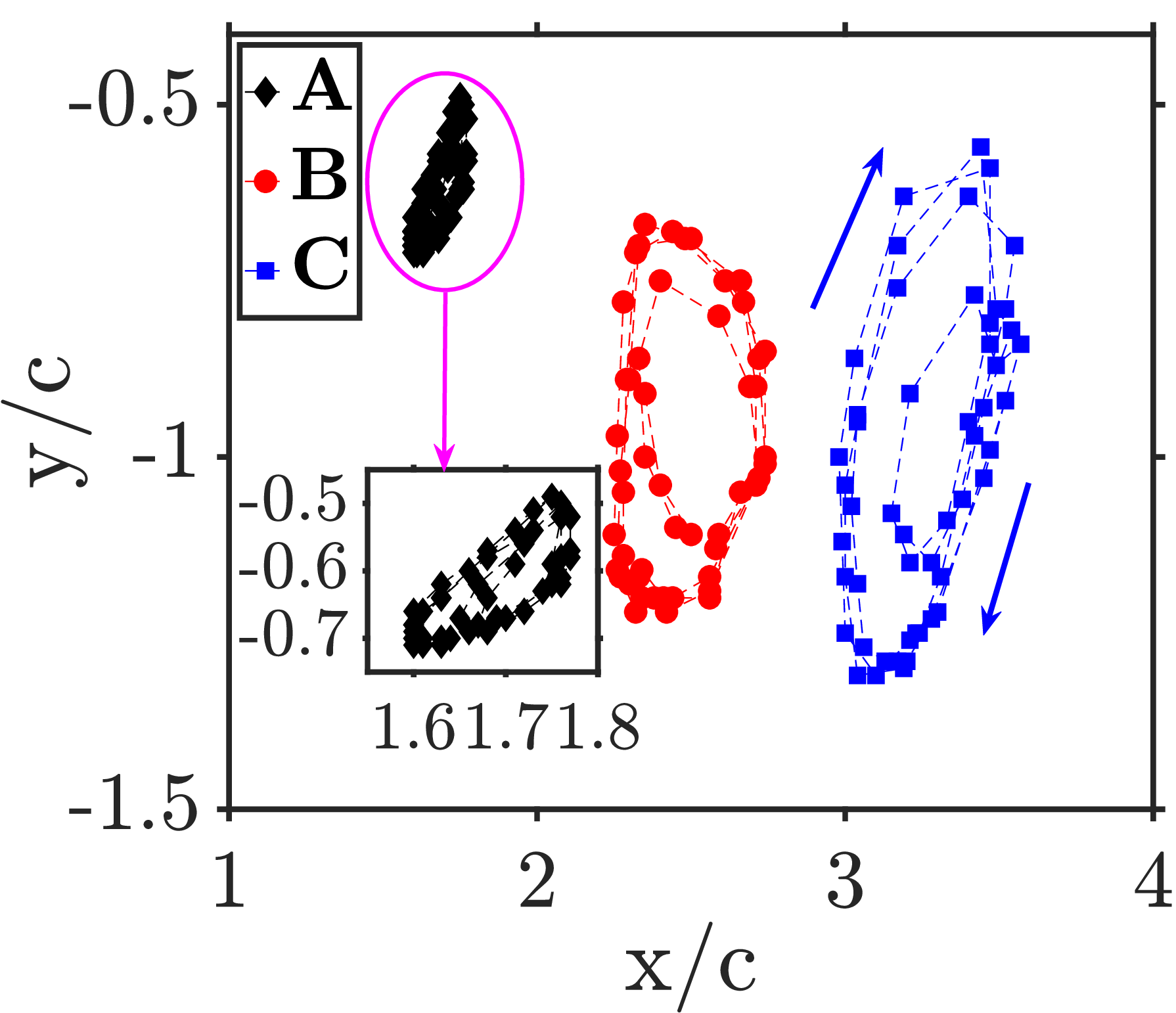}
			\caption{}
			\label{Vortex_core_plot}
		\end{subfigure}
		\caption{At $\kappa h=2.3 \, \& \, \gamma=0.38\,(\rm QP-3):$ (a) first CCW vortex, second CW vortex, and third CCW vortex are marked by $\mathbf{A}$, $\mathbf{B}$ and $\mathbf{C}$, respectively, (b) superposition of the vortex core locations of vortices $\mathbf{A}$, $\mathbf{B}$ and $\mathbf{C}$ for three consecutive cycles  $(t/T=73.25,\,74.25\,\&\,75.25)$, and (c) locations of vortex core centers of vortices $\mathbf{A}$, $\mathbf{B}$ and $\mathbf{C}$ for $51$ consecutive cycles $(t/T=70.25-120.25)$. The arrow shows the direction of shift in locations of vortex core centers with time forming a close loop, and the same direction is followed by other vortices also.}
		\label{Vortex_core_mark_plot}
	\end{figure}

%With further increase in flexibility, the system lost the ordered flow-field and became quasi-periodic again. This clearly suggests the presence of a critical threshold  for the chosen set of system parameters. 
The upper and lower vortex streets from the bifurcated wake was seen to come close to each other and the vortices from each intermingled in a not-so-ordered manner as flexibility increased further. There was a major manifestation of quasi-periodicity in the flow-field (Fig.~\ref{flow_field_gamma_38} for $\gamma = 0.38$). Although the formation of the vortex structures on the upper and lower vortex streets looked similar in successive cycles, they exhibited small shifts in their vortex core locations and strengths. This has been demonstrated in Fig.~\ref{Vortex_core_mark_plot}, where the first three vortices have been marked as $\mathbf{A}$, $\mathbf{B}$ and $\mathbf{C}$ as typical representatives of CCW and CW vortices in the near-field. They have been plotted together in Fig.~\ref{Vortex_core_mark_plot}(b) for three consecutive cycles  to highlight the shift in the vortex core locations. As the cores of $\mathbf{A}$, $\mathbf{B}$ and $\mathbf{C}$ were tracked, they were seen to deviate a little from their earlier occupied positions from one cycle to another in consecutive cycles. Their trajectory formed a closed loop, as shown in Fig.~\ref{Vortex_core_mark_plot}(c). This  pattern is a typical signature of quasi-periodic dynamics~\citep{heathcote2007jet}. The rigid foil also exhibited quasi-periodic dynamics  with a deflected wake. However, in this flexible case, the shifts in the near-wake vortices due to quasi-periodicity, caused the upper and the lowers streets of the bifurcated wake to come closer to each other at certain time instants, compared to the higher $\gamma$ cases. This facilitated the intermingling of the vortices from the upper and lower streets.

    \begin{figure}[b!]
        \begin{subfigure}{.5\textwidth}
			\centering
			\includegraphics[scale=0.25]{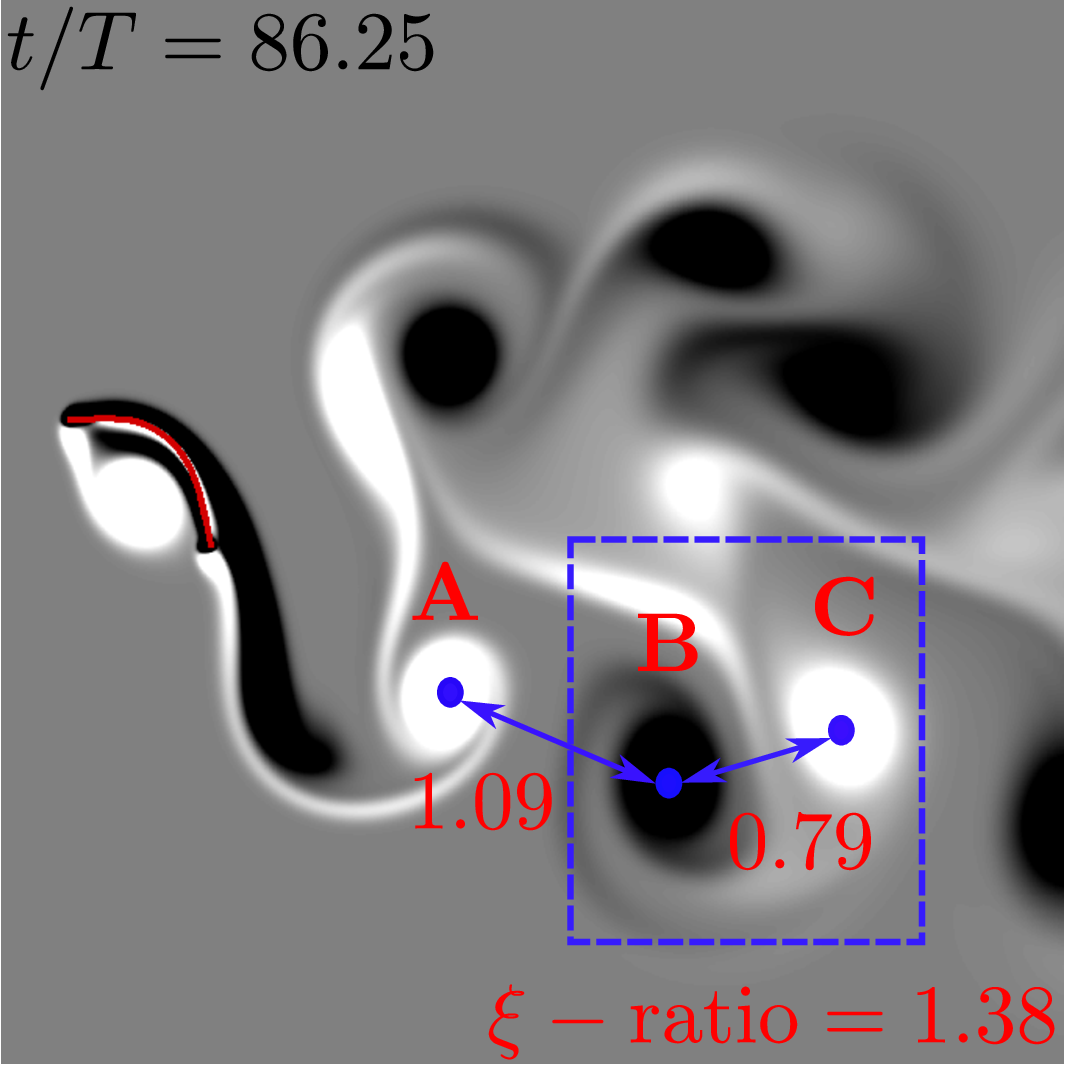}
			\caption{}
			\label{vortices_zeta_ratio_86_g_038}
		\end{subfigure}
        \vspace{6pt}
		\begin{subfigure}{.5\textwidth}
			\centering
			\includegraphics[scale=0.25]{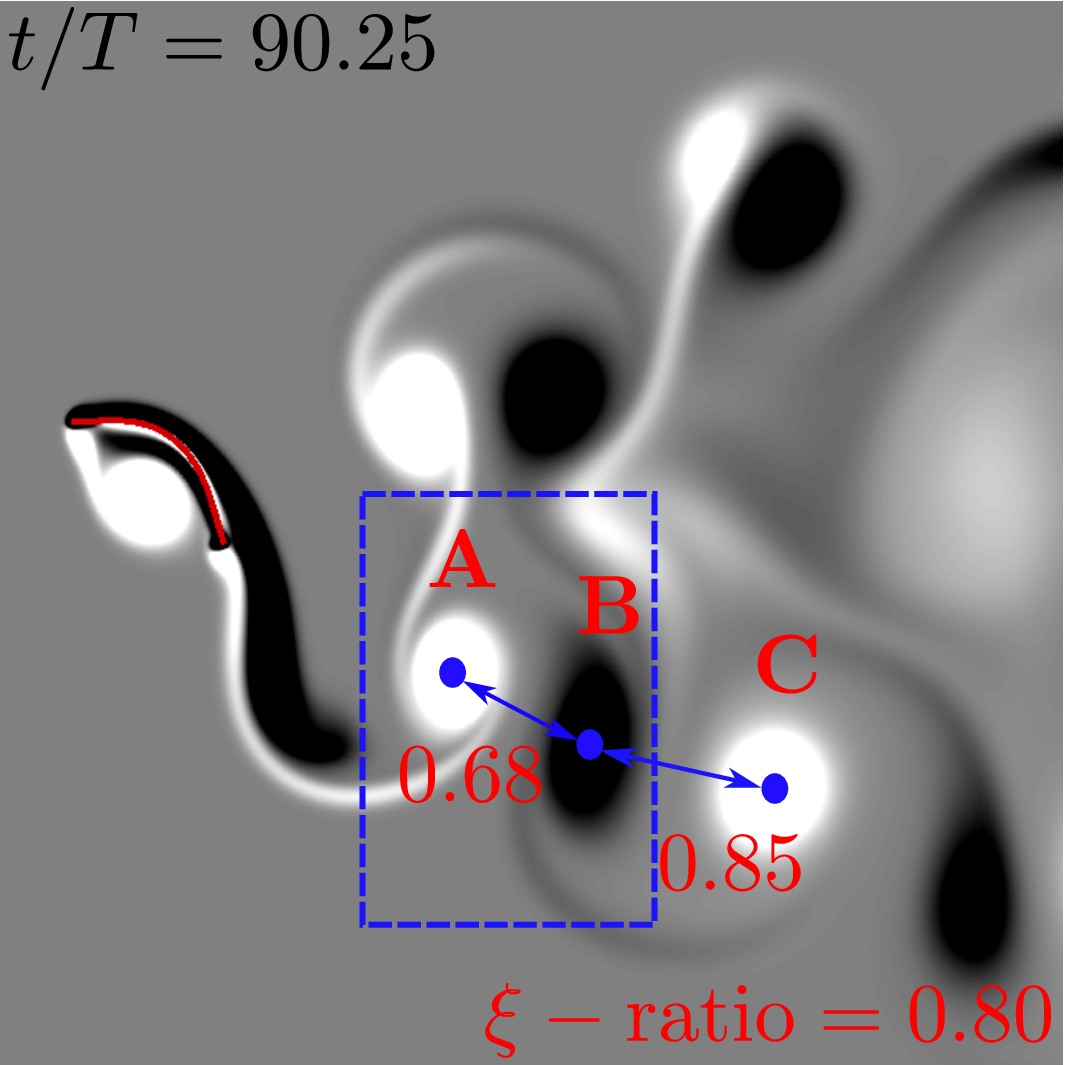}
			\caption{}
			\label{vortices_zeta_ratio_90_g_038}
		\end{subfigure}
        \vspace{6pt}
		% \begin{subfigure}{.5\textwidth}
		% 	\centering
		% 	\includegraphics[scale=0.18]{zeta_ratio_TH_lower_couple.eps}
		% 	\caption{}
		% 	\label{zeta_ratio_g_038}
		% \end{subfigure}
		\begin{subfigure}{1.0\textwidth}
			\centering
			\includegraphics[scale=0.2]{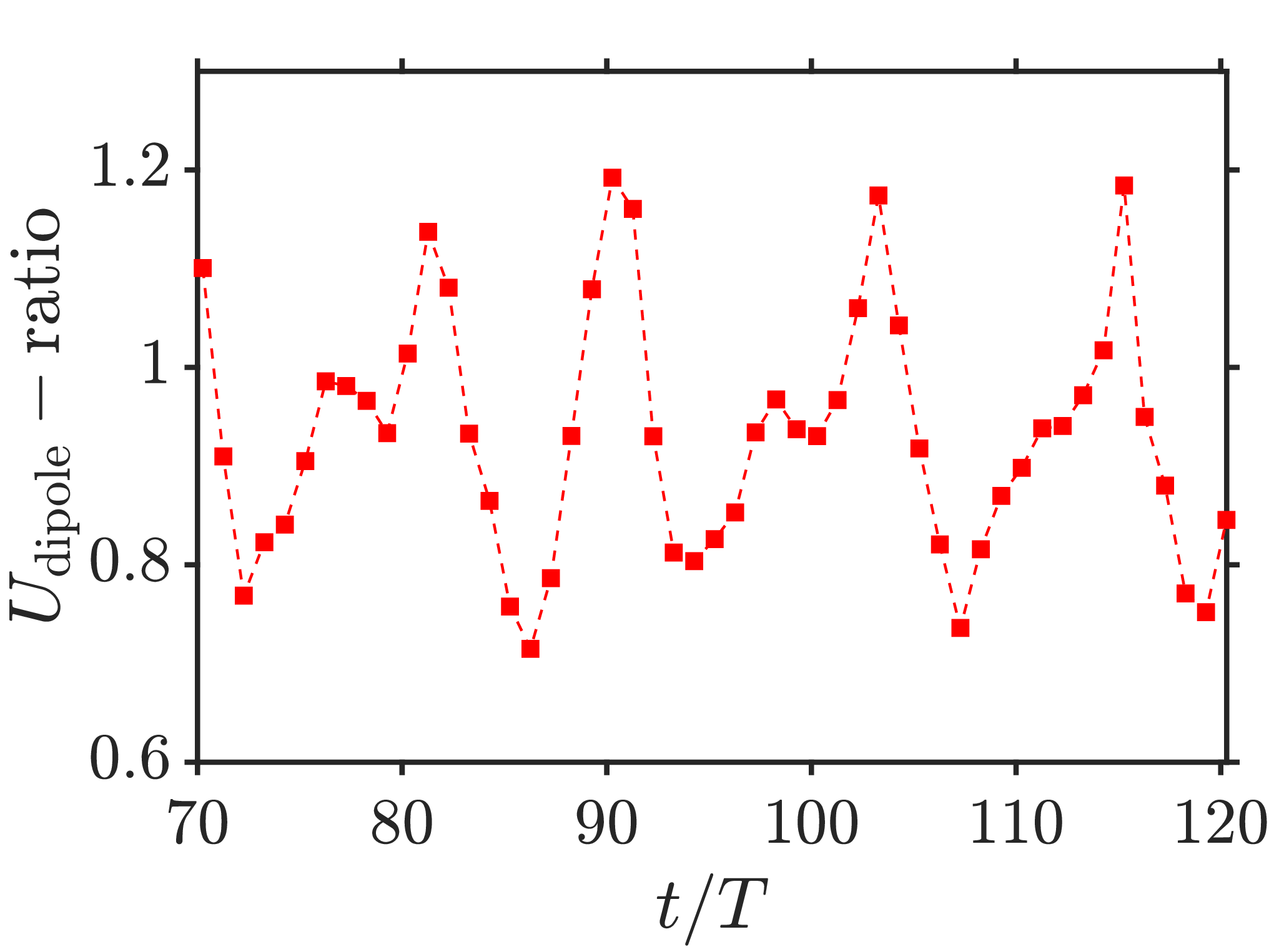}
			\caption{}
			\label{dipole_ratio_g_038}
		\end{subfigure}
		\caption{At $\kappa h=2.3 \, \& \, \gamma=0.38\,(\rm QP-3):$ (a) showing the vortices $\mathbf{B}$ and $\mathbf{C}$ forming a dominant couple, (b) vortices $\mathbf{A}$ and $\mathbf{B}$ forming a dominant couple, (c) variation of $\xi$-ratio, and (d) $U_{\rm dipole}$-ratio for $51$ consecutive cycles $(t/T=70.25-120.25)$.}
		\label{zeta_dipole_ratio_g_038}
	\end{figure}

\begin{figure}
		\centering
		\includegraphics[scale=0.25]{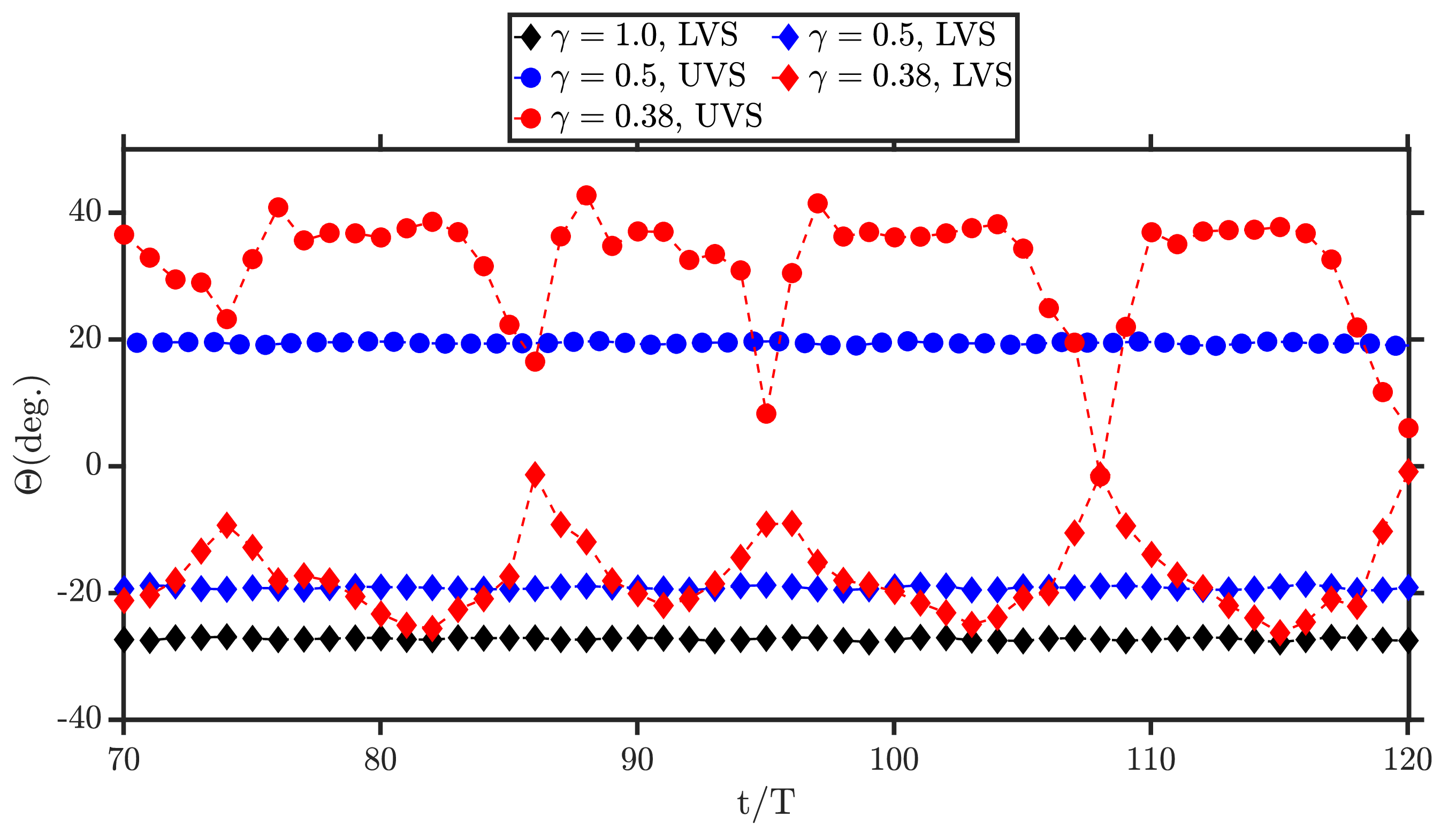}
		%\vspace{0pt}
		\caption{At $\kappa h=2.3 \, \& \, \gamma=0.38\,(\rm QP-3):$ variation of angle of deflection $(\Theta)$ with time for $\gamma=1.0,\,0.5$ and $0.38$; LVS $\rightarrow$ Lower Vortex Street and UVS $\rightarrow$ Upper Vortex Street.}
		\label{deflection_angle_g_038}
	\end{figure}

To intermingling of the streets is described in the following. The change in the vortex street gap distance due to shifts in the core locations was computed  in terms of the distance ratio of vortices \textbf{A}, \textbf{B} and \textbf{C} marked in Fig.~\ref{Vortex_core_mark_plot}(a), $\xi_{\rm ratio}$. Here, $\xi_{\rm ratio} = \xi_{\mathbf{AB}} / \xi_{\mathbf{BC}} = $ distance between \textbf{AB} / distance between \textbf{BC}. In due course, when \textbf{A} and \textbf{B} came close to each other (\textit{i.e.} $\xi_{\rm ratio}<1$), they formed an upward deflecting couple  \textbf{A-B},  dictated by their self-induced velocity (Fig.~\ref{zeta_dipole_ratio_g_038}(b) marking the $\xi_{\rm ratio}$ as 0.80 for the chosen time instant). Similarly, when \textbf{B} and \textbf{C} came close together (\textit{i.e.} $\xi_{\rm ratio}>1$), a downward deflecting couple \textbf{B-C} was formed (Fig.~\ref{zeta_dipole_ratio_g_038}(a) marking the $\xi_{\rm ratio}$ as 1.38 for the chosen time instant). The dominance of \textbf{AB} or \textbf{BC} was also quantified in terms of the ratio of their self-induced dipole velocity, $U_{\rm dipole}$-ratio ($= U_{\rm dipole}^{\mathbf{AB}} / U_{\rm dipole}^{\mathbf{BC}}$) \citep{godoy2008transitions}. Here, $U_{\rm dipole}$ is the self-induced dipole velocity given by,  $U_{\rm dipole} = \frac{\Gamma_{\rm avg}}{2\pi \xi}$ \citep{anderson2010fundamentals}, with $\Gamma_{\rm avg}$ being the average of the absolute circulation values of the two partners. Evidently,  $U_{\rm dipole}$-ratio $>1$ indicates the dominance of \textbf{AB}  and  $U_{\rm dipole}$-ratio $<1$ indicates the dominance of \textbf{BC}. The dipole velocity ratio has been chosen here to track the change  in the dominance of the couples with time (Fig.~\ref{zeta_dipole_ratio_g_038}(c)). A detailed description of the technique to compute  $\xi_{\rm ratio}$ and $U_{\rm dipole}$-ratio, is available in~\citep{majumdar2020effect, shah2022chordwise}. Dominant upward deflecting couple \textbf{AB} (Fig.~\ref{zeta_dipole_ratio_g_038}(b)) pulled the downward deflected street upwards, decreasing the distance between the upper and the lower streets, which eventually helped in intermingling of the vortices from the upper and lower streets. With further decrease in $\gamma$, this became more pronounced, resulting in irregular wake behaviour and chaos. The interaction between the upper and lower streets in the quasi-periodic case was captured in their corresponding wake deflection angles presented in Fig.~\ref{deflection_angle_g_038}. The deflection angles of both streets changed with time, and their intermingling was indicated by their fluctuations in sync with each other. This is in contrast to the periodic cases ($0.39\le \gamma \le 1.0$), where the deflection angle remained constant with time, as there were no shifts in the vortex core positions in different cycles, neither were there any changes in the relative dominance of the vortex couples (not shown here). The continuous temporal evolution of the flow-field for $\gamma = 0.38$ has been presented in supplementary video SV6.

\subsubsection{Intermittency (Quasi-periodic time windows interspersed with chaotic time windows)}

\begin{figure}[t!]
		\centering
		\includegraphics[scale=0.15]{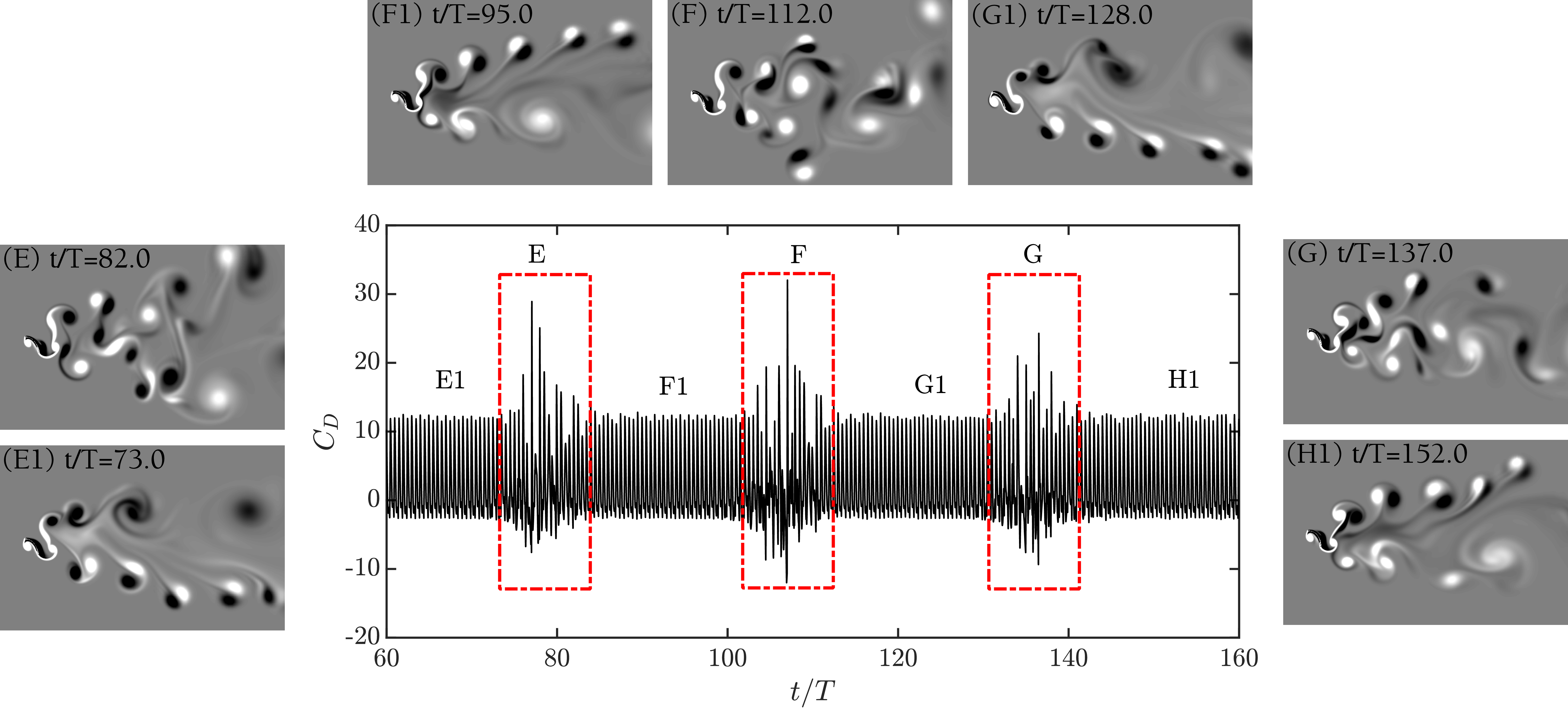}
		\vspace{6pt}
		\caption{At $\kappa h=2.3 \, \& \, \gamma=0.37\,(\rm INT):$ $C_D$ time history and instantaneous vorticity contours during quasi-period (`E1', `F1', `G1' \& `H1') and chaotic windows (`E', `F' \& `G') (dynamical state of intermittency).}
		\label{flow_field_gamma_37}
	\end{figure}

This was encountered at $\gamma = 0.37$. Irregular chaotic bursts were seen in between the quasi-periodic time windows of the load time history at $\gamma = 0.37$ (similar to $\gamma = 6.0$). A representative time length was chosen where the chaotic windows have been marked by rectangular boxes `E', `F' \& `G', and the quasi-periodic windows with `E1', `F1', `G1' \& `H1' in Fig.~\ref{flow_field_gamma_37}. The characteristic vorticity snapshots corresponding to different quasi-periodic and chaotic windows have  also been presented alongside. Note that intermingling between the upper and lower vortex streets disturbed the bifurcated wake pattern of the periodic case and it came back to a primary and a weaker secondary wake during the quasi-periodic time windows of intermittency. We considered the above-mentioned time windows which appeared in a chronological manner. During the `E1' quasi-periodic window, a downward deflected reverse K\'arm\'an primary street was seen, and the CW secondary vortices merged quasi-periodically. 
    \begin{figure}[t!]
		\centering
		\includegraphics[scale=0.26]{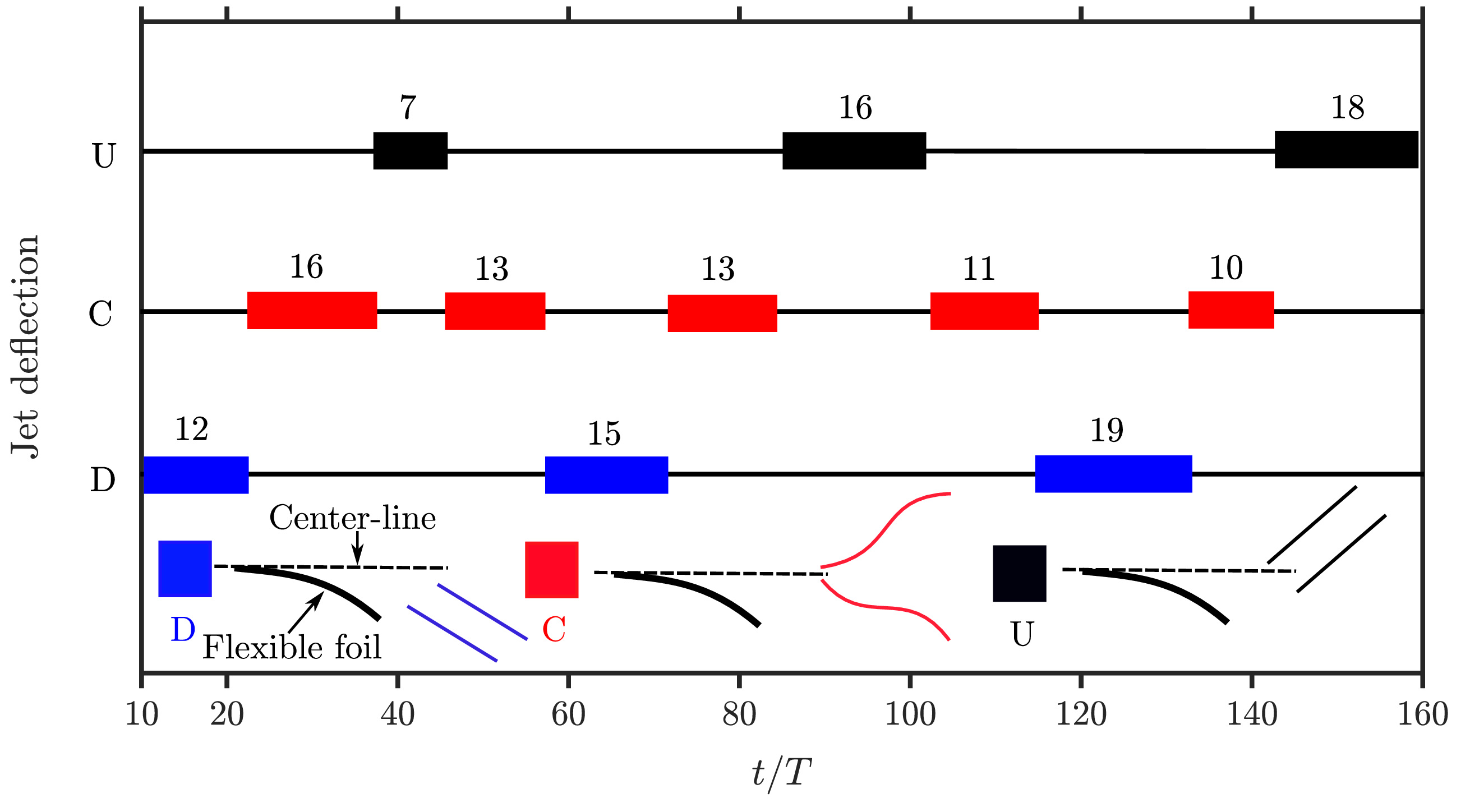}
		\vspace{6pt}
		\caption{At $\kappa h=2.3 \, \& \, \gamma=0.37\,(\rm INT):$ wake patterns with different jet deflection directions; $\rm U \rightarrow$ upward deflected wake, $\rm C \rightarrow$ chaotic wake and $\rm D \rightarrow$ downward deflected wake. Note that the number over the boxes denotes the duration of each window in terms of the number of flapping cycles.}
		\label{Intermittency_windows_gamma_37}
	\end{figure}
The merging process of the secondary vortices to form a bigger vortex continued for a few flapping cycles before it moved downstream  (Fig.~\ref{flow_field_gamma_37}(E1)). Subsequently, the flow transitions into a chaotic state during time window `E' (Fig.~\ref{flow_field_gamma_37}(E)) through a rapid intermingling of the streets and other vortex interaction processes. In the subsequent quasi-periodic window, `F1', the primary vortices were seen to be deflected in the upward direction, and secondary CCW vortices formed the secondary vortex street (Fig.~\ref{flow_field_gamma_37}(F1)). After chaotic window `F' (Fig.~\ref{flow_field_gamma_37}(F)), the primary reverse K\'arm\'an street got deflected downwards once again (Fig.~\ref{flow_field_gamma_37}(G1)), whereas after `G', the deflection was upwards (Fig.~\ref{flow_field_gamma_37}(H1)). Thus every intermittent chaotic burst reversed the direction of the primary wake. This is in contrast to the observation made for the rigid case at $\kappa h=1.88$ (Fig~\ref{Rigid_wake_pattern}(c)), where jet-switching did not occur after every chaotic burst. However, the mechanism for altering the deflection direction was similar in both rigid and flexible situations. Recall that at higher bending rigidity ($\gamma=6.0$), jet-switching was not observed even though the system exhibited dynamical state of intermittency. Recurrence-based proof of this alternating dynamical behaviour has been presented in Figs.~2(i) \& 2(j) of the supplementary document. The switchings in the deflection direction of the primary street and the time duration spent in any particular wake mode (upward deflected, downward deflected or chaotic wake) have been further presented in Fig.~\ref{Intermittency_windows_gamma_37}. The continuous temporal evolution of the flow-field for $\gamma = 0.37$, showing the deflection direction of the primary vortex street changing from downward to upward and vice-versa, has been  resented in supplementary video SV7. At this same  flexibility level the system exhibited periodicity with an undeflected wake for $\kappa h=1.0$, periodicity with a deflected wake for  $\kappa h=1.5$, and intermittency with near-wake switching for $\kappa h=1.88$; see Figs.~3 and 6(d) in the supplementary document.

\subsubsection{Back to sustained chaos (Unpredictable irregular flow-field)}

\begin{figure}[t!]
		\centering
		\includegraphics[scale=0.2]{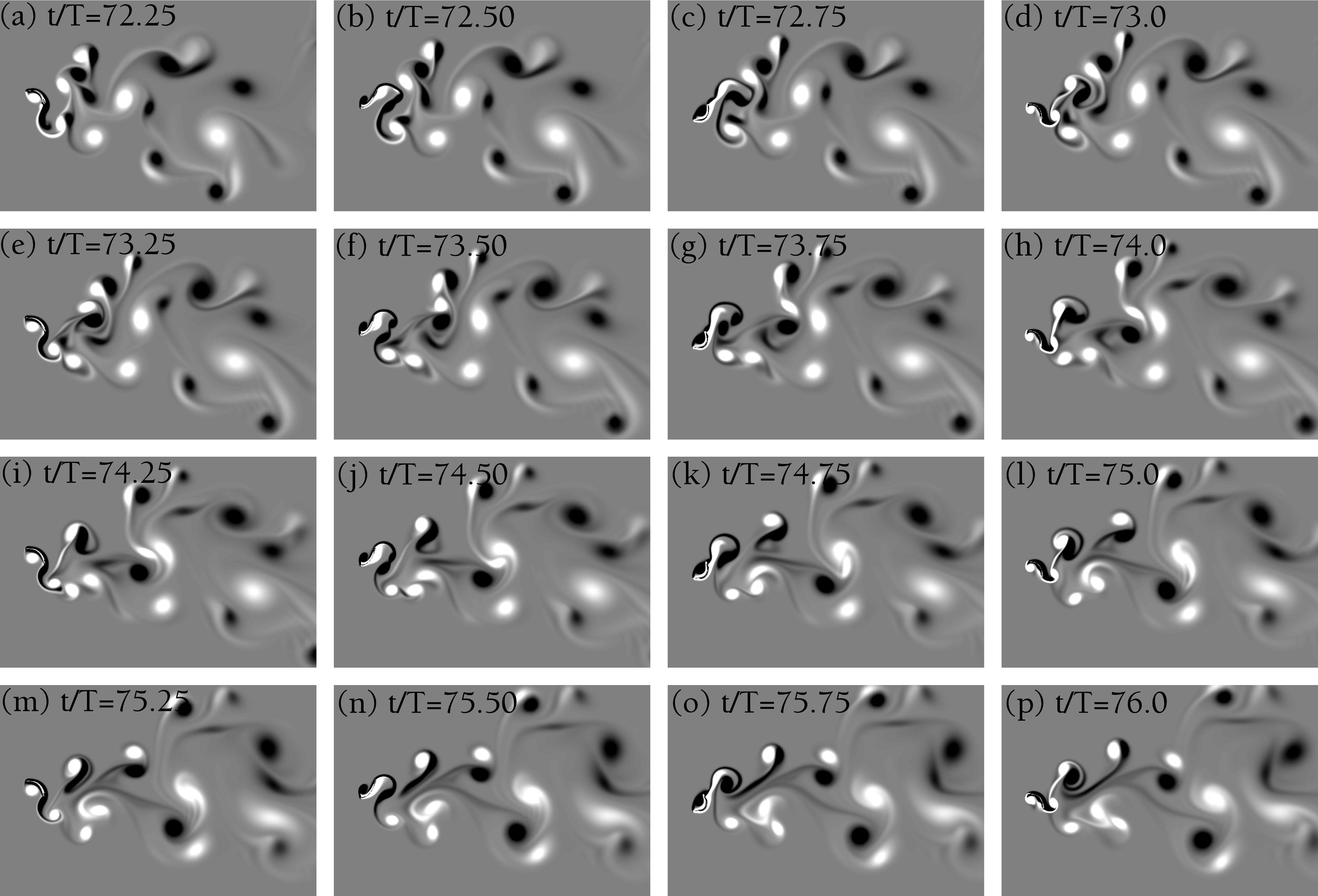}
		\vspace{6pt}
		\caption{At $\kappa h=2.3 \, \& \, \gamma=0.3\,(\rm SC):$ instantaneous vorticity contours during $73^{\rm rd}$--$76^{\rm th}$ cycles displaying sustained chaos.}
		\label{flow_field_gamma_03}
	\end{figure}

At the highly flexible case of $\gamma = 0.30$, the flow-field returned to chaos  (Fig.~\ref{Rigid_wake_pattern}(d)). The chronology of vortex interactions for four typical consecutive cycles $73^{\rm rd}$--$76^{\rm th}$ have been presented in Fig.~\ref{flow_field_gamma_03}. There was no correlation between the flow-field snapshots in different cycles, and the long-term behaviour of the flow-field became completely unpredictable and erratic. The width of the wake increased significantly due to the free movement of the moving couples and their subsequent collisions with other flow structures in the wake. The fundamental vortex interactions took place spontaneously at different time instances and spatial locations in quick succession without having any regularity in their occurrence which sustained chaos in the far-field. The continuous temporal evolution of the flow-field for $\gamma=0.30$ depicting sustained chaos is presented in supplementary video SV8. The flow-field mechanism behind the chaotic transition was  different from the rigid case and a comparison of the near-field interactions  have been included in the supplementary document (Section 7).

Note that  at this flexibility level the system exhibited periodicity with an undeflected wake at $\kappa h=1.0$, periodicity with a deflected wake at $\kappa h=1.5$ and  intermittency with near-wake switching at $\kappa h=1.88$. Moreover, quasi-periodicity was observed with a far-wake switch (shown in the supplementary for $\gamma = 0.1 and 0.05$) and  it entered into sustained chaos for $\gamma \le 0.01$, for $\kappa h=1.0$.  For $\kappa h=1.5$, quasi-periodic dynamics with a deflected wake was observed  at $\gamma=0.1$ (shown  in the supplementary) and chaotic dynamics was observed for $\gamma \le 0.05$. For $\kappa h=1.88$, sustained chaos was observed for $\gamma \le 0.1$. These results have been summarised in the supplementary document (Fig.~3). Thus, chaos was inevitably seen for other $\kappa h$ cases as well, albeit at different flexibility thresholds.

A few simulations were also carried out at $Re=500$ and $\kappa h=2.3$ for different bending rigidity values, to understand the robustness of the present qualitative trend of the results. The observations were similar: introduction of flexibility was seen to regularise  the flow-field up to a certain flexibility level, beyond which chaos reappeared. However, the quantitative transitional onsets were altered. Detailed flow-field snapshots for this case have been given in the supplementary document (Section 6, Fig. 15). Similar observations were also made with different flapper thicknesses and the corresponding results have been included in Section 5 of the supplementary document.

\textcolor{black}{As expected in a two-way coupled dynamical system, the structural deformation profile also followed the same   dynamical transitions observed in the flow-field  (regular/periodic for $\gamma \in (1.0,0.39)$, and irregular beyond the critical range of $0.39$). Note that very large deformation and irregular bending of the flapper at high flexibility could be expected to induce   strong feedback to the flow-field and instrumental in ushering aperiodicity in the system (discussed in more details in Section 7 of supplementary). However,  pinpointing the exact aperiodic trigger in a flow-structure coupled system is difficult. Changes in the deformation profile and the nature of the trailing-edge trajectories have been  presented in Section~8 of the supplementary document.}

\section{Propulsive performance and linking it to the dynamical changes}
\label{sec:propulsive_analysis}

The present flexible flapper could act as a simplified canonical model for animal locomotion observed in nature, and the present findings could be useful towards the design of biomimetic flapping devices. In the context of realistic engineering applications, it is also important to look into the role of flexibility on the overall propulsive characteristics. In this section, the propulsion characteristics of the system and the related efficiency have been analysed. One of the main ideas is to see whether periodicity and optimal propulsion go hand in hand.  In Section~\ref{sec:flexible_dynamical_transition} for the chosen $\kappa h$, a range of flexibility ($\gamma = 4.0-0.39$) was seen for which the wake was semi-regularized (quasi-periodic and periodic); otherwise, the wake exhibited strong aperiodic dynamics. The cycle-averaged thrust, the input power coefficients ($\overline{C}_T$ and $\overline{C}_P$, respectively), and the propulsive efficiency $(\eta)$ were calculated as given below,

 \begin{equation}
     \overline{C}_T=-\frac{\frac{1}{T}\int_{0}^{T}\int_{0}^{L} F_x(s,t)\,ds\, dt}{\frac{1}{2}\rho_f U_{\infty}^2L},
 \end{equation}
  \begin{equation}
     \overline{C}_P=\frac{\frac{1}{T}\int_{0}^{T} \int_{0}^{L} -\mathbf{F}(s,t).\frac{\partial \mathbf{X}(s,t)}{\partial t}\, ds\, dt}{\frac{1}{2}\rho_f U_{\infty}^3L},
 \end{equation}
  \begin{equation}
     \eta=\frac{\overline{C}_T}{\overline{C}_P}.
 \end{equation}
 
Here, $\mathbf{F}(s,t)$ is the fluid force acting on the body and $\frac{\partial \mathbf{X}(s,t)}{\partial t}$ is the velocity of any point on the body. $T$ is the time period $=1/f_s$, $f_s$ being the heaving frequency. $\overline{C}_T$ and $\overline{C}_P$ were calculated for over $50$ cycles discarding the initial transients. 

\begin{figure}[t!]
		\vspace{6pt}
		\begin{subfigure}{.5\textwidth}
			\centering
			\includegraphics[scale=0.19]{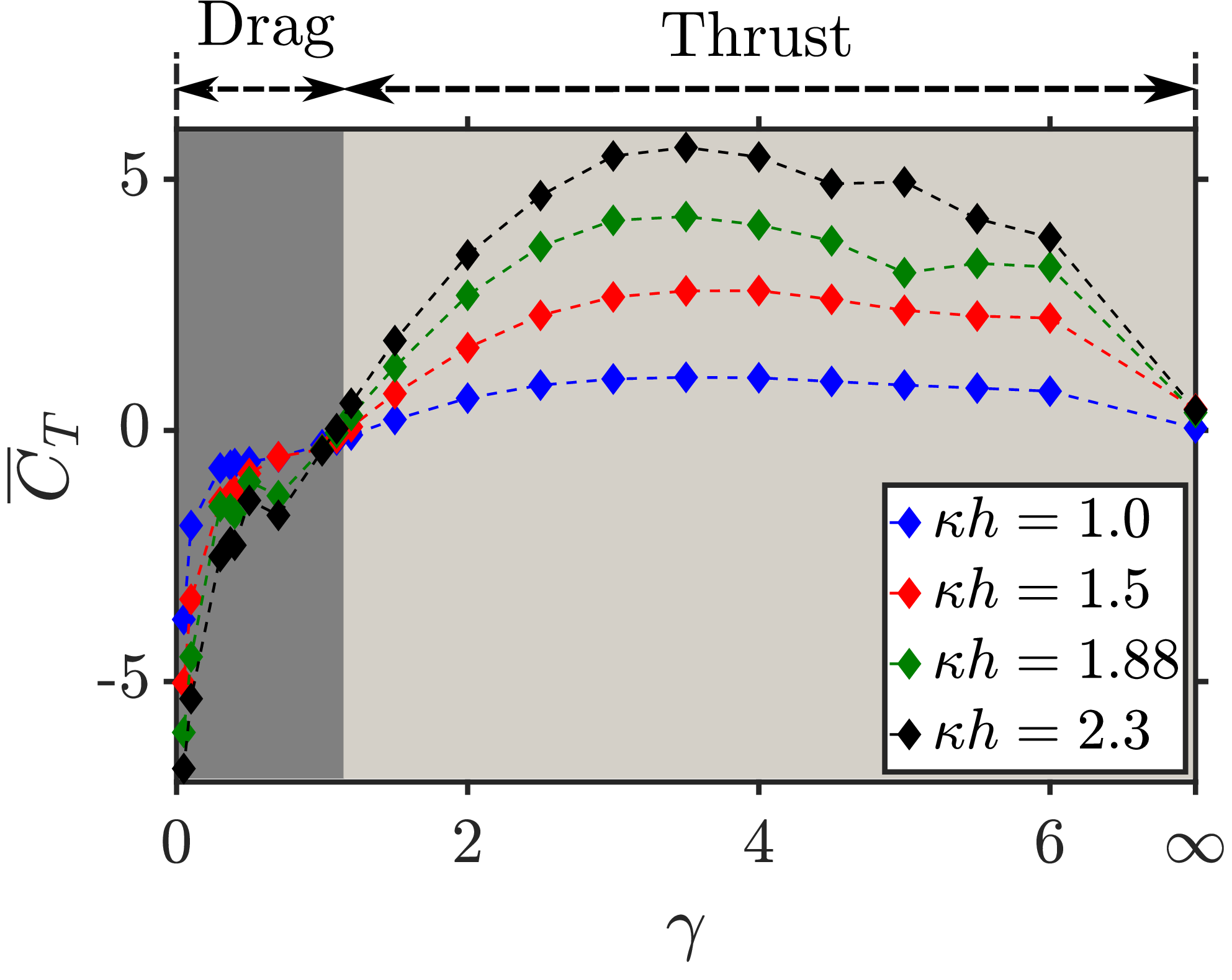}
			\caption{}
			\label{CT_gamma_alpha}
		\end{subfigure}
		\vspace{6pt}
		\begin{subfigure}{.5\textwidth}
			\centering
			\includegraphics[scale=0.19]{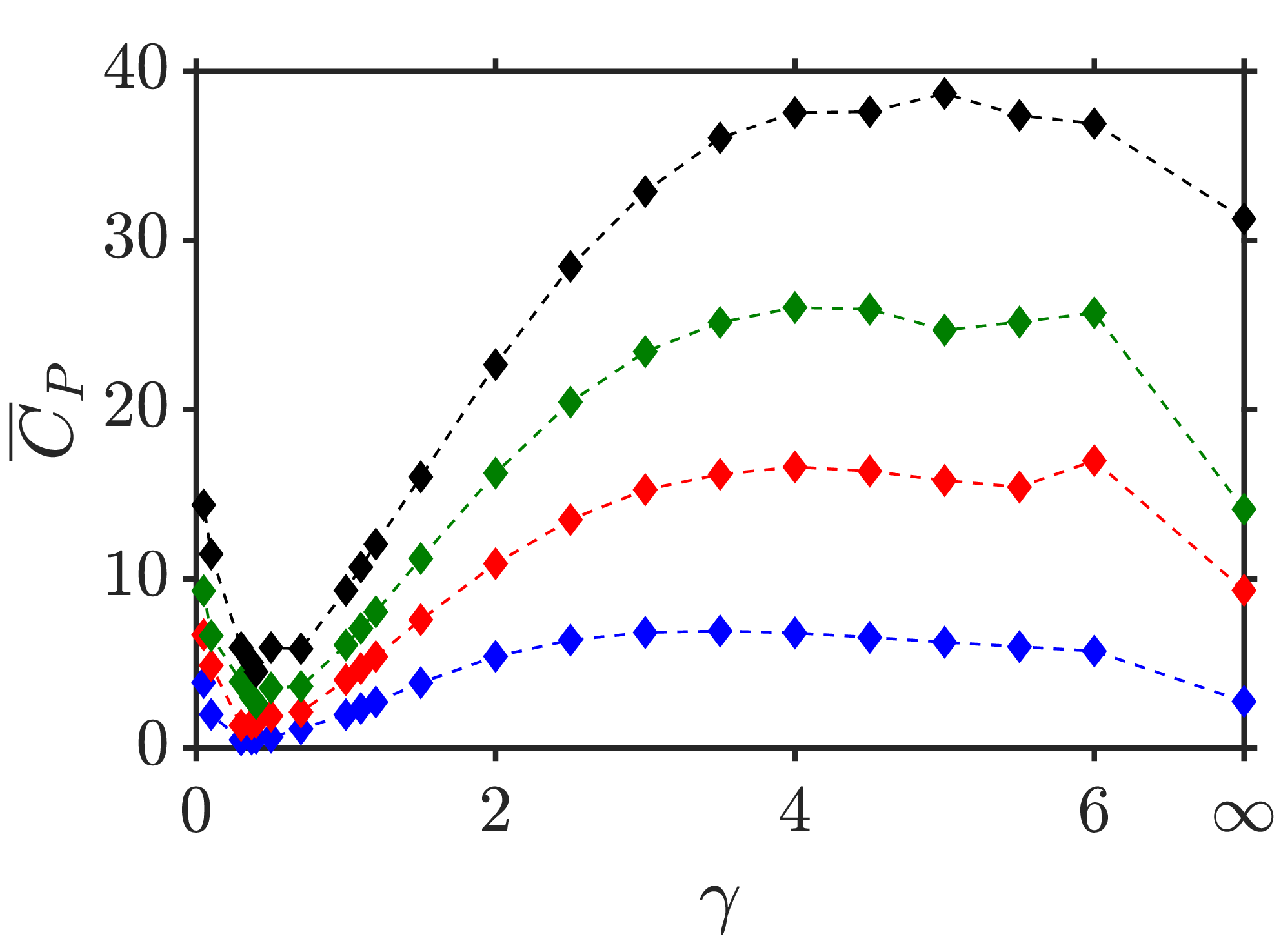}
			\caption{}
			\label{CP_gamma_alpha}
		\end{subfigure}
		\vspace{6pt}
        \begin{subfigure}{.5\textwidth}
			\centering
			\includegraphics[scale=0.19]{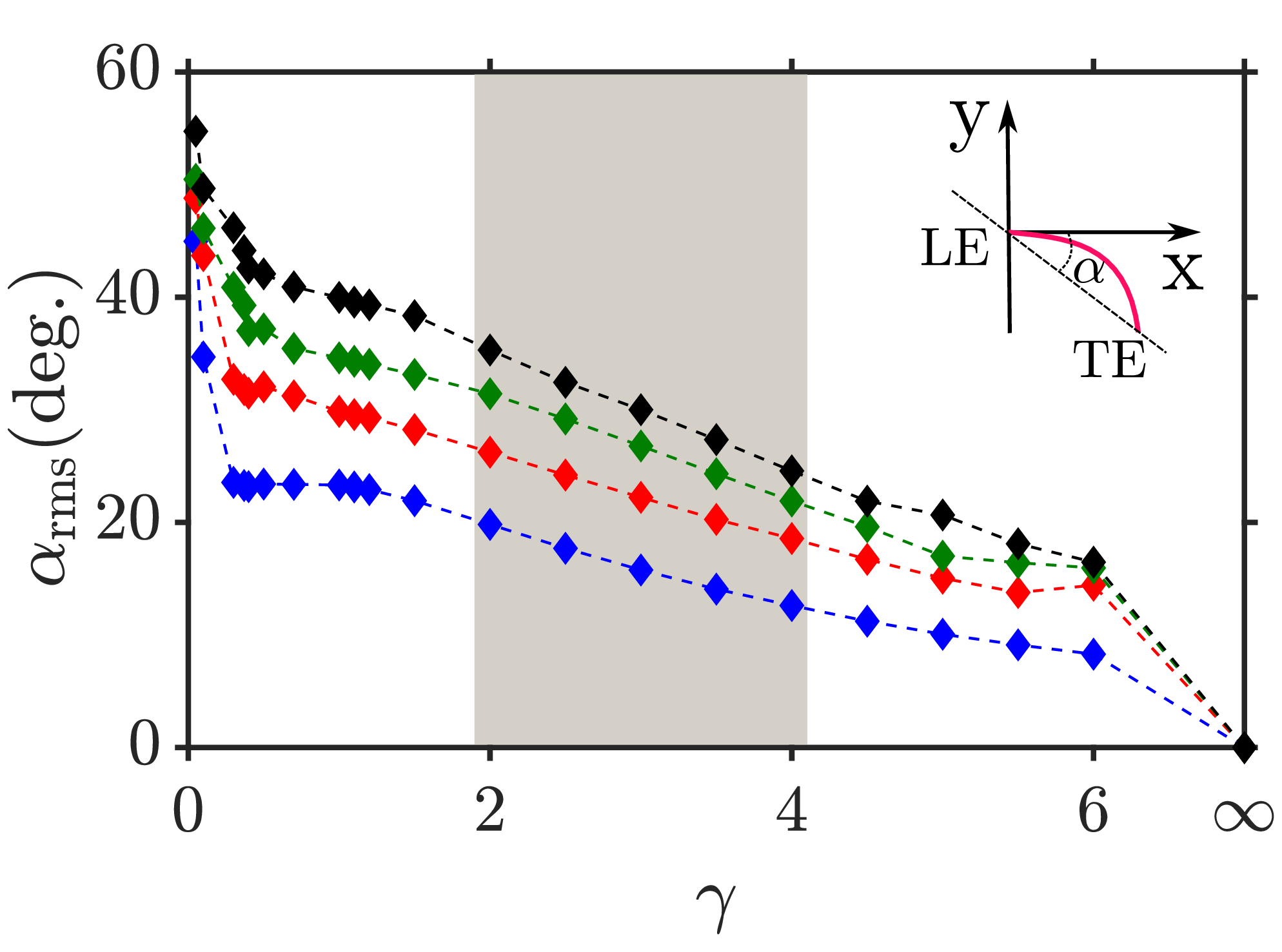}
			\caption{}
			\label{gamma_alpha}
		\end{subfigure}
        \vspace{6pt}
		\begin{subfigure}{.5\textwidth}
			\centering
			\includegraphics[scale=0.19]{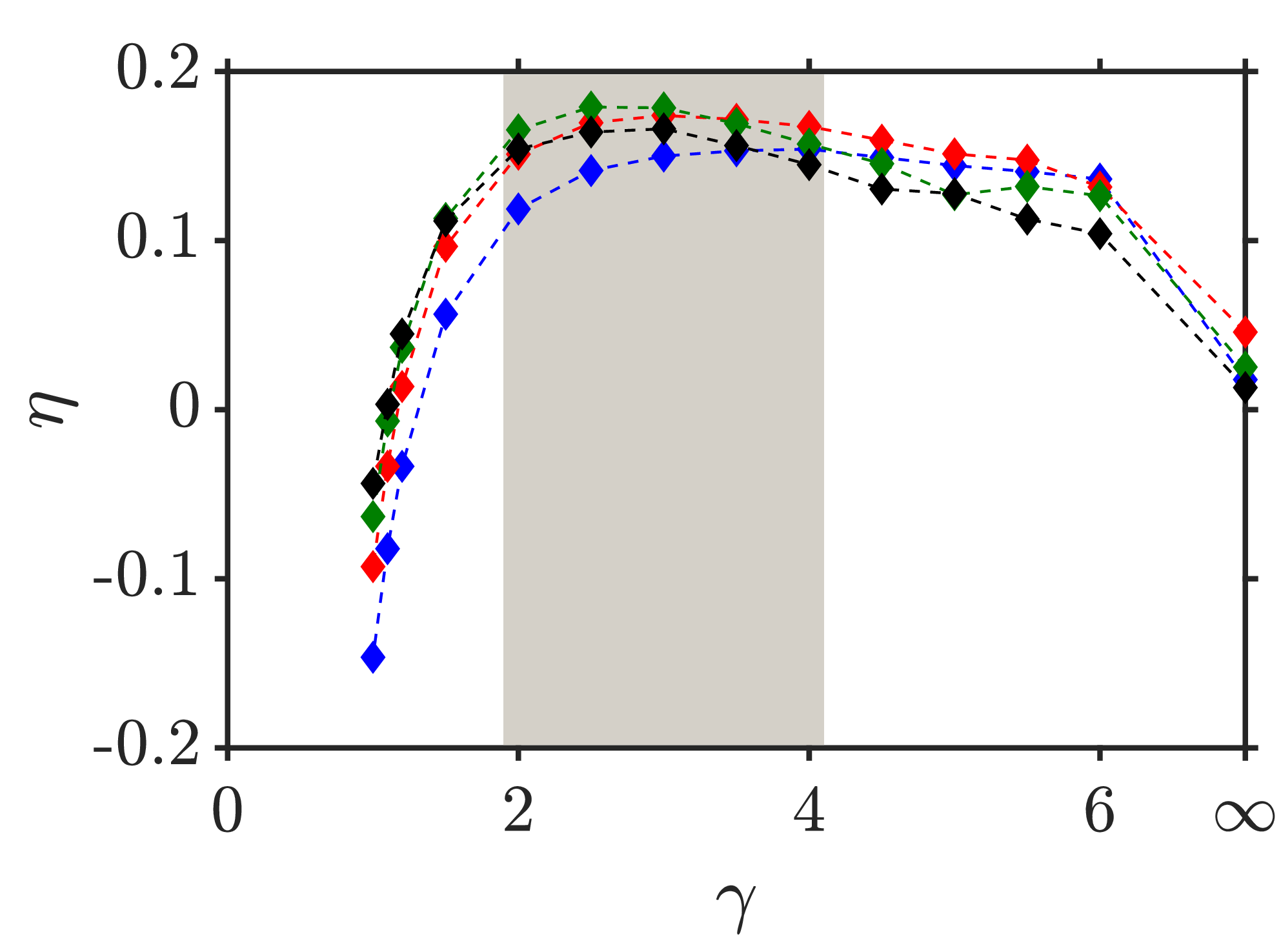}
			\caption{}
			\label{eta_gamma_alpha}
		\end{subfigure}
        \begin{subfigure}{1.0\textwidth}
			\centering
			\includegraphics[scale=0.155]{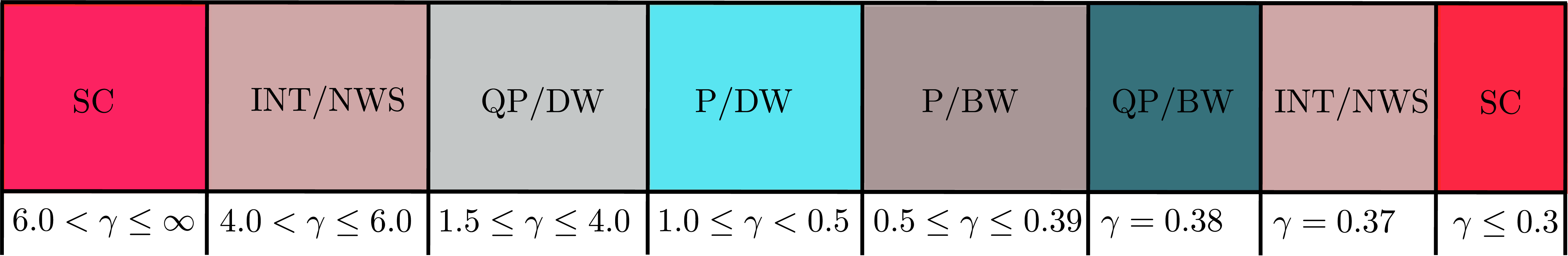}
			\caption{Vortex modes with change in $\gamma$ at $\kappa h=2.3.$}
			\label{vortex_modes}
		\end{subfigure}
		\caption{Variation in the (a) cycle-averaged thrust coefficient, (b) input power coefficient, (c) root-mean-square value of the induced passive pitch angle, (d) propulsive efficiency with respect to the foil flexibility, and (e) dynamical signatures and wake patterns for $\kappa h=2.3$; P/DW = periodic \& deflected wake, P/BW = periodic \& bifurcated wake, QP/DW = quasi-periodic \& deflected wake, QP/BW = quasi-periodic \& bifurcated wake, INT/NWS = intermittency \& near-wake switching and SC = sustained chaos.}
		\label{propulsive_efficiency}
	\end{figure}

The effect of flexibility on the propulsive performance has been  analysed here in a consolidated manner for different $\kappa h$s ($\kappa h=1.0,\,1.5,\,1.88\,\&\,2.3$) discussed in Section~4 and 5. With increasing flexibility (decreasing $\gamma$), initially, the thrust production increased for $\gamma \in (\infty, 3.5)$ (Fig.~\ref{propulsive_efficiency}(a)). However, it dropped rapidly for $\gamma \le 2.0$ and soon the system entered into a net drag-producing regime. This drag production could be attributed to the significant level of tip deformation of the body. This deformation was estimated in terms of the induced passive pitch angle, $\alpha$, computed as the angle between the $x$-axis and the straight line connecting the leading and trailing-edges (see Fig.~\ref{propulsive_efficiency}(c) inset). Note that there were no active pitching actuations and only heaving actuation was given at the leading-edge, while the rest of the foil oscillated passively. The root-mean-square (rms) value of $\alpha$ over a flapping cycle was quite high ($\alpha_{\rm rms}\in(22^{\rm o},39^{\rm o})$) for $\gamma \approx 1.5$ at all $\kappa h$ cases (Figure~\ref{propulsive_efficiency}(c)). The value of $\alpha_{\rm rms}$ increased even more as $\gamma$ was lowered further. The overall high drag generation could be attributed to the high deformation in the trailing-edge part of the structure. This effect was characterised by $\alpha_{\rm rms}$,  resulting in strong trailing-edge separations, and was the reason behind net drag production despite having relatively weaker LEVs at lower $\gamma$ values.

Similar to the trend of thrust production, the propulsive efficiency was optimal for $\gamma = 2.0 - 4.0$ for all $\kappa h$ cases (Fig.~\ref{propulsive_efficiency}(d)). This optimal propulsive efficiency was observed for $\alpha_{\rm rms} = 13^{\rm o}-35^{\rm o}$ for all the $\kappa h$ presented here, which belongs to the range reported by~\citet{fish1993power,anderson1998oscillating,zhang2010locomotion,hua2013locomotion}. The shaded region has been marked in Figs.~\ref{propulsive_efficiency}(c) and~\ref{propulsive_efficiency}(d) to indicate the range of maximum efficiency. 
Evidently, the inclusion of flexibility in the system enhanced the propulsive efficiency compared to the rigid configuration.  In the optimal efficiency range of $\gamma = 2.0 - 4.0$,  the flow-field was not irregular/chaotic but quasi-periodic for $\kappa h = 2.3$. The dynamical signatures and their corresponding wake patterns/vortex modes observed for $\kappa h=2.3$ have been summarised in Fig.~\ref{propulsive_efficiency}e. For $\kappa h$ values of $1.0,\,1.5\,\&\,1.88$, in the optimum propulsive efficiency of $\gamma$ $\approx2.0-4.0$, the flow-field was perfectly  periodic (as shown in Fig.~3 of the supplementary document). Thus one can say that a range of $\gamma$ of $2.0-4.0$ would be a good choice for an efficient propulsive design without a serious compromise on the periodicity of the flow-field.

\section{Conclusions} \label{sec:conclusions}

The effect of chord-wise flexibility on the dynamical transitions of the wake behind a fully flexible flapper was investigated using an IBM-based in-house FSI solver. The results were compared with the rigid counterpart. The rigid system exhibited a quasi-periodic intermittency route to chaos and a robust chaotic state was observed at a high $\kappa h$ value. With the introduction of flexibility, the system came out of the chaotic state and gradually became ordered. Perfectly periodic behaviour was seen for a range of flexibility during which a variety of wake patterns were observed. The most noticeable was a symmetrically bifurcated wake. At higher flexibilities, the order was lost and the system returned to chaotic dynamics again. The interactions of the vortices during the  transitions were tracked. The various wake patterns and their associated flow-dynamics with change in flexibility  were compared with its rigid counterpart. %Structural behaviour recorded in terms of the deformation profile and trajectories of the trailing-edge exhibited similar dynamics to that of the flow-field and the aerodynamic loads. 
A narrow range of flexibility was identified where optimum propulsive performance was observed. 
%This study focused strongly on the underlying flow-field interaction mechanisms with the corresponding dynamical signatures of the wake. %That optimal  flexibility can inhibit chaos and reinstate periodicity in the wake has been revealed for the first time in this study. 
The novelty of the study lies 
in scrutinising the nonlinear dynamical states/bifurcations  whose presence is unique  in the nonlinear system  in question. Also, optimal  flexibility inhibiting chaos  and reinstate periodicity in the wake through a specific dynamical route has been reported for the first time, to the best of the authors' knowledge. The study also concluded that fully flexible bio-mimetic devices could be operated at significantly high ranges of amplitude and frequencies without any serious compromise with the periodicity/regularity, by carefully choosing the level of flexibility of the system.
	
\section{Declaration of interests}
The authors report no conflict of interest.

\bibliographystyle{elsarticle-harv} 
\bibliography{cas-refs}

\end{document}